\documentclass[aps,prd,11pt,a4paper,nofootinbib,superscriptaddress,twoside]{revtex4-1}
\pdfoutput = 1

\usepackage{xspace,amsmath,amssymb,color,float,cancel,xcolor}
\usepackage{graphicx,tensor,slashed,paralist,cases,mathrsfs}

\usepackage[colorlinks=true,urlcolor=cerisepink,linkcolor=red,citecolor=blue,linktocpage=true]{hyperref}

\definecolor{cerisepink}{rgb}{0.93, 0.23, 0.51}
\definecolor{ceruleanblue}{rgb}{0.16, 0.32, 0.75}

\newcommand{\ovr}{\overline}
\newcommand{\asigv}{\langle \sigma v_{\mathrm{rel}} \rangle}

\newcommand{\lagr}{\mathcal{L}}
\newcommand{\lagrsm}{\mathcal{L}_{\mathrm{SM}}}
\newcommand{\vo}{v_0}
\newcommand{\lhep}{\texttt{LanHEP}\xspace}

\newcommand{\micro}{\texttt{micrOMEGAs}\xspace}
\newcommand{\lcalc}{\texttt{LUXCalc}\xspace}
\newcommand{\gamLike}{\texttt{gamLike}\xspace}

\begin{document}

\title{Combined analysis of effective Higgs portal dark matter models}

\author{Ankit Beniwal}
\email[Electronic address: ]{ankit.beniwal@adelaide.edu.au}
\affiliation{ARC Center of Excellence for Particle Physics at the Terascale \& CSSM, Department of Physics, University of Adelaide, Adelaide, South Australia 5005, Australia}

\author{Filip Rajec}
\email[Electronic address: ]{filip.rajec@adelaide.edu.au}
\affiliation{ARC Center of Excellence for Particle Physics at the Terascale \& CSSM, Department of Physics, University of Adelaide, Adelaide, South Australia 5005, Australia}

\author{Christopher Savage} 
\email[Electronic address: ]{chris@savage.name}
\affiliation{Nordita, KTH Royal Institute of Technology and Stockholm University, Stockholm, Sweden}
\affiliation{Department of Physics \& Astronomy, University of Utah, Salt Lake City, Utah, USA}

\author{Pat Scott}
\email[Electronic address: ]{p.scott@imperial.ac.uk}
\affiliation{Department of Physics, Imperial College London, Blackett Laboratory, Prince Consort Road, London SW7 2AZ, United Kingdom} 

\author{Christoph Weniger}
\email[Electronic address: ]{c.weniger@uva.nl}
\affiliation{GRAPPA, University of Amsterdam, Science Park 904, 1090 GL Amsterdam, Netherlands}

\author{Martin White}
\email[Electronic address: ]{martin.white@adelaide.edu.au}
\affiliation{ARC Center of Excellence for Particle Physics at the Terascale \& CSSM, Department of Physics, University of Adelaide, Adelaide, South Australia 5005, Australia}

\author{Anthony G. Williams}
\email[Electronic address: ]{anthony.williams@adelaide.edu.au}
\affiliation{ARC Center of Excellence for Particle Physics at the Terascale \& CSSM, Department of Physics, University of Adelaide, Adelaide, South Australia 5005, Australia}

\date{\today}                              

\begin{abstract}
We combine and extend the analyses of effective scalar, vector, Majorana and Dirac fermion Higgs portal models of dark matter (DM), in which DM couples to the Standard Model (SM) Higgs boson via an operator of the form $\mathcal{O}_{\textrm{DM}}\, H^\dagger H$. For the fermion models, we take an admixture of scalar $\overline{\psi} \psi$ and pseudoscalar $\overline{\psi} i\gamma_5 \psi$ interaction terms. For each model, we apply constraints on the parameter space based on the Planck measured DM relic density and the LHC limits on the Higgs invisible branching ratio. For the first time, we perform a consistent study of the indirect detection prospects for these models based on the WMAP7/Planck observations of the cosmic microwave background, a combined analysis of 15 dwarf spheroidal galaxies by Fermi-LAT and the upcoming Cherenkov Telescope Array (CTA). We also perform a correct treatment of the momentum-dependent direct search cross section that arises from the pseudoscalar interaction term in the fermionic DM theories. We find, in line with previous studies, that current and future direct search experiments such as LUX and XENON1T can exclude much of the parameter space, and we demonstrate that a joint observation in both indirect and direct searches is possible for high mass weakly interacting massive particles. In the case of a pure pseudoscalar interaction of a fermionic DM candidate, future gamma-ray searches are the only class of experiment capable of probing the high mass range of the theory.
\end{abstract}

\maketitle


\section{Introduction}
Evidence for dark matter (DM) in our Universe has been accumulating for decades \cite{1937ApJ86217Z}. As none of the particles of the Standard Model (SM) adequately explains a wide range of astrophysical observations of DM, we are forced to look for viable candidates in theories that lie beyond the SM (BSM). It has been noted that a GeV-scale DM candidate can accurately reproduce the observed relic density of DM provided that it has a weak interaction cross section, leading to the so-called ``weakly interacting massive particle'' (WIMP) miracle \cite{Bergstrom:2000pn,Bertone:2004pz}.

Within the WIMP scenario, there are typically multiple complementary experimental probes that exploit the couplings between the WIMP and SM fermions/gauge bosons. In addition, there are two primary theoretical approaches for specifying the SM-WIMP couplings. One is the top-down approach, where a WIMP is included amongst the new particle content of a BSM theory at high energies, and one can derive the WIMP interactions from the details of this new theory (e.g., supersymmetry with a neutralino WIMP). The other is the bottom-up approach in which effective field theories (EFT) are constructed from the lowest-dimensional operators allowed in a weak-scale Lagrangian. In this approach, one assumes a DM-SM interaction Lagrangian of the form
\begin{equation}
\lagr_{\textrm{int}} \supset \Lambda^{-n} \, \mathcal{O}_{\textrm{DM}} \,\mathcal{O}_{\textrm{SM}},
\end{equation}
where $\Lambda$ is the EFT cutoff scale and $\mathcal{O}_{\textrm{DM}}$ ($\mathcal{O}_{\textrm{SM}}$) are the DM (SM) operators that are singlets under the SM gauge groups. An EFT ceases to be valid when the momentum transfer in an experimental process of interest approaches the interaction mediator mass. An option that has generated much interest in the literature is that in which the DM interacts with SM particles via a Higgs portal \cite{Patt:2006fw,Kim:2006af,Barger:2007im,Kim:2008pp,Yu:2014pra,Kanemura:2010sh,Low:2011kp,Englert:2011aa,Batell:2011pz,Fox:2011pm,Lebedev:2011iq,Mambrini:2011ik,Pospelov:2011yp,Englert:2011yb,Kamenik:2012hn,Cline:2012hg,Farzan:2012hh,Tsai:2013bt,Greljo:2013wja,Walker:2013hka,Carpenter:2013xra,Dienes:2013xya,Busoni:2013lha,Craig:2014lda,Ko:2015eaa,Baek:2014jga,Ghorbani:2014qpa,Ko:2014gha,Duerr:2015mva,Han:2015dua,Duerr:2015aka,Queiroz:2014pra}. This is motivated partly by the simplicity of the model in terms of the required BSM particle content, and partly by the fact that the operator $H^\dagger H$ is one of the two lowest-dimensional gauge-invariant operators that one can write in the SM (the other being the hypercharge field strength tensor $B_{\mu\nu}$). The recent discovery of the Higgs boson by the ATLAS \cite{Aad:2012tfa} and CMS \cite{Chatrchyan:2012xdj} experiments of the Large Hadron Collider gives us an extra experimental window into the possible connections between the Higgs boson and DM. 

In this paper, we revisit four simple effective field theories for Higgs portal dark matter, namely those in which the DM is assumed to be a scalar, vector or Majorana/Dirac fermion. For the fermion models, one may consider parity-conserving scalar couplings $(\mathcal{O}_{\mathrm{DM}} \propto \ovr{\psi} \psi)$ and/or parity-violating pseudoscalar couplings $(\mathcal{O}_{\mathrm{DM}}  \propto \ovr{\psi} i\gamma_5 \psi)$. There are a few omissions in the literature that make a new combined study of these scenarios worthwhile. The first is that most studies have not properly summarized the impacts of current and future indirect search experiments on the parameter space of the models. In some cases (e.g., fermionic dark matter with scalar interactions) this is because the indirect signatures are velocity suppressed, while in others it has been assumed that the ability of direct detection to probe much of the parameter space renders indirect detection less important. We argue instead that it is vital to establish which regions of the model space are visible in multiple experiments, since this might ultimately present us with a much deeper understanding of the DM problem (e.g., one could imagine using a more detailed knowledge of the particle physics obtained in one experiment to better measure astrophysical factors such as the DM halo distribution, as proposed in Ref.~\cite{Baltz:2006fm}). With the approaching availability of the Cherenkov Telescope Array (CTA), it is an opportune moment to explore its expected reach. It is also the case that studies of the direct detection constraints on fermionic models have not always treated the momentum suppression of the direct detection cross section correctly in the case of a pseudoscalar coupling. We remedy this in our study by rederiving the limits from the XENON100 and LUX experiments using a dedicated code that includes an accurate treatment of the momentum-dependent scattering cross sections. However, in such cases, the results do not change significantly and the impact of the direct detection experiments remains negligible. Finally, we perform the same comprehensive study across all of the four models, and thus are able to present a consistent set of detailed results for the first time.

Our paper can be viewed as an extension of the scalar singlet DM model study presented in Ref.~\cite{Cline:2013gha}. We leave consideration of UV completions of EFTs for future work. A global fit of the same model was presented in Ref.~\cite{Cheung:2012xb}. The scalar, vector and Majorana fermion Higgs portal models we consider have previously been studied in Ref.~\cite{Djouadi:2011aa} in light of the WMAP and XENON100, as well as the Higgs invisible width and XENON1T prospects. Current LHC constraints from a CMS vector boson fusion analysis and LHC monojet and mono-Z analyses are shown to be very weak for the scalar singlet DM model in Ref.~\cite{Endo:2014cca}. Monojet constraints on all Higgs portal models (scalar, fermion and vector) are shown to be weak in Ref.~\cite{Djouadi:2012zc}. Constraints on the parameter space of the scalar model from perturbativity and electroweak vacuum stability in the early Universe  were imposed first in Ref.~\cite{Kahlhoefer:2015jma} and recently in Ref.~\cite{Han:2015hda}. Limits on the scalar model from gamma-ray line searches in the Higgs resonance region  were considered recently in Ref.~\cite{Feng:2014vea}. The LUX limits on  the scalar DM model were first considered in Ref.~\cite{Urbano:2014hda}. The same paper also evaluated the limits from antiproton data, which were shown to be important in the region of the Higgs resonance and competitive with the LUX limits at higher DM masses.  For the fermion models, the corresponding antiproton limits in scalar interactions are weak due to the velocity suppression of the annihilation cross section $\sigma v_{\textrm{rel}}$. In all cases, the results are strongly dependent on the propagation model. A combination of the parity-conserving and parity-violating terms in fermion models were first considered in Ref.~\cite{LopezHonorez:2012kv} and recently in Refs.~\cite{Fedderke:2014wda,Matsumoto:2014rxa} where it was noted that the parity-violating term can significantly generate a parity-conserving coupling after electroweak symmetry breaking (EWSB). 

We aim to provide a complete description of the current and projected future limits on the parameter space of the effective scalar, vector, Majorana and Dirac fermion models in this paper. For each model, we apply currently available constraints from the dark matter relic density, Higgs invisible width, indirect and direct detection limits. For the indirect searches, the current limits we impose come from the WMAP 7-year observations of the cosmic microwave background (WMAP7) and a combined analysis of 15 dwarf galaxy observations by Fermi-LAT; for projected future limits, we use the Planck polarization data, projected improvements in Fermi dwarf galaxy observations and prospects for the planned CTA. Our direct detection data are taken from XENON100 (2012) and LUX (2013). 

The paper is organized as follows. In Sec.~\ref{sec2}, we introduce our four Higgs portal models based on the DM spin. Section \ref{sec3} contains a detailed description of the model constraints and their implementations. Our model results and their prospects for detection at current and/or future based experiments are presented in Sec.~\ref{sec4}. Our conclusions are given in Sec.~\ref{sec5}. A detailed derivation of the physical mass basis required to understand the fermionic EFTs after EWSB is given in the Appendix.

\section{Models}\label{sec2}
The DM fields are assumed to be SM gauge singlets. We consider four cases for the spin of the DM: scalar $(S)$, vector $(V_\mu)$, Majorana $(\chi)$ and Dirac $(\psi)$ fermion. Lagrangians for the corresponding DM fields, invariant under the symmetries of the SM, are\footnote{We adopt our scalar and vector models from Ref.~\cite{Kanemura:2010sh} with the following substitutions for \texttt{Scalar}, $\phi \rightarrow S$, $M_S \rightarrow \mu_S$, $c_S \rightarrow \lambda_{hS}$, $d_S \rightarrow \lambda_S$, and \texttt{Vector}, $V_{\mu\nu} \rightarrow W_{\mu\nu}$, $M_V \rightarrow \mu_V$, $c_V \rightarrow \lambda_{hV}$, $d_V \rightarrow \lambda_V$. The Dirac and Majorana models are taken instead from Ref.~\cite{Fedderke:2014wda} with the following substitutions for \texttt{Dirac}, $\chi \rightarrow \psi$, $M_0 \rightarrow \mu_\psi$, $1/\Lambda \rightarrow -\lambda_{h\psi}/\Lambda_\psi$, and \texttt{Majorana}, $\chi \rightarrow \chi/\sqrt{2}$, $M_0 \rightarrow \mu_\chi$, $1/\Lambda \rightarrow -\lambda_{h\chi}/\Lambda_\chi$.}
\begin{align}
  \lagr_{S} &= \lagrsm + \frac{1}{2} (\partial_\mu S)(\partial^\mu S) - \frac{1}{2} \mu_S^2 S^2 - \frac{1}{4!}\lambda_S S^4 - \frac{1}{2} \lambda\indices{_{hS}} \, S^2 H^\dagger H, \label{scalar} \\
  \lagr_{V} &= \lagrsm -\frac{1}{4} W\indices{_{\mu\nu}} W\indices{^{\mu\nu}} + \frac{1}{2} \mu_V^2 V_\mu V^\mu -\frac{1}{4!} \lambda_{V} (V_\mu V^\mu)^2 + \frac{1}{2} \lambda_{hV} \,  V_\mu V^\mu H^\dagger H, \label{vector} \\
  \lagr_{\chi} &= \lagrsm + \frac{1}{2} \ovr{\chi} \left(i\slashed{\partial} - \mu_\chi \right)\chi - \frac{1}{2}\frac{\lambda\indices{_{h\chi}}}{\Lambda_\chi} \, \Big(\cos\theta \, \ovr{\chi}\chi + \sin\theta \, \ovr{\chi}i\gamma_5 \chi \Big) H^\dagger H, \label{majorana} \\
  \lagr_{\psi} &= \lagrsm + \ovr{\psi} \left(i \slashed{\partial} - \mu_\psi \right) \psi - \frac{\lambda\indices{_{h\psi}}}{\Lambda_\psi} \, \Big(\cos\theta \, \ovr{\psi}\psi + \sin\theta \, \ovr{\psi}i\gamma_5 \psi \Big) H^\dagger H, \label{dirac} 
\end{align}
where $\lagrsm$ is the SM Lagrangian, $W\indices{_{\mu\nu}} \equiv \partial_\mu V_\nu - \partial_\nu V_\mu$ is the field-strength tensor of the vector field $V_\mu$ and $H$ is the SM Higgs doublet. The parameter $\theta$ determines the type of interaction between DM and the Higgs field: $\cos\theta = 1$ corresponds to a pure scalar interaction and $\cos\theta = 0$ corresponds to a pure pseudoscalar interaction.

DM particles are required to be stable on cosmological time scales. For the models introduced above, their stability is guaranteed by imposing an assumed $\mathbb{Z}_2$ symmetry: $X \rightarrow -X$ for $X \in (S,V_\mu,\chi,\psi)$. Under the $\mathbb{Z}_2$ symmetry, the DM fields $(S,V_\mu,\chi,\psi)$ are assumed to be \emph{odd} while the SM fields are assumed to be \emph{even}. The decay of a DM particle into SM particles is therefore forbidden. Furthermore, the imposed symmetry also prohibits linear and cubic terms in the scalar field Lagrangian as well as the kinetic mixing terms in the vector field-strength tensor $W_{\mu\nu}$. As an explicit bare mass term for the DM field is allowed by the $\mathbb{Z}_2$ symmetry, we have included it in our Lagrangians above for completeness. 

The scalar model in Eq.~\eqref{scalar} is the simplest extension of the SM and was first introduced 30 years ago \cite{Silveira1985136}. It was later studied by the authors of Refs.~\cite{PhysRevD.50.3637,Burgess:2000yq}. It is both theoretically and phenomenologically satisfactory as long as the $\mathbb{Z}_2$ symmetry remains unbroken. It is renormalizable and hence valid up to high energy scales as long as the Landau pole is not reached. The vector model in Eq.~\eqref{vector} is very simple, compact and appears renormalizable due to the presence of dimension-2 and dimension-4 operators only. However, in reality, it is nonrenormalizable and violates unitarity (in a similar fashion to the four-fermion interaction of the preelectroweak theory). Therefore, it is an effective model which needs to be UV completed. UV complete models will be the subject of future work and hence are considered to be beyond the scope of this study. Simple UV completions can be found in Refs.~\cite{Baek:2012se,Duch:2015jta}. The Majorana and Dirac fermion models in Eqs.~\eqref{majorana} and \eqref{dirac} respectively are also nonrenormalizable, and a suggested UV completion can be found in Ref.~\cite{Baek:2011aa}.

After EWSB, the SM Higgs doublet acquires a vacuum expectation value (VEV). In the unitary gauge, the SM Higgs doublet reduces to 
\begin{equation}
  H = \frac{1}{\sqrt{2}}
  \begin{pmatrix}
    0 \\
    v_0 + h
  \end{pmatrix}
\end{equation}
where $h$ is the physical SM Higgs field and $v_0 = 246.22$\,GeV is the SM Higgs VEV. Consequently, the $H^\dagger H$ term in the above Lagrangians expands to 
\begin{equation}
  H^\dagger H = \frac{1}{2} \vo^2 + \vo h + \frac{1}{2} h^2
\end{equation}
which generates mass and interaction terms for the DM fields $X \in (S,V_\mu,\chi,\psi)$. After substituting the expanded expression for $H^\dagger H$ into the above Lagrangians, we find
\begin{align}
  \lagr_S &= \mathcal{L}_{\textrm{SM}} + \frac{1}{2} (\partial_\mu S)(\partial^\mu S) - \frac{1}{2} m_S^2 S^2 -\frac{1}{4!}\lambda_S S^4 - \frac{1}{2} \lambda\indices{_{hS}} \, S^2 \left(\vo h + \frac{1}{2}h^2 \right), \\
  \lagr_{V} &= \lagrsm -\frac{1}{4} W\indices{_{\mu\nu}} W\indices{^{\mu\nu}} + \frac{1}{2} m_V^2 V_\mu V^\mu - \frac{1}{4!} \lambda_{V} (V_\mu V^\mu)^2 + \frac{1}{2} \lambda_{hV} \,  V_\mu V^\mu \left(\vo h + \frac{1}{2}h^2 \right), \\
  \lagr_{\chi} &= \lagrsm + \frac{1}{2} \ovr{\chi} i\slashed{\partial}\chi - \frac{1}{2} \left[\mu_\chi \ovr{\chi}\chi + \frac{1}{2}\frac{\lambda_{h\chi}}{\Lambda_\chi}\vo^2 \Big(\cos\theta \, \ovr{\chi}\chi + \sin\theta \, \ovr{\chi}i\gamma_5 \chi \Big) \right] \notag \\
  &\hspace{5cm} -\frac{1}{2}\frac{\lambda\indices{_{h\chi}}}{\Lambda_\chi} \, \Big(\cos\theta \, \ovr{\chi}\chi + \sin\theta \, \ovr{\chi}i\gamma_5 \chi \Big) \left(\vo h + \frac{1}{2}h^2 \right) , \label{majorana2} \\
  \lagr_{\psi} &= \lagrsm + \ovr{\psi} i\slashed{\partial}\psi - \left[\mu_\psi \ovr{\psi}\psi + \frac{1}{2}\frac{\lambda_{h\psi}}{\Lambda_\psi}\vo^2 \Big(\cos\theta \, \ovr{\psi}\psi + \sin\theta \, \ovr{\psi}i\gamma_5 \psi \Big) \right] \notag \\
  &\hspace{5cm} -\frac{\lambda\indices{_{h\psi}}}{\Lambda_\psi} \, \Big(\cos\theta \, \ovr{\psi}\psi + \sin\theta \, \ovr{\psi}i\gamma_5 \psi \Big) \left(\vo h + \frac{1}{2}h^2 \right), \label{dirac2} 
\end{align}
where the physical masses of the scalar and vector fields are
\begin{equation}
  \emph{Scalar:} \quad m_S^2 = \mu_S^2 + \frac{1}{2}\lambda\indices{_{hS}} \vo^2, \quad
  \emph{Vector:} \quad m_V^2 = \mu_V^2 + \frac{1}{2} \lambda\indices{_{hV}} \vo^2. 
\end{equation}
For the Majorana and Dirac models, we have non-mass-type contributions that are purely quadratic in the DM fields when $\sin\theta \neq 0$. As explained in the Appendix, it is convenient to perform a chiral rotation and field redefinition after EWSB to have a properly defined real mass. The chiral rotation of the fields is given by
\begin{align}\label{chiral-transform}
\begin{split}
  \emph{Majorana:} \quad \chi &\rightarrow \exp(i\gamma_5 \alpha/2) \,\chi, \quad \ovr{\chi} \rightarrow \ovr{\chi}\exp(i\gamma_5 \alpha/2), \\
  \emph{Dirac:} \quad \psi &\rightarrow \exp(i\gamma_5 \alpha/2) \,\psi, \quad \ovr{\psi} \rightarrow \ovr{\psi}\exp(i\gamma_5 \alpha/2) ,
\end{split}
\end{align}
where $\alpha$ is a real parameter, independent of space-time coordinates. 

After substituting the redefined, rotated Majorana and Dirac fields from Eq.~\eqref{chiral-transform} into Eqs.~\eqref{majorana2} and \eqref{dirac2}, we demand that the coefficients of the terms $\ovr{\chi}i\gamma_5 \chi$ and $\ovr{\psi}i\gamma_5 \psi$ vanish in order to go to the real mass basis. Consequently, this imposes a constraint on the allowed values of the parameter $\alpha$. Using our derived results for the Dirac and Majorana models in the Appendix, the allowed values of the parameter $\alpha$ are
\begin{equation*}
  \tan\alpha = \left(-\frac{1}{2} \frac{\lambda_{h\chi,h\psi}}{\Lambda_{\chi,\psi}}\,\vo^2 \sin\theta \right) \left(\mu_{\chi,\psi} + \frac{1}{2}\frac{\lambda_{h\chi,h\psi}}{\Lambda_{\chi,\psi}} \, \vo^2 \cos\theta \right)^{-1}.  
\end{equation*}
Consequently, the post-EWSB Lagrangians for the Majorana and Dirac models are
\begin{align}
  \lagr_\chi &= \mathcal{L}_{\textrm{SM}} + \frac{1}{2}\ovr{\chi} i\slashed{\partial}\chi - \frac{1}{2}m_\chi \ovr{\chi}\chi - \frac{1}{2}\frac{\lambda_{h\chi}}{\Lambda_\chi} \Big[\cos\xi \, \ovr{\chi} \chi + \sin\xi \, \ovr{\chi} i\gamma_5 \chi \Big] \left(\vo h + \frac{1}{2} h^2 \right), \label{majoranaFinal} \\
  \lagr_\psi &= \mathcal{L}_{\textrm{SM}} + \ovr{\psi} i\slashed{\partial}\psi - m_\psi \ovr{\psi}\psi -\frac{\lambda_{h\psi}}{\Lambda_\psi} \Big[\cos\xi \, \ovr{\psi} \psi + \sin\xi \, \ovr{\psi} i\gamma_5 \psi \Big] \left(\vo h + \frac{1}{2} h^2 \right), \label{diracFinal}
\end{align} 
where $\xi \equiv \theta + \alpha$ and
\begin{align*}
  \cos\xi &= \frac{\mu_{\chi,\psi}}{m_{\chi,\psi}} \left(\cos\theta + \frac{1}{2}\frac{\lambda_{h\chi,{h\psi}}}{\Lambda_{\chi,\psi}} \frac{\vo^2}{\mu_{\chi,\psi}} \right), \quad \sin\xi = \frac{\mu_{\chi,\psi}}{m_{\chi,\psi}}\sin\theta, \\
  m_{\chi,\psi} &= \sqrt{\left(\mu_{\chi,\psi} + \frac{1}{2}\frac{\lambda_{h\chi,h\psi}}{\Lambda_{\chi,\psi}} \vo^2 \cos\theta \right)^2 + \left(\frac{1}{2}\frac{\lambda_{h\chi,h\psi}}{\Lambda_{\chi,\psi}}\vo^2 \sin\theta \right)^2}.
\end{align*}
The parameters $m_\chi$ and $m_\psi$ are the physical masses of the Majorana $(\chi)$ and Dirac $(\psi)$ DM fields respectively.

\section{Constraints}\label{sec3}
For the scalar and vector models after EWSB, the remaining free model parameters are the DM mass $m_{S,V}$ and its coupling $\lambda_{hS,hV}$ with the SM Higgs boson.\footnote{For our analysis, we ignore the quartic self-coupling $\lambda_{S,V}$ since it plays no observable role in the DM phenomenology. However, it is important when the constraints from electroweak vacuum stability and model perturbativity are imposed (see, e.g., Refs.~\cite{Kahlhoefer:2015jma,Gonderinger:2012rd,Gonderinger:2009jp,Kadastik:2011aa,Khan:2014kba,Lerner:2009xg,Xing:2011aa,EliasMiro:2011aa}.)} The Majorana and Dirac fermion models in Eqs.~\eqref{majoranaFinal} and \eqref{diracFinal} respectively yield three model parameters: $\cos\xi$, $m_{\chi,\psi}$ and $\lambda_{h\chi,h\psi}/\Lambda_{\chi,\psi}$. For simplicity, we choose to keep the parameter $\cos\xi$ fixed when imposing fermion model constraints. More specifically, we study the cases when $\cos\xi = 1$ (pure scalar interaction), $\cos\xi = 1/\sqrt{2}$ (equally mixed scalar-pseudoscalar interaction) and $\cos\xi = 0$ (pure pseudoscalar interaction). Hence, the  remaining free parameters in the fermion models are the DM mass $m_{\chi,\psi}$ and its coupling $\lambda_{h\chi,h\psi}/\Lambda_{\chi,\psi}$ with the SM Higgs boson.\footnote{In the model Lagrangians, the scalar/vector DM coupling $\lambda_{hS,hV}$ has mass dimension $M^0$ while the Majorana/Dirac fermion DM coupling $\lambda_{h\chi,h\psi}/\Lambda_{\chi,\psi}$ has mass dimension $M^{-1}$.}

As we impose identical sets of constraints on each model parameter space, we will instead refer to parameters $m_X$ and $\lambda_{hX}$ where $X \in (S,V,\chi,\psi)$. In situations where a given constraint is model specific, we impose those constraints on the model parameters directly. 

\subsection{Relic density}
Currently, the best-known value of the DM relic density (or abundance) comes from the Planck satellite measurement of the cosmic microwave background (CMB) temperature and lensing-potential power spectra \cite{Ade:2013zuv} 
\begin{equation}\label{relic}
  \Omega_{\textrm{DM}} h^2 = 0.1199 \pm 0.0027,
\end{equation}
where $\Omega_{\mathrm{DM}} \equiv \rho\indices{_{\textrm{DM}}}/\rho\indices{_{\textrm{c}}}$ is the fraction of the DM mass density relative to the critical density $\rho_c = 3H_0^2/8\pi G$, and the parameter $h = H_0/(100$\,km\,s$^{-1}$\,Mpc$^{-1})$ is the reduced Hubble constant. 

In general, WIMPs in thermal equilibrium in the early Universe that annihilate purely via the $s$-wave have a relic density that varies inversely with the velocity-averaged annihilation cross section $\asigv$ through
\begin{equation}
\label{relicDform}
\Omega_{\mathrm{DM}} h^2 \sim \frac{3 \times 10^{-27} \, \textrm{cm}^3\,\textrm{s}^{-1}}{\asigv}.
\end{equation}
Consequently, a smaller $\asigv$ produces an \emph{overabundance} of WIMPs, whereas a larger $\asigv$ produces an \emph{underabundance} of WIMPs in the Universe today. 

For our Higgs portal models, the model relic density is mostly determined by the Higgs-mediated $s$-channel annihilation into SM particles. A subdominant role is played by the annihilation into $hh$ through the direct $h^2 X^2$ vertex as well as the Higgs-mediated $t$-channel annihilation of $X$. Since $\asigv$ scales as the \emph{square} of the SM Higgs-DM coupling $\lambda_{hX}$, larger (smaller) values of $\lambda_{hX}$ lead to a suppressed (enhanced) relic density $\Omega_X h^2$ in agreement with Eq.~\eqref{relicDform}. 

We use the publicly available software \texttt{LanHEP3.2.0} \cite{Semenov:2014rea,Semenov:2008jy} to implement the scalar model in Eq.~\eqref{scalar}, vector model in Eq.~\eqref{vector} and the redefined, chiral-rotated Majorana and Dirac models in Eqs.~\eqref{majoranaFinal} and \eqref{diracFinal} respectively.\footnote{\lhep homepage: \url{http://theory.sinp.msu.ru/~semenov/lanhep.html}.} For the calculation of the relic density $\Omega_X h^2$, annihilation cross section $\sigma v_{\textrm{rel}}$ and the gamma-ray yields per annihilation, we use \texttt{micrOMEGAs3.6.9.2} \cite{Belanger:2013oya,Belanger:2014hqa} to automate calculations in each model.\footnote{\micro homepage: \url{https://lapth.cnrs.fr/micromegas/}.} In computing the relic density, we also take into account the annihilation of DM into virtual gauge bosons.

The Planck measured relic density in Eq.~\eqref{relic} restricts the allowed values of the SM Higgs-DM coupling. For each of our models, we scan along the DM mass axis and for each mass, find the coupling that gives the correct relic density using Brent's method \cite{Brent73book}.\footnote{Brent's method is a root-finding algorithm which combines the bisection method, the secant method and inverse quadratic interpolation.} These couplings are roots (or zeros) of the function
\begin{equation}\label{root-relic}
  f_\Omega(m_X,\lambda_{hX}) = \Omega_{\textrm{DM}}h^2 - \Omega_{X} h^2 ,
\end{equation}
where $\Omega_X h^2$ is the model relic density computed in \micro. For our analysis, we take $\Omega_{\textrm{DM}} h^2 = 0.1199$ corresponding to the central value of Eq.~\eqref{relic}. 

The possibility of a multicomponent dark sector where a given model constitutes a fraction of the total DM density is a strong one. To address this possibility, we define a relic abundance parameter 
\begin{equation}
  f_{\textrm{rel}} \equiv \frac{\Omega_X}{{\Omega_{\textrm{DM}}}},
\end{equation}
which quantifies the contribution of $X$ towards the total DM density in our Universe. For plotting purposes, we take $f_{\textrm{rel}}$ = 1, 0.1 and 0.01 corresponding to $X$ relic abundances of 100\%, 10\% and 1\% respectively.

At DM masses below $m_h/2$, we found multiple values of the coupling $\lambda_{hX}$ at fixed DM masses that were consistent with the Planck measured value. Naively, one expects a monotonic decrease in $\Omega_X h^2$ at larger values of $\lambda_{hX}$ due to an increased $\asigv$, as evident in Eq.~\eqref{relicDform}. However, $\asigv$ has an additional dependence on $\lambda_{hX}$ through the full Higgs boson width $\Gamma_h(\sqrt{s})$ as a function of the center-of-mass energy $\sqrt{s}$. Therefore, as $\lambda_{hX}$ increases, the increasing Higgs boson width counteracts the increase in $\asigv$ from $\lambda_{hX}$ alone and at some point outstrips it, such that $\asigv$ starts decreasing with increasing $\lambda_{hX}$, causing a minimum relic density $\Omega_{X} h^2$ for any given DM mass. When these features appear, we always found that only one root was consistent with the Higgs invisible width constraint and this is the one presented in our results. Indeed, the same feature was also seen in the preparation of Ref.~\cite{Cline:2013gha}, but was not explicitly discussed in the final manuscript. 

Furthermore, roots of Eq.~\eqref{root-relic} for $f_{\textrm{rel}} = 0.1, 0.01$ and DM masses below $m_h/2$ do not exist as the minimum relic density at these masses is larger than the value of the contour being drawn. Consequently, gaps in our relic density contours for $f_{\textrm{rel}} = 0.1, 0.01$ will appear in our plots. 

For DM masses above $m_h/2$, the relic density monotonically decreases with larger couplings $\lambda_{hX}$. This results in a single root (or coupling) at each DM mass that gives the Planck measured relic density. The Higgs invisible width $\Gamma_{\textrm{inv}} (h \rightarrow \ovr{X} X)$ vanishes in this region due to  the decay $h \rightarrow \ovr{X} X$ becoming kinematically forbidden. 

\subsection{Higgs invisible width}
When $m_X < m_h/2$, the decay $h \rightarrow \ovr{X} X$ is kinematically allowed. This contributes to the invisible width $(\Gamma_{\mathrm{inv}})$ of the SM Higgs boson. The LHC constraints on $\Gamma_{\textrm{inv}}$ continue to improve as precise measurements of the properties of the Higgs boson are shown to be increasingly in agreement with the SM expectations. 

An upper limit of 19\% on the SM Higgs invisible branching ratio $\mathcal{BR}(h \rightarrow \ovr{X} X)$ at $2\sigma$\,C.L. is obtained in Ref.~\cite{Belanger:2013xza} through combined fits to all Higgs production and decay channels probed by ATLAS, CMS and the Tevatron.\footnote{The Higgs invisible branching ratio is defined as $\mathcal{BR}(h \rightarrow \ovr{X} X) \equiv \Gamma_{\mathrm{inv}}/(\Gamma_{\mathrm{vis}} + \Gamma_{\mathrm{inv}})$ where $\Gamma_{\textrm{vis}}$ is the visible (SM) contribution to the SM Higgs width.} An additional projected limit of 5\% on the Higgs invisible branching ratio at no more than $1\sigma$\,C.L. can be further imposed if no additional Higgs decay is detected at the 14\,TeV LHC run after 300\,fb$^{-1}$ of luminosity data is collected \cite{Peskin:2012we}.

For our Higgs portal models, the Higgs invisible widths are given by \cite{Kanemura:2010sh,Fedderke:2014wda}
\begin{align}
  \Gamma_{\textrm{inv}} (h \rightarrow S S) &= \frac{\lambda_{hS}^2 \vo^2}{32 \pi m_h} \left(1- \frac{4m_S^2}{m_h^2}\right)^{1/2}, \label{invS} \\
  \Gamma_{\textrm{inv}} (h \rightarrow VV)  &= \frac{\lambda_{hV}^2 \vo^2 m_h^3}{128 \pi m_V^4} \left(1 - \frac{4m_V^2}{m_h^2} + \frac{12 m_V^4}{m_h^4} \right) \left(1-\frac{4m_V^2}{m_h^2}\right)^{1/2}, \label{invV} \\
  \Gamma_{\textrm{inv}} (h \rightarrow \ovr{\chi} \chi) &=  \frac{m_h \vo^2}{16\pi} \left(\frac{\lambda_{h\chi}}{\Lambda_\chi}\right)^2 \left(1 - \frac{4m_\chi^2}{m_h^2} \right)^{1/2} \left(1- \frac{4 m_\chi^2 \cos^2 \xi}{m_h^2}\right), \label{invX} \\
  \Gamma_{\textrm{inv}} (h \rightarrow \ovr{\psi} \psi) &= \frac{m_h \vo^2}{8\pi} \left(\frac{\lambda_{h\psi}}{\Lambda_\psi}\right)^2 \left(1 - \frac{4m_\psi^2}{m_h^2} \right)^{1/2}\left(1- \frac{4 m_\psi^2 \cos^2 \xi}{m_h^2}\right). \label{invP} 
\end{align}

Using an upper limit of $Y$ ($Y$ = 0.19 or 0.05 from above) on $\mathcal{BR}(h \rightarrow \ovr{X} X)$, the Higgs invisible width $\Gamma_{\textrm{inv}}$ can be expressed in terms of the visible contribution $\Gamma_{\textrm{vis}}$ through
\begin{equation}\label{invbranch}
  \Gamma_{\textrm{inv}} \leq \frac{Y}{1-Y} \Gamma_{\textrm{vis}}.
\end{equation}
Using the Higgs invisible width expressions from Eqs.~\eqref{invS} to \eqref{invP}, an upper limit on $\Gamma_{\textrm{inv}}$ in Eq.~\eqref{invbranch} provides an upper limit on the SM Higgs-DM coupling as a function of the DM mass
\begin{align*}
  \emph{Scalar:} \quad \lambda_{hS} &\leq \left[\frac{Y}{1-Y} \frac{32 \pi m_h \, \Gamma_{\textrm{vis}}}{\vo^2} \left(1- \frac{4m_S^2}{m_h^2}\right)^{-1/2}\right]^{1/2}, \\
  \emph{Vector:} \quad \lambda_{hV} &\leq \left[\frac{Y}{1-Y} \frac{128 \pi m_V^4 \, \Gamma_{\textrm{vis}}}{\vo^2 m_h^3} \left(1- \frac{4m_V^2}{m_h^2} + \frac{12 m_V^4}{m_h^4} \right)^{-1} \left(1- \frac{4m_V^2}{m_h^2}\right)^{-1/2}\right]^{1/2}, \\
  \emph{Majorana:} \quad \frac{\lambda_{h\chi}}{\Lambda_\chi} &\leq \left[\frac{Y}{1-Y} \frac{16 \pi \, \Gamma_{\textrm{vis}}}{m_h \vo^2} \left(1- \frac{4m_\chi^2}{m_h^2}\right)^{-1/2} \left(1-\frac{4m_\chi^2 \cos^2 \xi}{m_h^2}\right)^{-1}\right]^{1/2}, \\
  \emph{Dirac:} \quad \frac{\lambda_{h\psi}}{\Lambda_\psi} &\leq \left[\frac{Y}{1-Y} \frac{8\pi \, \Gamma_{\textrm{vis}}}{m_h \vo^2} \left(1- \frac{4m_\psi^2}{m_h^2}\right)^{-1/2} \left(1-\frac{4m_\psi^2 \cos^2 \xi}{m_h^2}\right)^{-1}\right]^{1/2},
\end{align*}
where we take the visible contribution $\Gamma_{\text{vis}} \equiv \Gamma_{\textrm{SM}} = 4.07$\,MeV for $m_h = 125$\,GeV \cite{Dittmaier:2011ti}.

\subsection{Indirect detection}
Indirect detection of DM involves searches for fluxes of $\gamma$-rays, $e^{\pm}$, $p^\pm$ and neutrinos produced from DM annihilations in distant astrophysical sources. As the annihilation flux $(\Phi_{\textrm{ann}})$ scales as the \emph{square} of the DM mass density (i.e., $\Phi_{\textrm{ann}} \propto \rho_{\textrm{DM}}^2$), natural places to look for DM annihilations are those with high DM content such as the dwarf spheroidal (dSph) galaxies, the Galactic Center (GC) or our Sun. Current indirect experiments searching for signs of DM annihilations include the Fermi Large Area Telescope (Fermi-LAT) \cite{2009ApJ697.1071A}, H.E.S.S. \cite{2011PhRvL.106p1301A} and AMS-02 \cite{Corti:2014ria}. The upcoming Cherenkov Telescope Array \cite{2009arXiv0912.3742W} is one of the next generation of ground-based gamma-ray telescopes. 

To impose parameter space constraints from indirect search experiments, we use a combined log-likelihood function that depends on the model parameters
\begin{equation}\label{lfunc}
  \ln \lagr_{\textrm{total}} (m_X,\lambda_{hX}) = \ln \lagr_{\textrm{CMB}} (m_X,\lambda_{hX}) + \ln \lagr_{\textrm{dSphs}} (m_X,\lambda_{hX}) + \ln \lagr_{\textrm{CTA}} (m_X,\lambda_{hX}). 
\end{equation}
In general, the contribution to the total log-likelihood function in Eq.~\eqref{lfunc} comes from all three indirect searches. However, the CTA log-likelihood function enters in our calculation only when we discuss \emph{projected} limits. Each log-likelihood function depends on $m_X$ and on the Higgs-DM coupling $\lambda_{hX}$ via the zero-velocity annihilation cross section\footnote{Dark matter annihilations are assumed to occur in the nonrelativistic limit since the particles involved have speeds $v \sim10^{-3} c$. In the zero-velocity limit, the center-of-mass energy $\sqrt{s} \rightarrow 2m_X$.} $\langle \sigma v_{\textrm{rel}} \rangle_0$, the branching ratio $\mathcal{B}_f$ into the SM final state $f$ and the model relic density $\Omega_{X}h^2$. The final states included in our calculations involve DM annihilations into $W^+ W^-$, $Z Z$, $h h$, $\mu^+ \mu^-$, $\tau^+ \tau^{-}$ and $\ovr{q} q$ for $q = b,c,t$. 

We scale all indirect detection signals for each combination of $m_X$ and $\lambda_{hX}$ by $f_{\textrm{rel}}^2 = (\Omega_X/\Omega_{\textrm{DM}})^2$, thereby suppressing signals where $X$ constitutes only a fraction of the total dark matter. In regions where the model relic density is \emph{larger} than the observed DM relic density, we simply rescale in the same way, thereby increasing the expected signals. This is done for the sake of simplicity and is of no practical consequence given that the relevant region is robustly excluded by the relic density constraints. 

In complete analogue with the implementation of the relic density constraints, we perform scans in each model parameter space using Brent's method to impose the indirect search limits. The function used for the root-finding algorithm is
\begin{equation}\label{root-id}
  f_{\textrm{ID}}(m_X,\lambda_{hX}) = \Delta \ln \lagr_{\textrm{total}} (m_X,\lambda_{hX}) - \Delta \ln \lagr_{\textrm{C.L.}} (m_X, \lambda_{hX}),
\end{equation}
where $\Delta \ln \lagr_{\textrm{total}}$ is the combined delta log-likelihood, defined such that it is zero for the case of no DM signal. The term $\Delta \ln \lagr_{\textrm{C.L.}}$ is the delta log-likelihood for a given C.L. For our purposes, we implement $1\sigma$ and $90\%$\,C.L. by taking
\begin{equation*}
\Delta \ln \lagr_{\textrm{C.L.}} = 
\begin{cases}
  -0.500000 &\mbox{for } \textrm{1$\sigma$\,C.L.,} \\
  -1.352771 &\mbox{for } \textrm{90\%\,C.L.} 
\end{cases}
\end{equation*}
When implementing \emph{current} indirect search limits, the combined log-likelihood function is
\begin{equation}\label{current}
  \ln \lagr_{\textrm{total}} (m_X,\lambda_{hX}) = \ln \lagr_{\textrm{WMAP}} (m_X,\lambda_{hX}) + \ln \lagr_{\textrm{dSphs}} (m_X,\lambda_{hX}),
\end{equation}
whereas for the \emph{projected} future limits, the combined log-likelihood function is
\begin{equation}\label{future}
  \ln \lagr_{\textrm{total}} (m_X,\lambda_{hX}) = \ln \lagr_{\textrm{Planck}} (m_X,\lambda_{hX}) + \ln \lagr_{\textrm{dSphs}}^{\textrm{projected}} (m_X,\lambda_{hX}) + \ln \lagr_{\textrm{CTA}} (m_X,\lambda_{hX}).
\end{equation}
Here $\ln \lagr_{\textrm{dSphs}}^{\textrm{projected}} (m_X,\lambda_{hX})$ differs from $\ln \lagr_{\textrm{dSphs}} (m_X,\lambda_{hX})$ due to the  projected improvements of Fermi-LAT in adding more dwarf galaxies to its search, and observing for a longer duration. 

When $m_X < m_h/2$, we again encounter situations where multiple values of $\lambda_{hX}$ satisfy Eq.~\eqref{root-id}. When this occurs, we ignore the root at larger $\lambda_{hX}$ and choose the smaller coupling $\lambda_{hX}$ due to the fact that larger couplings will be robustly excluded by the Higgs invisible width constraints. 

\subsubsection{CMB likelihood}
It is well known that the temperature fluctuations and polarization of the CMB are sensitive to the redshift of recombination $z \sim 1100$ as it determines the surface of last scattering. If the decay or annihilation of dark matter deposits extra electromagnetic radiation after $z \sim 1100$, it can delay the time taken for recombination and/or produce distortions in the CMB. 

A key quantity of interest in determining the CMB bounds on dark matter annihilations is the efficiency $f(z)$ for producing ionizing radiation as a function of redshift $z$. For annihilations, $f(z)$ is determined in terms of the electromagnetic power injected per unit volume \cite{Cline:2013fm}
\begin{equation}
\frac{dE}{dt \, dV} = f(z) \frac{\asigv}{m_\textrm{X}} \Omega_{X}^2 \rho_c^2\, c^2 (1+z)^6,
\end{equation} 
where $\Omega_X = \rho_X/\rho_c$ and $\rho_c$ is the critical mass density of the Universe today.

Bounds on $\asigv$ can be encoded in terms of an integral involving $f(z)$ and a set of \emph{principle component} basis functions $e_i(z)$ \cite{Finkbeiner:2011dx,Slatyer:2012yq}. In terms of these basis functions, $f(z)$ can be expanded as
\begin{equation*}
  \varepsilon f(z) = \sum_{i=1}^\infty \varepsilon_i e_i(z)
\end{equation*}
where $\varepsilon \equiv \asigv/m_X$ and $\varepsilon_i = \varepsilon f(z)\cdot e_i(z)/e_i(z)\cdot e_i(z)$. The inner product is an integral over $z$ with the integration limits $z_1 = 86.83$ and $z_2 = 1258.2$. For annihilating DM, these basis functions are chosen to maximize sensitivity to a generally expected $z$-dependence of the energy injection from annihilating dark matter in such a way that most important contributions are described by the lowest components. 

It is useful to consider a quantity $f_{\textrm{eff}}$ defined in terms of a ``universal WIMP annihilation'' curve $e_W(z)$ as $f_\textrm{eff} \sim (f\cdot e_W)/(e_W \cdot e_W)$, which has the interpretation that $f_{\textrm{eff}} < 1$ denotes the average efficiency of energy injection for the annihilation channel under consideration. With the expansion $e_W = \sum_i c_i e_i$ for WMAP7 \cite{2011ApJS..192...18K}, only the first principle component is dominant. Hence, $f_\textrm{eff}$ for WMAP7 is given by
\begin{equation*}
  f_{\textrm{eff}} \equiv \frac{(f\cdot e_1)}{c_1(e_1\cdot e_1)} \quad (\textrm{WMAP})
\end{equation*}
where numerically $c_1 = 4.64$. For Planck \cite{Ade:2013zuv}, the contribution from all three principle components must be included and therefore 
\begin{equation*}
  f_\textrm{eff} \equiv \frac{1}{\sqrt{\lambda_1}c_1} \left[\sum_i \lambda_i \left(\frac{f\cdot e_i}{e_i\cdot e_i} \right)^2 \right]^{1/2} \quad (\textrm{Planck})
\end{equation*}
where $\lambda_i$ and $e_i$ are parameters appropriate for Planck.

We use the tabulated values of the effective efficiency $(f_{\text{eff}})$ for various SM final states and DM masses from Ref.~\cite{Cline:2013fm}. At intermediate masses, we interpolate the values of $f_{\textrm{eff}}$ in terms of $\log_{10} m_X$. For DM annihilation into multiple channels, the total $f_\textrm{eff}$ is a weighted sum of effective efficiency $f_{\textrm{eff},f}$ over final states $f$ with branching ratio $\mathcal{B}_f$ as
\begin{equation}
f_{\text{eff}} = \sum_f \mathcal{B}_f \, f_{\text{eff},f}.
\end{equation} 

To impose CMB constraints at arbitrary C.L., we use a log-likelihood function for the annihilation cross section $\langle \sigma v_{\mathrm{rel}} \rangle$, assuming a DM mass $m_X$ and a branching ratio $\mathcal{B}_f$ into final state $f$. For the Planck experiment, the log-likelihood is given by \cite{Cline:2013fm}
\begin{equation}\label{planck}
\ln \lagr_{\text{Planck}}(\asigv | m_X,\mathcal{B}_f) = -\frac{1}{2} f_{\text{eff}}^2 \, \lambda_1 c_1^2 \left(\frac{\asigv}{2 \times 10^{-27} \text{cm}^3 \, \text{s}^{-1}} \right)^2 \left(\frac{\text{GeV}}{m_X} \right)^2,
\end{equation}
where $c_1 = 4.64$ and $\lambda_1 = 3.16$.\footnote{This is in fact a projected log-likelihood for the Planck polarization data, but it agrees very well with the actual constraints in Ref.~\cite{Slatyer:2015jla}.} The above equation assumes a linear response on the CMB against the deposited energy, which is not accurate for WMAP. This however can be corrected by the replacement $2 \times 10^{-27} \textrm{cm}^3 \, \textrm{s}^{-1} \rightarrow 3.2 \times 10^{-27} \textrm{cm}^3 \, \text{s}^{-1}$ and $\lambda_1 = 0.279$. Therefore, for the WMAP experiment, the log-likelihood is given by \cite{Cline:2013fm}
\begin{equation}\label{wmap}
\ln \lagr_{\text{WMAP}}(\asigv | m_X,\mathcal{B}_f) = -\frac{1}{2} f_{\text{eff}}^2 \,\lambda_1 c_1^2 \left(\frac{\asigv}{3.2 \times 10^{-27} \text{cm}^3 \, \text{s}^{-1}} \right)^2  \left(\frac{\text{GeV}}{m_X} \right)^2 ,
\end{equation}
where $c_1 = 4.64$ and $\lambda_1 = 0.279$. 

\subsubsection{Fermi dwarfs likelihood}
To place upper limits on $\asigv$, one must quantify how it influences the flux of gamma rays detected by the Large Area Telescope aboard the Fermi satellite \cite{Ackermann:2013uma}. The nonobservation of gamma rays from the dwarf spheroidal satellite galaxies of the Milky Way (dSphs) can be used to place strong upper limits on $\asigv$ in various SM final states. 

The differential gamma-ray flux from DM annihilation in a region $\Delta \Omega$ towards a dwarf spheroidal galaxy is given by
\begin{equation}\label{annflux}
  \frac{d\Phi_{\textrm{ann}}}{dE} = \underbrace{\kappa \, \frac{\asigv}{4\pi m_X^2} \frac{dN_\gamma}{dE}}_{\Phi_{\textrm{PP}}} \underbrace{\int_{\Delta \Omega} d\Omega \int_{\textrm{l.o.s.}} \rho_{\textrm{DM}}^2 \, (r) \, dl}_{J_{\textrm{ann}}},
\end{equation}
where $\kappa$ equals a $1/2$ for self-conjugate DM (due to the phase-space considerations) and $1/4$ for non-self-conjugate DM (due to the density of DM particles and antiparticles, each being half of the total DM density $\rho_{\textrm{DM}}$ in the $J_{\textrm{ann}}$ term) \cite{Toma:2013hva,Ko:2014loa}. The first term $\Phi_{\textrm{PP}}$ describes the particle physics aspect of DM annihilation: the velocity-averaged annihilation cross section $\asigv$, the differential gamma-ray spectrum $dN_\gamma/dE$ per annihilation\footnote{Specifically, the differential gamma-ray spectrum per annihilation is a sum over the differential spectrum from all possible final states: $dN_\gamma/dE = \sum_f \mathcal{B}_f \, dN_\gamma^f/dE$, where $\mathcal{B}_f$ is the branching ratio into the SM final state $f$.} and the DM mass $m_X$. The second term $J_{\textrm{ann}}$ is commonly known as the ``$J$-factor,'' and describes the astrophysical aspects of a dwarf galaxy. It is given by the line-of-sight (l.o.s.) integral of the DM distribution in a dwarf galaxy, integrated over the solid angle $\Delta\Omega$. 

The main advantage of the form of the differential gamma-ray flux in Eq.~\eqref{annflux} is that the terms describing the particle physics of DM annihilation (specifically $\Phi_{\textrm{PP}}$) and the astrophysics of a dwarf galaxy (specifically $J_{\textrm{ann}}$) separate into two independent factors. Since the prefactor $\Phi_{\textrm{PP}}$ is constant for a given set of model parameters, only the $J$-factors have to be determined individually for each dwarf galaxy. A combined analysis based on the observations of multiple dwarf satellite galaxies is therefore straightforward.

The expected number of gamma rays produced from DM annihilations between energies $E_1$ and $E_2$ in a specific region of the sky is
\begin{equation}
  N_{\textrm{ann}} = T_{\textrm{obs}} \, \kappa \, \frac{\asigv J_{\textrm{ann}}}{4\pi m_X^2} \int_{E_1}^{E_2} dE \, \frac{dN_\gamma}{dE} A_{\textrm{eff}}(E) \phi(E),
\end{equation}
where $T_{\textrm{obs}}$ is the experimental observation time, $A_{\textrm{eff}} (E)$ is the energy-dependent effective area and $\phi(E)$ is the detector energy resolution. 

The Fermi-LAT is a powerful tool in searching for signs of DM annihilations from distant astrophysical sources. Currently, it provides the strongest upper bounds on $\asigv$, based on a combined analysis of 15 dwarf spheroidal galaxies in the Milky Way using 6 years of Fermi-LAT data, processed with the new \texttt{Pass-8} event-level analysis \cite{Ackermann:2015zua}. For each of the 15 dwarf galaxies included in the analysis, the results are publicly available in the form of tabulated values of the energy times integrated gamma-ray flux (i.e., $E \, \Phi_{\textrm{ann}}$ in units of MeV\,cm$^{-2}$\,s$^{-1}$) and  log-likelihoods $(\ln \lagr_{\textrm{dSphs}})$ in 24 energy bins between 500\,MeV and 500\,GeV.\footnote{\url{https://www-glast.stanford.edu/pub_data/1048/}.} 

In our analysis, we calculate the Fermi dwarf log-likelihood $(\ln \lagr_{\textrm{dSphs}})$ using \texttt{gamLike\,1.1} (C. Weniger \emph{et al}., to be published), a package designed for the evaluation of log-likelihoods for gamma-ray searches. It is primarily written in \texttt{C++} and uses various integration routines to  integrate the input arrays of $\Phi_\textrm{PP}(E)$ in each of the energy bins relevant for a given experiment. The integrated $\Phi_\textrm{PP}(E)$'s are combined with the $J$-factors, giving an array of integrated gamma-ray fluxes $(\Phi_{\textrm{ann}})$ in each energy bin. The integrated flux times energy ($E \Phi_\textrm{ann}$) arrays are compared with the tabulated log-likelihood values. Finally, a log-likelihood value at a fixed set of model parameters is returned by summing over the interpolated log-likelihood values in each energy bin. 

The inputs to the \gamLike package are arrays of gamma-ray energies $E$ (GeV) and $\Phi_{\textrm{PP}}(E)$ (cm$^3$\,s$^{-1}$\,GeV$^{-3}$) where 
\begin{equation}
  \Phi_{\textrm{PP}}(E) = \kappa \, \frac{\asigv}{4\pi m_X^2} \frac{dN_\gamma (E)}{dE}.
\end{equation}
Since the tabulated results of Ref.~\cite{Ackermann:2015zua} are given between gamma-ray energies of 500\,MeV and 500\,GeV, we interpolate the differential gamma-ray spectrum $dN_\gamma/dE$ calculated within \micro between 500\,MeV and the DM mass $m_X$.\footnote{The differential gamma-ray spectrum $dN_\gamma/dE$ is zero after $E = m_X$.} The $\ln \lagr_{\textrm{dSphs}}(m_X,\lambda_{hX})$ entering in Eq.~\eqref{current} is then a difference of the log-likelihoods for a DM-signal and background-only hypothesis ($\Phi_{\textrm{PP}} = 0$).

For \emph{projected} indirect search limits, we assume that Fermi operates for at least 10 years in its current survey mode and is able to add as many southern dwarf galaxies in the future as there are mostly northern dwarf galaxies now. Assuming that the improvements in the Fermi-LAT reach are dominated by the statistical uncertainty (and thus the limits on $\asigv$ scale as $\sqrt{N}$, where $N$ is the number of dwarfs), we calculate the projected Fermi sensitivities by scaling the current limits by a factor of $\sqrt{2\times 10/6} = \sqrt{20/6} \sim 1.83$.

\subsubsection{CTA likelihood}
The CTA is a multinational project to build the next generation of ground-based gamma-ray instruments and to have sensitivity over energies from a few tens of\,GeV to 100\,TeV. It is intended to improve the flux sensitivities of the current generation of Imaging Atmospheric Cherenkov Telescopes (IACTs) such as MAGIC \cite{2012APh....35..435A}, H.E.S.S. \cite{Aharonian:2006pe} and VERITAS \cite{Holder:2008ux} by an order of magnitude. 

CTA will consist of several tens of telescopes of 2--3 different types, with sizes varying between 5 and 25 meters, distributed over an area of several square kilometers. The sensitivity will be a factor of 10 better than current instruments, and the field of view (FoV) will be up to 10$^\circ$ in diameter. It is envisaged as a two part telescope, with southern and northern sites. CTA South is most relevant for DM searches towards the GC. Its final design is not yet fixed. Apart from the construction and maintenance questions, the relevant remaining design choices are the relative emphasis on the higher or lower energies, the angular and energy resolution and the FoV. A first detailed Monte Carlo (MC) analysis was presented in Ref.~\cite{2013APh....43..171B} where 11 different array configurations for the CTA South were discussed.

We use the array configuration known as ``Array I,'' which has a balanced configuration with 3 large ($\sim$\,24\,m aperture), 18  medium ($\sim$\,12\,m) and 56 small telescopes ($\sim$\,4--7\,m). This configuration also provides a good compromise in sensitivity between lower and higher energies. Extensive information on the effective area, background rates, and angular and energy resolution of Array I is also available. Furthermore, previous DM sensitivity studies have used a very similar array: Array E in Ref.~\cite{2013APh43189D} and the Paris-MVA analysis of Array I in Ref.~\cite{Pierre:2014tra}. The point-source sensitivities of Arrays E and I agree very well at energies $\lesssim 1$\,TeV, whereas at higher energies Array I is more sensitive (only by a factor of less than 2). 

The main performance aspects of Array I include an effective area of 100 m$^2$ at its threshold energy of 20\,GeV, which then increases quickly to about $4 \times  10^5$\,m$^2$ at 1\,TeV and $3 \times 10^6$\,m$^2$ at 10\,TeV. The angular resolution in terms of the 68\% containment radius is about $r_{68} \simeq 0.3^\circ$ at threshold and drops to below $0.06^\circ$ at energies above 1\,TeV. The energy resolution is relatively large at threshold, $\sigma(E)/E \sim 50\%$, but drops to below 10\% at energies above 1\,TeV.

\begin{figure}
  \begin{minipage}[t]{0.48\textwidth}  
    \centering  
	\includegraphics[width=\textwidth]{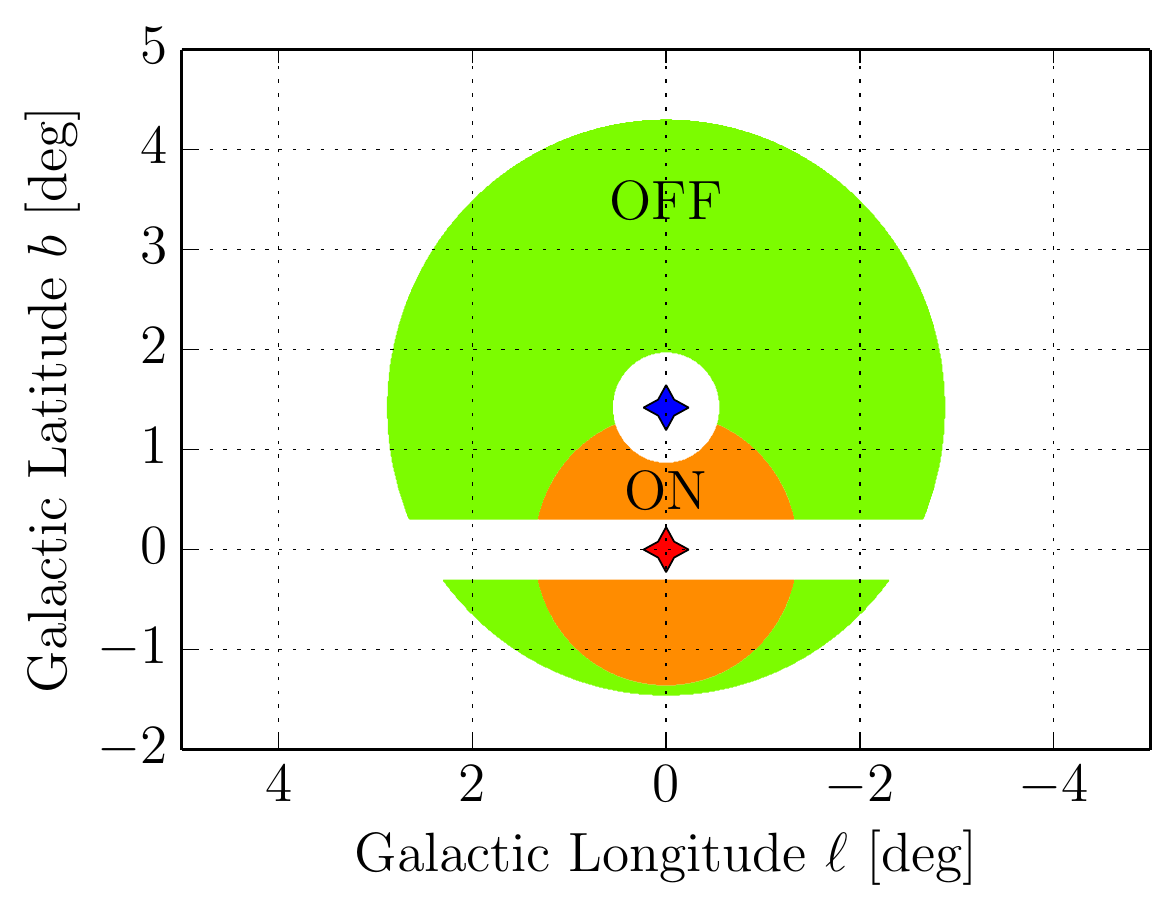}
  \end{minipage}
  \hfil
  \begin{minipage}[t]{0.48\textwidth}  
    \centering  
	\includegraphics[width=\textwidth]{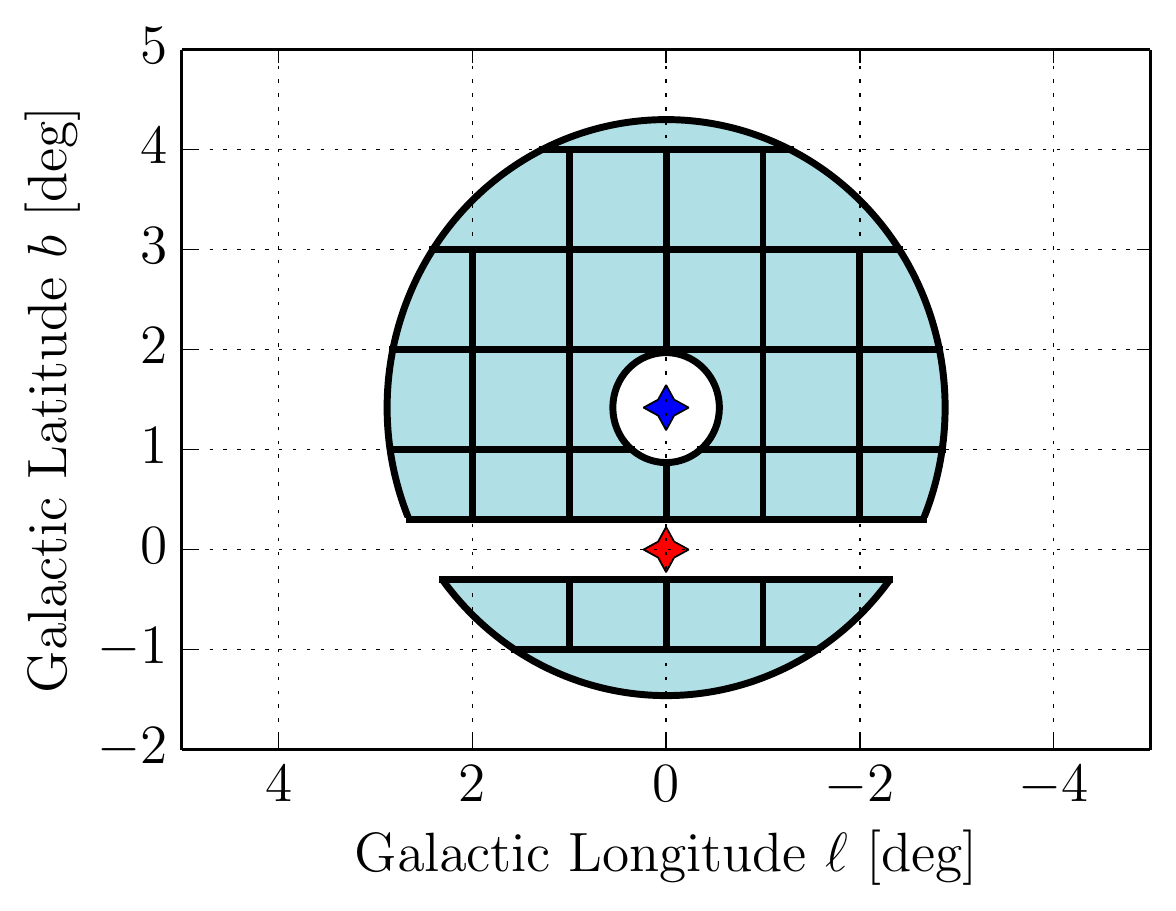}    
  \end{minipage} 
  \caption{Figures adopted with permission from Ref.~\cite{Silverwood:2014yza}. Left: The ``signal'' and ``background'' regions of interest (RoIs) used in the ring method of Ref.~\cite{2013APh....43..189D}. Right: Separation of the signal and background RoIs into 28 sub-RoIs for the morphological analysis of Ref.~\cite{Silverwood:2014yza}.}
  \label{ringAndMorph}
\end{figure}

Following the analysis of Ref.~\cite{Silverwood:2014yza}, we use their tabulated values of the integrated gamma-ray flux times energy (i.e., $\Phi_{\textrm{ann}} E$ in units of MeV\,cm$^{-2}$\,s$^{-1}$) and delta log-likelihoods $(\Delta \ln \lagr_{\textrm{CTA}})$ between gamma-ray energies of 25\,GeV and 10\,TeV. The main features of their study are
\begin{itemize}
  \item Assessing the impacts of all backgrounds, including protons and electrons in cosmic rays hitting the atmosphere and diffuse astrophysical emissions. Galactic diffuse emission (GDE) substantially degrades the CTA differential sensitivity (see Fig. 4 of Ref.~\cite{Silverwood:2014yza} for a comparison). 
  \item Introducing a statistical framework that accounts for the impacts of differential acceptance uncertainties from sources such as event reconstruction, Monte Carlo (MC) determination of the effective areas and the uncertainty in atmospheric conditions \cite{2011ExA....32..193A} within a FoV on DM limits from CTA. Specifically, the tabulated results correspond to a systematic uncertainty of 1\%. 
  \item Using the Einasto profile \cite{Navarro:2003ew} to calculate the $J$-factors for the GC. The Einasto profile best fits the DM density profiles seen in the $N$-body simulations of Milky Way type galaxies 
  \begin{equation}\label{einasto}
    \rho_{\textrm{DM}} (r) \propto \exp\left(-\frac{2}{\alpha} \left[\left(\frac{r}{r_s}\right)^\alpha - 1 \right]\right),
  \end{equation}
  where the profile is normalized to a local DM density of $\rho_\odot \equiv \rho_{\textrm{DM}} (r_\odot) = 0.4$\,GeV\,cm$^{-3}$ by choosing $\alpha$ = 0.17, $r_s = 20$ kpc and $r_\odot$ = 8.5\,kpc \cite{2011PhRvD..83b3518P}. 
  \item Using a slightly contracted generalized Navarro-Frenk-White (NFW) profile \cite{Navarro:1996gj} to indicate how the limits improve with a more optimistic DM distribution. The profile is parametrized as 
  \begin{equation}\label{NFW}
  \rho_{\textrm{DM}}(r) \propto \frac{1}{r^\gamma (r_s + r)^{3-\gamma}},
  \end{equation}
  where $\gamma = 1.3$ is the inner slope of the profile and $r_s$ = 20\,kpc is the scale radius. It is normalized in the same way as the Einasto profile (i.e., $\rho_\odot$ = 0.4\,GeV\,cm$^{-3}$).   
  \item Performing a ``morphological'' analysis by covering the area occupied by the two regions of interest (RoIs) in the left panel of Fig.~\ref{ringAndMorph} and dividing it into $1^\circ \times 1^\circ$ squares, giving a total of 28 RoIs as shown in the right panel of Fig.~\ref{ringAndMorph}. The morphological analysis allows for a proper exploitation of the shape differences between the GDE, which is concentrated along the Galactic plane, and the DM annihilation signal, which is spherically distributed around the GC. The resultant constraints are found to be more stringent by a factor of a few when compared against traditional ``ring'' analyses.
\end{itemize}

In summary, the tabulated results of Ref.~\cite{Silverwood:2014yza} are based on a morphological analysis over the 28 RoIs, assuming $\gtrsim$ 100\,h of GC observation by CTA,\footnote{Since the limits of Ref.~\cite{Silverwood:2014yza} are systematics dominated, they are more or less independent of any increase in the CTA observation time beyond 100\,h. For our portal models, we derive the indirect search limits based on 100\,h of CTA observation time.} 1\% instrumental systematics, $J$-factors for the GC using the Einasto profile [Eq.~\eqref{einasto}] and inclusion of all known backgrounds (cosmic-ray electrons/protons, and the galactic diffusion emissions).

In analogy with the Fermi dwarf log-likelihood calculation, we use the \texttt{gamLike\,1.1} package to calculate the CTA log-likelihood from the tabulated results of Ref.~\cite{Silverwood:2014yza}. As these results are for gamma-ray energies between 25\,GeV and 10\,TeV, the corresponding input parameters to the \gamLike package are arrays of gamma-ray energies $E$ (GeV) (between 25\,GeV and $m_X$) and $\Phi_{\textrm{PP}}(E)$ (cm$^3$\,s$^{-1}$\,GeV$^{-3}$). 

Limits on $\asigv$ from CTA observation of the GC are presented in Fig.~7 of Ref.~\cite{Silverwood:2014yza} where different annihilation channels and DM halo profiles are assumed. For the DM annihilation into a $\ovr{b} b$ final state, a contracted generalized NFW profile [Eq.~\eqref{NFW}] yields a factor of 6 better limits on $\asigv$ when compared against the Einasto profile [Eq.~\eqref{einasto}]. This is mainly due to an increase in the GC $J$-factors by a factor of 2.9 (when summed over all RoIs). To extend the expected search capability of CTA towards higher WIMP masses, we apply these improved limits in order to see the effect of using a more optimistic DM distribution.  

\subsection{Direct detection}
Direct detection experiments aim to measure the recoil of a nucleus in a collision with a DM particle. After an elastic collision between DM of mass $m_X$ and a nucleus of mass $M$, the nucleus recoils with energy $E = \mu^2 v^2 (1-\cos\theta)/M$, where $\mu \equiv m_X M/(m_X + M)$ is the DM-nucleus reduced mass, $v$ is the speed of DM particle relative to the nucleus rest frame and $\theta$ is the scattering angle in the center-of-mass (COM) frame. 

The differential rate per day (given in units of cpd kg$^{-1}$ keV$^{-1}$ where cpd is counts per day) from primarily spin-independent (SI) interactions is given by \cite{2010pdmo.book..347C,Freese:2012xd}
\begin{equation}\label{rate}
  \frac{dR}{dE} = \frac{\rho_\odot}{m_X M} \int d^3v \, v f(\mathbf{v},t) \frac{d\sigma_X^{\textrm{SI}}}{dE},
\end{equation}
where $\rho_\odot$ is local DM mass density and $f(\mathbf{v},t)$ is the time-dependent DM velocity distribution. The term $d\sigma_X^{\textrm{SI}}/dE$ is the momentum-dependent differential SI cross section. In the typical case where the target material contains more than one isotope, the differential rate is given by a mass-fraction weighted sum over contributions from the isotopes, each of the forms given by Eq.~\eqref{rate}. 

The momentum-dependent differential SI cross section is given by
\begin{equation}\label{ds-dE}
  \frac{d\sigma_X^{\textrm{SI}}}{dE} =  \frac{1}{E_{\textrm{max}}} \sigma_0 F^2(q) \Theta(q_{\textrm{max}} - q),
\end{equation}
where $E_\textrm{max} = q_{\textrm{max}}^2/2M =  2\mu^2 v^2/M$ is the maximum energy transfer in a collision at a relative velocity $v$, $\sigma_0$ is the SI cross section in the zero-velocity limit\footnote{Generally, $\sigma_{\textrm{SI}}$ and $\sigma_{\textrm{SD}}$ are used to represent this term for nuclear spin-independent (SI) and spin-dependent (SD) interactions respectively. To avoid confusion with existing literatures, we will represent the SI contribution to $\sigma_0$ as $\sigma_{\textrm{SI}}$ in our models.} and  $\Theta$ is the Heaviside step function. The term $F (q)$ is the form factor accounting for the finite size of the nucleus. When DM coherently scatters off the entire nucleus with low momentum transfer, $F^2(q) \rightarrow 1$. However, when the de Broglie wavelength of the momentum transfer $q$ becomes comparable to the size of the nucleus, DM becomes sensitive to the internal structure of the nucleus and $F^2(q) < 1$, with $F^2(q) \ll 1$ at high momentum transfers. 

It is traditional to define an \emph{effective} SI cross section corrected with the form factor as
\begin{equation}\label{effsig}
  \sigma(q) \equiv \sigma_0 F^2(q), 
\end{equation}
such that the dependency on the momentum transfer $q$ is contained entirely within the form factor $F^2(q)$. The \emph{actual} cross section is $\int dq^2 \, d\sigma_X^{\textrm{SI}}(q^2,v)/dq^2 $ for a given DM-nucleus relative speed $v$. With the chosen convention in Eq.~\eqref{effsig}, the expression for the momentum-dependent differential SI cross section in Eq.~\eqref{ds-dE} simplifies to
\begin{equation}\label{ds-dE2}
  \frac{d\sigma_X^{\textrm{SI}}}{dE} =  \frac{M}{2\mu^2 v^2} \sigma (q) \Theta(q_{\textrm{max}} - q).
\end{equation}

With the form of the the momentum-dependent differential SI cross section in Eq.~\eqref{ds-dE2}, the differential rate per day in Eq.~\eqref{rate} becomes
\begin{equation}\label{rate2}
  \frac{dR}{dE} = \frac{1}{2m_X \mu^2} \sigma(q) \, \rho_\odot\, \eta(v_\textrm{min}(E),t),
\end{equation}
where 
\begin{equation}\label{etaForm}
  \eta(v_\textrm{min}(E),t) = \int_{v > v_{\textrm{min}}(E)} d^3v \, \frac{f(\mathbf{v},t)}{v}
\end{equation}
is the mean inverse speed and 
\begin{equation*}
  v_{\textrm{min}}(E) = \sqrt{\frac{ME}{2\mu^2}}
\end{equation*}
is the minimum DM speed that results in a nuclear recoil with energy $E$. The requirement $q < q_{\textrm{max}}$ in the Heaviside step function of Eq.~\eqref{ds-dE} results in a lower limit $v > v_{\textrm{min}}(E)$ on the integral in Eq.~\eqref{etaForm}. 

The main advantage of writing the recoil spectrum in the form of Eq.~\eqref{rate2} instead of Eq.~\eqref{rate} is that the particle physics and astrophysics aspects separate into two distinct factors: $\sigma(q)$ describes the particle physics aspect; while $\rho_\odot$ and $\eta(v_\text{min}(E),t)$ describe the astrophysical aspect. 

For our Higgs-mediated scalar and/or pseudoscalar DM coupling, the SI cross section $(\sigma_{\textrm{SI}})$ takes the  form 
\begin{equation}\label{sigSI}
  \sigma_\textrm{SI} = \frac{\mu^2}{\pi} \Big[Z \, G_{p}^{\textrm{SI}} + (A-Z) \, G_{n}^\textrm{SI} \Big]^2 + \frac{q^2}{4m_X^2}\frac{\mu^2}{\pi} \Big[Z \, \widetilde{G}_{p}^{\textrm{SI}} + (A-Z) \, \widetilde{G}_{n}^\textrm{SI} \Big]^2,
\end{equation}
where $Z$ and $A-Z$ are the number of protons and neutrons in the nucleus respectively and $G_N^{\textrm{SI}} \, (\widetilde{G}_N^{\textrm{SI}})$ for $N \in (p, n)$ are the effective scalar (pseudoscalar) DM-nucleon couplings in analogue with the $G_F$-like effective four-fermion coupling constants \cite{Gondolo:1996qw}. For the scalar and vector models, $\widetilde{G}_{N}^{\textrm{SI}} = 0$, whereas in fermion models both terms contribute to $\sigma_\textrm{SI}$. Due to the $q^2/4m_X^2$ factor arising from the pseudoscalar DM-nucleon interaction in fermion models \cite{Berlin:2014tja}, the corresponding detection rates are momentum suppressed. 

When the effective scalar and pseudoscalar couplings are approximately equal for protons and neutrons, leading to $\sigma_{\textrm{SI},p} \simeq \sigma_{\textrm{SI},n}$, the SI cross section in Eq.~\eqref{sigSI} is enhanced by a factor of $A^2$ because the matrix elements for the cross section are a coherent sum over the individual protons and neutrons in the nucleus. 

For the SI interaction, the form factor is a Fourier transform of the nucleus mass distribution. As a reasonably accurate approximation, we use the Helm form factor \cite{Lewin199687}, which was first introduced as a modification to the form factor for a uniform sphere with a Gaussian function to account for the soft edge of the nucleus \cite{Duda:2006uk}. It is given by
\begin{equation}\label{helm}
  F(q) = 3 e^{-q^2 s^2/2} \, \left[\frac{\sin(qr_n) - qr_n \cos(qr_n)}{(qr_n)^3}\right],
\end{equation}
where $s \simeq 0.9$\,fm and $r_n^2 = c^2 + \frac{7}{3}\pi^2 a^2 - 5s^2$ is the effective nuclear radius with $a \simeq 0.52$\,fm and $c \simeq 1.23 A^{1/3} - 0.60$\,fm.

We use the standard astrophysical parameters, namely the local DM density $\rho_\odot = 0.4$\,GeV\,cm$^{-3}$ and the local galactic disk rotation speed $v_{\textrm{rot}}$ = 220\,km\,s$^{-1}$, with the same value for the most probable speed $\ovr{v}$ of the Maxwell-Boltzmann velocity distribution. We take the Galactic escape speed as $v_{\textrm{esc}} = 544$\,km\,s$^{-1}$ (see Ref.~\cite{2011JCAP...04..012A} for a discussion). 

In the standard analysis where only a single component of DM constitutes the total DM relic density, the differential rate of detection is proportional to $(\rho_\odot \,\sigma_{\textrm{SI}})/m_X$ as evident in Eq.~\eqref{rate2}. To address the multicomponent dark matter scenario in our models, we rescale the limiting value of $\sigma_{\textrm{SI}}$ by the fraction $f_{\textrm{rel}} = \Omega_{X}/\Omega_{\textrm{DM}}$ of the energy density contributed by $X$ to the total DM density. This results in the local $X$ energy density of $f_{\textrm{rel}} \, \rho_\odot$. 

\subsubsection{Nonfermion models}
The SI cross section in scalar and vector models is given by
\begin{align}
  \emph{Scalar:} \quad \sigma_{\textrm{SI}} &= \frac{\mu^2}{\pi} \frac{\lambda_{hS}^2 f_N^2 m_N^2}{4 m_S^2 m_h^4} = \frac{m_N^4}{4 \pi (m_S + m_N)^2} \frac{\lambda_{hS}^2 f_N^2 }{m_h^4}, \label{scalarSIcx} \\
  \emph{Vector:} \quad \sigma_{\textrm{SI}} &= \frac{\mu^2}{\pi} \frac{\lambda_{hV}^2 f_N^2 m_N^2}{4 m_V^2 m_h^4} = \frac{m_N^4}{4 \pi (m_V + m_N)^2} \frac{\lambda_{hV}^2 f_N^2 }{m_h^4}, \label{vectorSIcx}
\end{align}
where $m_N = 0.931$\,GeV and $f_N = 0.30$ \cite{Cline:2013gha} (see also Ref.~\cite{Alarcon:2011zs}). The term $f_N$ is related to the quark matrix elements inside nucleons through
\begin{equation}
  f_N = \sum_{\textrm{quarks}} f_{Tq}^{(N)} = \sum_{\textrm{quarks}} \frac{m_q}{m_N} \langle N| \ovr{q}q | N \rangle = \frac{2}{9} + \frac{7}{9} \sum_{q = u,d,s} f_{Tq}^{(N)},
\end{equation}
where the last equality follows from the heavy-quark expansion \cite{Shifman1978}.

For the XENON1T \cite{2012arXiv1206.6288A,Beltrame:2013bba,Aprile:2015uzo} experiment, we apply the 90\%\,C.L. from XENON100 \cite{Aprile:2012nq} appropriately weighted by the relic abundance parameter $f_{\textrm{rel}} = \Omega_{X}/\Omega_{\textrm{DM}}$. Assuming that the sensitivity as a function of the DM mass scales relative to that of XENON100 simply by the exposure $\varepsilon$, we demand for every value of $\{m_X, \lambda_{hX}|X \in S,V\}$ that
\begin{equation}\label{sigeff}
  \sigma_{\textrm{eff}}(m_X,\lambda_{hX}) \equiv \varepsilon f_{\textrm{rel}} \,\sigma_{\textrm{SI}} (m_X,\lambda_{hX}) \leq \sigma_{\textrm{Xe}}
\end{equation}
where $\sigma_{\textrm{Xe}}$ is the 90\%\,C.L. from XENON100. For the projected XENON1T, the expected improvement in sensitivity over XENON100 is $\varepsilon = 100$.

For the LUX \cite{2013arXiv1306.0065W, 2013JPhCS.460a2005F} experiment, we construct the log-likelihood function from the Poisson distribution \cite{Feldman:1997qc} in the observed number of signal events $N$ by
\begin{equation}\label{likelihood}
  \lagr_{\textrm{Poisson}} (s|N) = P (N|s) = \frac{(b+s)^N e^{-(b+s)}}{N!},
\end{equation}
where $b$ is the expected number of background events and
\begin{equation}\label{events}
  s = MT \int_0^\infty dE \, \phi(E) \frac{dR}{dE}
\end{equation}
is the expected number of signal events, $MT$ is the detector mass $\times$ exposure time and $\phi(E)$ is a global efficiency factor incorporating trigger efficiencies, energy resolution and analysis cuts. 

We perform the LUX log-likelihood calculations using the \texttt{LUXCalc\,1.0.1} \cite{Savage:2015xta} package.\footnote{\lcalc homepage: \url{http://www.nordita.org/~savage/LUXCalc/index.html}.} For the LUX analysis region, we use $N = 1$ and $b = 0.64$ \cite{Akerib:2013tjd}; while for XENON100, we use $N = 2$ and $b = 1$ \cite{Aprile:2012nq}. The efficiency curves $\phi(E)$ are generated by \texttt{TPMC} \cite{Savage:2015tpcmc} using the \texttt{NEST} \cite{Szydagis:2011tk,Szydagis:2013sih} model.\footnote{\texttt{NEST} homepage: \url{http://nest.physics.ucdavis.edu/site/.}} 

The \lcalc package requires as inputs the effective SI scalar DM-nucleon coupling $G_{N}^{\textrm{SI}}$ and the pseudoscalar DM-nucleon coupling $\widetilde{G}_N^{\textrm{SI}}$ to calculate the SI cross section in Eq.~\eqref{sigSI}. For the scalar and vector models, the effective DM-nucleon couplings are
\begin{align}
  \emph{Scalar:} \quad G_{N}^{\textrm{SI}} = \frac{\lambda_{hS} f_N m_N}{2 m_S m_h^2}, \quad \widetilde{G}_{N}^{\textrm{SI}} = 0, \\
  \emph{Vector:} \quad G_{N}^{\textrm{SI}} = \frac{\lambda_{hV} f_N m_N}{2 m_V m_h^2}, \quad \widetilde{G}_{N}^{\textrm{SI}} = 0,
\end{align}
which reproduce the SI cross sections in Eqs.~\eqref{scalarSIcx} and \eqref{vectorSIcx} respectively. The pseudoscalar effective DM-nucleon couplings are zero because the interaction between DM and the nucleons proceeds through a spin-0 boson.

\subsubsection{Fermion models}
Due to the mediator being a spin-0 boson, the interaction between DM and nucleons are nucleon-spin independent. From the Majorana and Dirac model Lagrangians in Eqs.~\eqref{majoranaFinal} and \eqref{diracFinal} respectively, a mixing between a pure scalar ($\cos\xi = 1$) and a pure pseudoscalar ($\cos\xi = 0$) interaction occurs. 

For a pure pseudoscalar interaction, the SI cross section is suppressed by a factor of $q^2/4m_X^2$ \cite{Berlin:2014tja}, where $q$ is the momentum transfer. A direct comparison between the analytical SI cross section expressions and the limits imposed by XENON100 or LUX experiments is not accurate. We therefore use a generalized and augmented version of \lcalc to include the $q^2/4m_X^2$ factor in the differential rate per day $dR/dE$.

For the Majorana and Dirac fermion models, the effective scalar and pseudoscalar DM-nucleon couplings required by \lcalc  are
\begin{align*}
  \emph{Majorana:} \quad G_{N}^{\textrm{SI}} &= \frac{\lambda_{h\chi}}{\Lambda_\chi} \frac{f_N m_N \cos\xi}{m_h^2} , \quad \widetilde{G}_{N}^{\textrm{SI}} = \frac{\lambda_{h\chi}}{\Lambda_\chi} \frac{f_N m_N \sin\xi}{m_h^2}, \\
  \emph{Dirac:} \quad G_{N}^{\textrm{SI}} &= \frac{\lambda_{h\psi}}{\Lambda_\psi} \frac{f_N m_N \cos\xi}{m_h^2}, \quad
  \widetilde{G}_{N}^{\textrm{SI}} = \frac{\lambda_{h\psi}}{\Lambda_\psi} \frac{f_N m_N \sin\xi}{m_h^2}.
\end{align*}
For the XENON1T experiment, the expected improvement in sensitivity over XENON100 is $\varepsilon = 100$, whereas for the LUX experiment, limits are derived with $\varepsilon = 1$.  

With the inclusion of the momentum-suppressed term in the differential rate per day, we perform root-finding scans in the $(m_X,\lambda_{hX}/\Lambda_X)$ plane for $X \in (\chi,\psi)$ using Brent's method. The function for the root-finding algorithm is
\begin{equation}\label{root-dd}
  f_{\textrm{DD}}(m_X,\lambda_{hX}/\Lambda_X) = \Delta \ln \lagr_{\textrm{Poisson}} (m_X,\lambda_{hX}/\Lambda_X) - \Delta \ln \lagr_{\textrm{90\%\,C.L.}} (m_X, \lambda_{hX}/\Lambda_X)
\end{equation}
where $\Delta \ln \lagr_{\textrm{Poisson}}$ is the difference of $\ln \lagr_{\textrm{Poisson}}$ [Eq.~\eqref{likelihood}] between a DM signal $(s \neq 0)$ and no DM signal $(s = 0)$, whereas $\Delta \ln \lagr_{\textrm{90\%\,C.L.}} = -1.352771$ is the delta log-likelihood for a 90\%\,C.L. 

\section{Results}\label{sec4}
Having outlined an implementation of our model constraints from various experiments, we now present our results. As our current study is a generalization of the scalar singlet analysis in Ref.~\cite{Cline:2013gha} to nonscalar models, we first start by presenting our scalar model results obtained using \micro to validate their consistency. 

Although QCD corrections for quark final states at low scalar masses were included in Ref.~\cite{Cline:2013gha}, they are absent in our analysis because their inclusion in either \micro or \texttt{LanHEP} is rather nontrivial, given the fact that \micro relies heavily on autogenerated \lhep codes. 

The relic density from \micro shows a local step-function reduction of 5\%--12\% in a small range of couplings over the mass range $100\lesssim m_{S,V}/\mathrm{GeV} \lesssim 400$ and $5 \lesssim m_{\chi,\psi}/\mathrm{TeV} \lesssim 180$. This feature is not reproduced in the annihilation cross section, relic densities from the \micro routines employing the freeze-out approximation, or in our previous calculations for the scalar model \cite{Cline:2013gha}. Away from the feature, the full \micro relic densities are in good agreement with the freeze-out approximation and our previous results \cite{Cline:2013gha}.  The drop appears symptomatic of a numerical error in the full \micro Boltzmann solver, possibly due to poor convergence properties (it is not solved by the \emph{accurate} setting).  As the regions where this occurs are only a few tenths of a unit wide in the log of $\lambda_{hX}$ or the log of $(\lambda_{hX}/\Lambda_X)$ as appropriate, we simply omit them from our limit curves, interpolating across the small resulting gap. 

Another important consideration is the validity of our EFT approximation for the fermionic models, which involve dimension-5 effective operators. In general the validity of an EFT relies on the momentum exchange $q$ of the interaction being below the mass of the mediator $\phi$ involved in the underlying interaction described by the effective vertex.  This means that $m_\phi$ dominates the denominator of the internal propagator, allowing $q$ to be neglected and $\phi$ therefore to be ``integrated out.'' A given nonrenormalizable effective operator is therefore valid only if its dimensionful coupling implies a mediator mass below the interaction scale of the process it is used to describe.  For DM annihilation, this is simply $q \sim \sqrt{s} = 2m_X$, where $X \in (\chi,\psi)$, implying that the EFT is valid for indirect detection when $m_\phi > 2m_X$. For scattering of halo DM particles with nuclei in detectors on Earth, the momentum exchange occurs deep in the nonrelativistic regime, and $q$ is of order a few tens of MeV.  This implies that the EFT is valid for direct detection when $m_\phi > \mathcal{O}$(MeV).

To see what values of $\lambda_{hX}/\Lambda_X$ imply mediator masses consistent with the EFT approximation, it is instructive to consider a simple UV completion of the fermion EFTs.  Assume the $\phi$ is a scalar that interacts with the fermion field $X$ and the Higgs doublet $H$ through dimensionless couplings $g\indices{_{\phi X}}$ and $g\indices{_{\phi H}}$ respectively, as in the scenario considered by Ref.~\cite{Baek:2011aa}.\footnote{In this specific UV completion, the new scalar $\phi$ mixes with the SM Higgs boson $h$ and modifies the SM-like Higgs coupling to SM gauge bosons and fermions. In general this mixing should be strongly suppressed for $m_\phi \gg m_h$ (i.e. when the EFT is valid), but the presence and strength of this interaction is ultimately highly dependent on the details of the UV completion.  For our analysis in this paper, in line with the EFT assumption, we assume that such modifications are absent and the Higgs production cross section and visible decay widths are the same as in the SM.} In the $q^2 \ll m_\phi^2$ limit, the coupling $\lambda_{hX}/\Lambda_X$ in the EFT approach can be identified with $(g\indices{_{\phi X}} \tilde{g}\indices{_{\phi H}})/m^2_\phi$ where $\tilde{g}\indices{_{\phi H}} = m' g\indices{_{\phi H}}$, such that $m'$ is some characteristic mass scale of the new coupling (e.g. from a new Yukawa-type interaction). To place an upper limit on the values of $\lambda_{hX}/\Lambda_X$ where the EFT would be valid in \emph{some} UV completion, we therefore require $m_\phi$ to be as small as possible (i.e., $2m_X$ for DM annihilations) and the product $m' g\indices{_{\phi X}} g\indices{_{\phi H}}$ as large as possible.  The largest value $m'$ can realistically take is simply $m_\phi$ (as it would otherwise have already been integrated out), leaving $\lambda_{hX}/\Lambda_X \sim (g\indices{_{\phi X}} g\indices{_{\phi H}})/m_\phi$.  For the EFT itself to remain perturbative, we require  the product $g\indices{_{\phi X}} g\indices{_{\phi H}} \lesssim 4\pi$. Hence, the approximate values of $\lambda_{hX}/\Lambda_X$ for which we generically expect to start getting corrections to the EFT from UV effects in DM annihilation are $\lambda_{hX}/\Lambda_X \gtrsim 4\pi/2m_X$.  We stress that this is the value at which the EFT is \emph{sure} to break down for indirect detection and relic density calculations; depending on the UV completion, e.g. for weakly coupled theories, this scale could be a lot lower.  In contrast, the EFT approximation is perfectly valid for direct detection so long as the mediator is more massive than a few MeV, implying $\log_{10}(\lambda_{hX}/\Lambda_X \, \textrm{GeV}) \lesssim 2$.

In the fermion model plots, we show the regions $\lambda_{hX}/\Lambda_X > 4\pi/2m_X$ where the EFT approximation breaks down for DM annihilation. Whilst the EFT is valid for direct searches at all couplings that we show, by choosing to rescale the limits by the relic abundance parameter $f_{\textrm{rel}} \equiv \Omega_X/\Omega_{\textrm{DM}}$ which is itself subject to UV corrections, our direct search limits are indirectly affected when $\lambda_{hX}/\Lambda_X > 4\pi/2m_X$.

\subsection{Scalar model}
\begin{figure}
    \begin{minipage}[t]{0.48\textwidth}  
      \centering  
      \includegraphics[width=\textwidth]{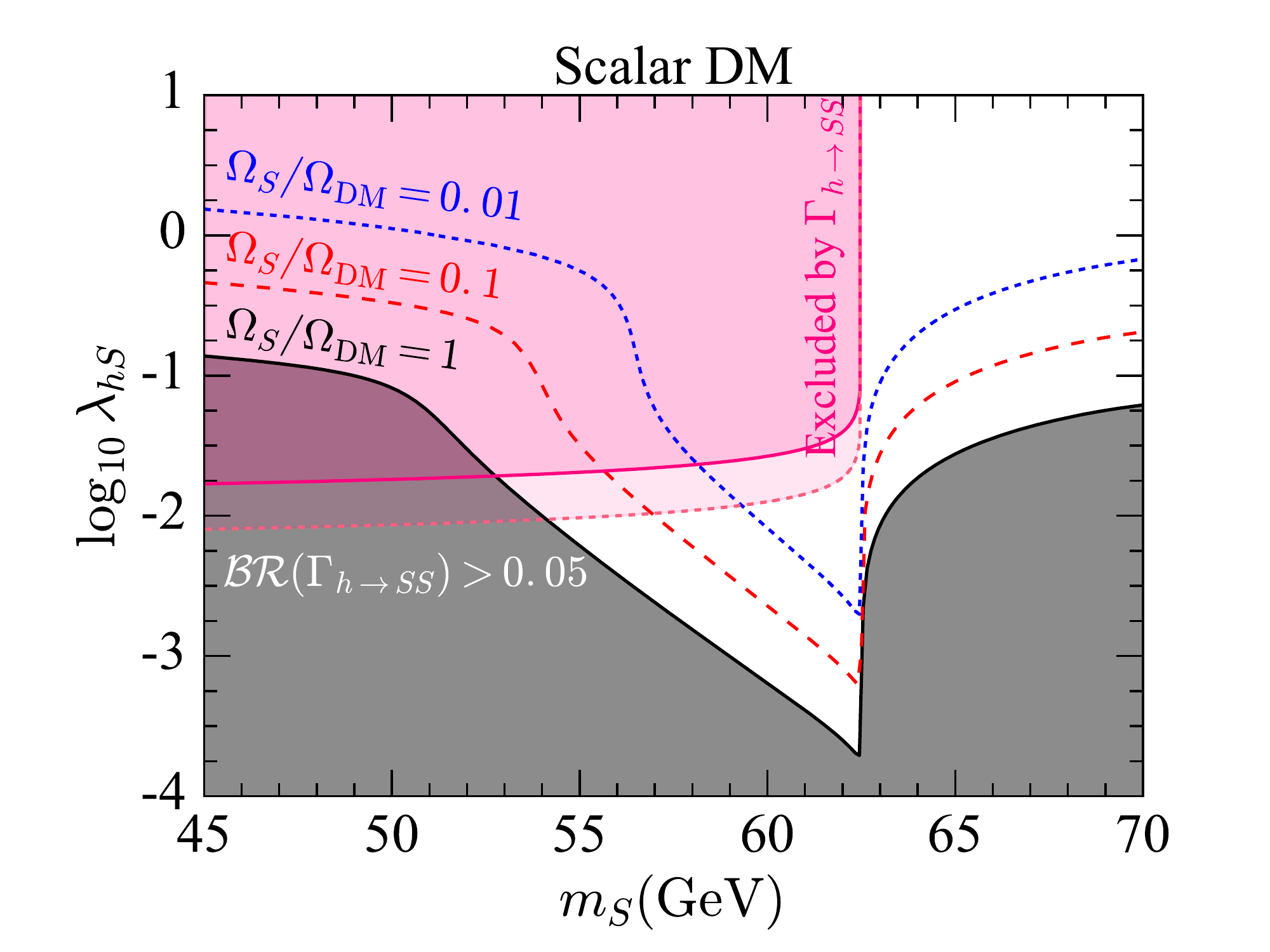}
    \end{minipage}
    \hfil
    \begin{minipage}[t]{0.48\textwidth}
      \centering
      \includegraphics[width=\textwidth]{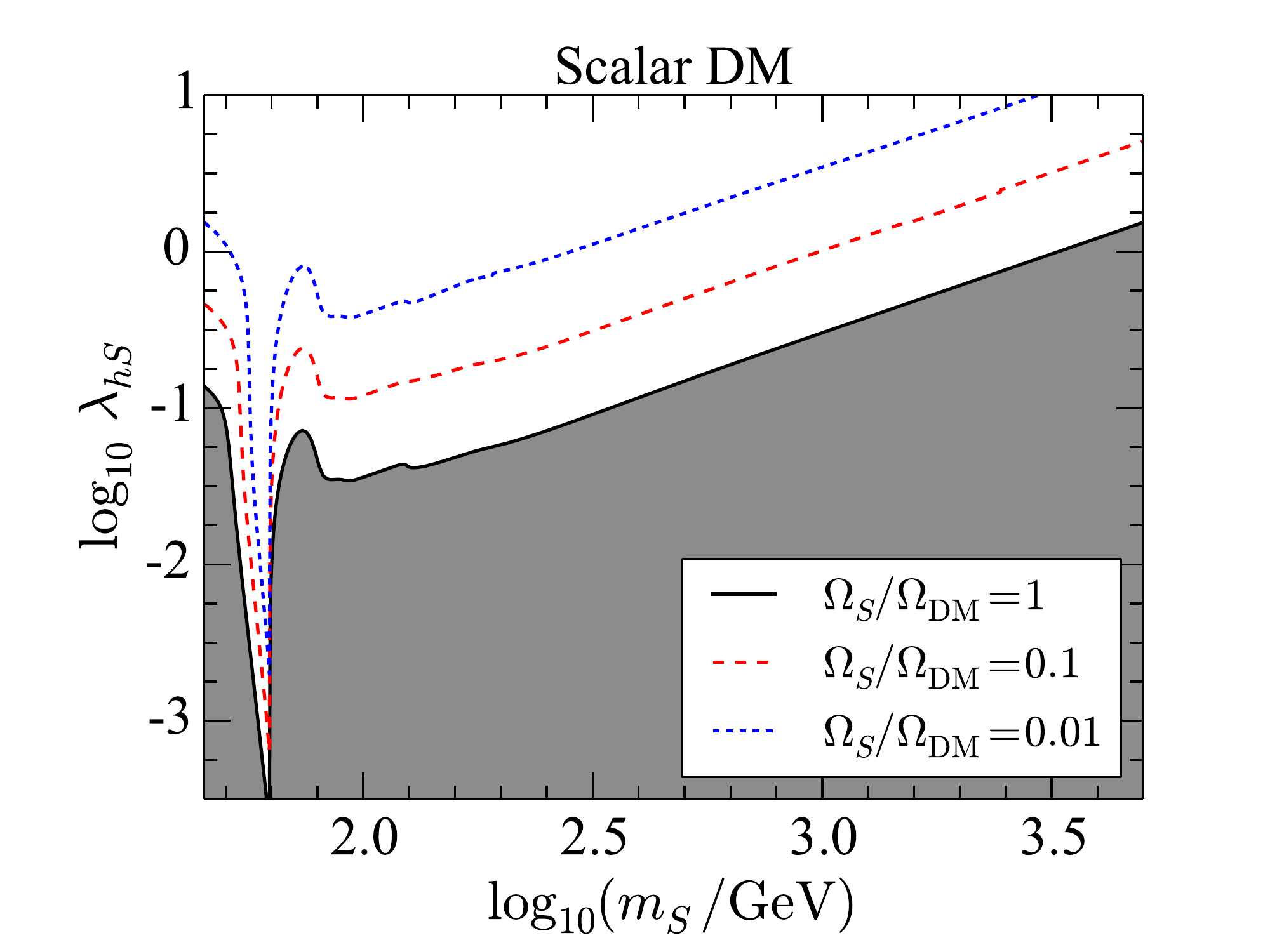}
    \end{minipage}
    \caption{Contours of fixed scalar relic density for $f_{\textrm{rel}} = 1$ (black solid), 0.1 (red dashed) and 0.01 (blue dotted). The grey shaded region is excluded due to an overabundance of dark matter. Left: A close-up of the resonantly enhanced annihilation region, $m_S \sim m_h/2$. Larger values of $\lambda_{hS}$ are excluded by an upper limit of 19\% (pink solid) at 2$\sigma$\,C.L. or 5\% (pink dotted) at 1$\sigma$\,C.L. on $\mathcal{BR}(h \rightarrow SS)$. Right: Relic density contours for the full range of $m_S$.}    
    \label{scalar-relic} 
	\vspace{2.5mm}    
	\begin{minipage}[t]{0.48\textwidth}        
      \centering
      \includegraphics[width=\textwidth]{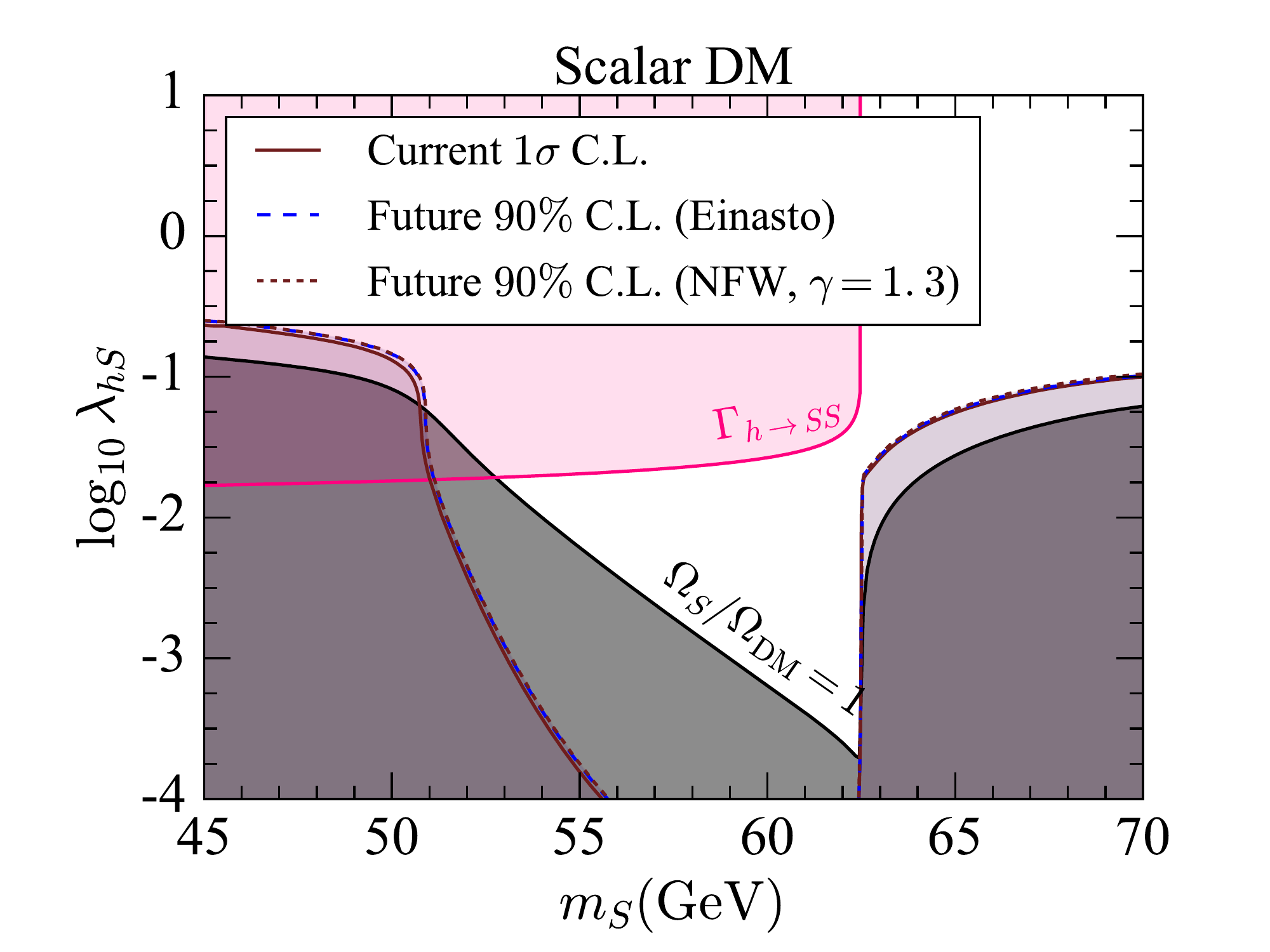}
    \end{minipage}
    \hfil
    \begin{minipage}[t]{0.48\textwidth}
      \centering
      \includegraphics[width=\textwidth]{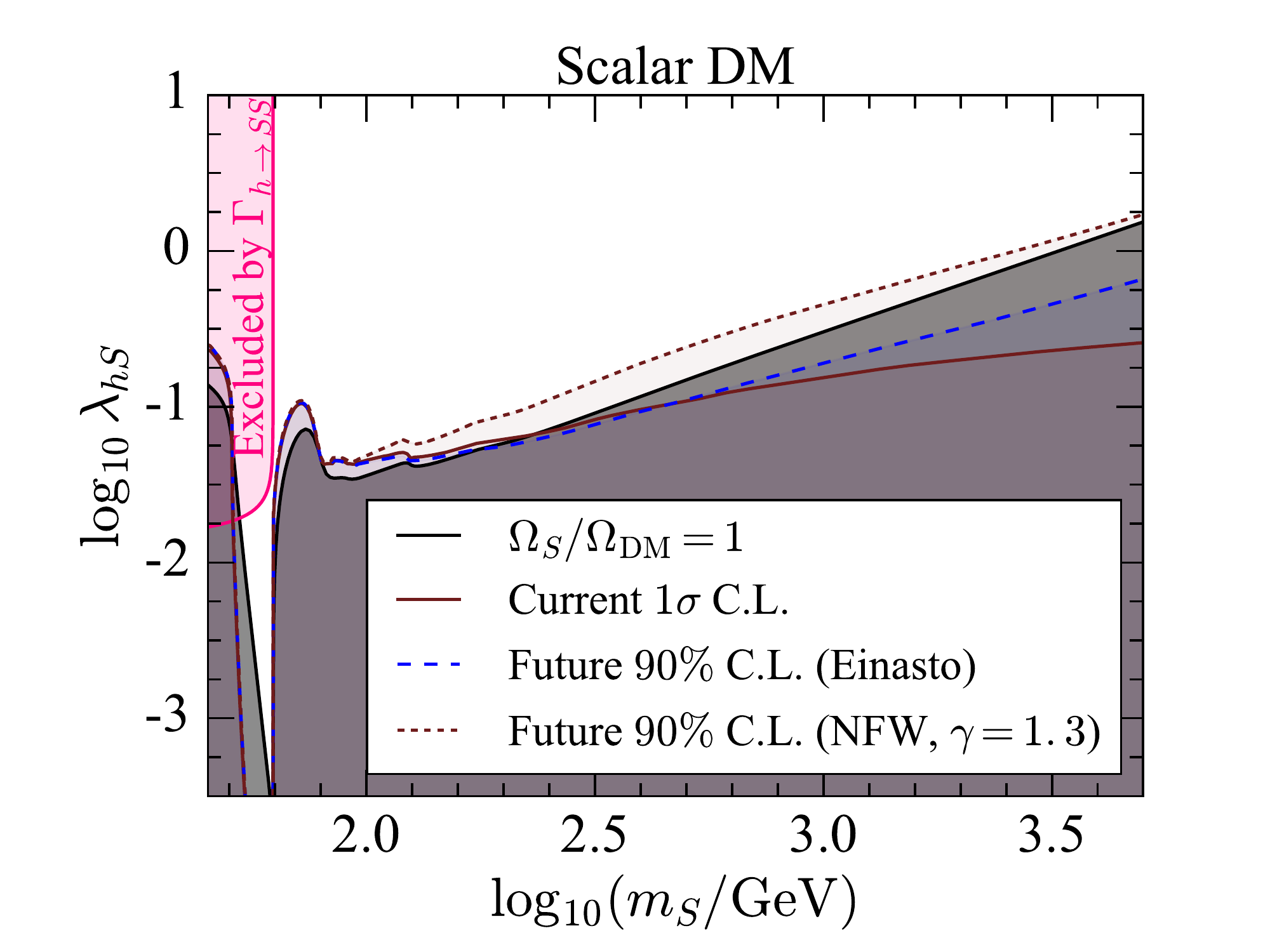}
    \end{minipage}
    \caption{Indirect search limits on the scalar model parameter space. The grey and pink shaded regions are excluded respectively by the observed DM relic density and an upper limit of 19\% on $\mathcal{BR}(h \rightarrow SS)$ at $2\sigma$\,C.L. Values of $\lambda_{hS}$ below the current 1$\sigma$\,C.L. (brown solid) curve are excluded at more than $1\sigma$\,C.L. Regions below the future 90\%\,C.L. curve with the Einasto (blue dashed) and contracted NFW (brown dotted) profile will be excluded. Left: A close-up of the resonantly enhanced annihilation region, $m_S \sim m_h/2$. Right: The full range of $m_S$.}
    \label{scalar-indirect}	
\end{figure}

We perform scans in the $(m_S, \lambda_{hS})$ plane and generate contours of fixed scalar relic density $(\Omega_S h^2)$ for $f_{\textrm{rel}}$ = 1 (black solid), 0.1 (red dashed) and 0.01 (blue dotted) as shown in Fig.~\ref{scalar-relic}. Values of $\lambda_{hS}$ in the grey shaded region are excluded by the relic density constraint. This is due to the fact that a lower value of $\lambda_{hS}$ gives smaller $\asigv$. Following Eq.~\eqref{relicDform}, there is therefore an overabundance of the scalar $S$ relic density. 

In the region $m_S < m_h/2$, an upper limit of 19\% (pink solid) at $2\sigma$\,C.L. and 5\% (pink dotted) at $1\sigma$\,C.L. on the Higgs invisible branching ratio $\mathcal{BR}(h \rightarrow SS)$ excludes coupling larger than $\log_{10}\lambda_{hS}\sim-1.75$ and $\log_{10} \lambda_{hS} \sim -2.1$ respectively. The combined constraints on the scalar relic density and the Higgs invisible width exclude lower scalar masses apart from a small triangle between 54\,GeV and $m_h/2$.

Near the resonance $m_S \sim m_h/2$, the annihilation cross section $\sigma v_{\textrm{rel}}$ is enhanced. Consequently, the relic density contours move to lower values of $\lambda_{hS}$ to compensate for the enhancement. Above $m_S > m_h/2$, the relic density contours scale essentially linearly with $\log_{10} m_S$. 

In Fig.~\ref{scalar-indirect}, we show the combined sensitivity of indirect searches to various regions of the scalar model parameter space. For the current limits, which include the combined analysis of 15 dwarf galaxies by Fermi-LAT and 7-year observations of the CMB (WMAP7), we only present $1\sigma$\,C.L. (brown solid). The region $m_h/2 \lesssim m_S < 70$\,GeV with $\log_{10}\lambda_{hS} \in [-2.5,-1]$ can be seen to be in tension with the current indirect searches at slightly more than $1\sigma$\,C.L. The same is true for scalar masses below $\sim$\,51\,GeV; however this region is currently excluded by the Higgs invisible width constraint at more than 2$\sigma$\,C.L. In extending the current indirect search limits to higher scalar masses, we find that scalar masses up to $\sim$\,232\,GeV are excluded by the current indirect DM searches at more than 1$\sigma$\,C.L., if $S$ makes up all of the dark matter.

The combined future limits incorporate the Planck polarization data, $\gtrsim$\,100\,h GC observation by CTA and extended improvements in the Fermi-LAT data from the addition of more southern dwarf galaxies in its search. Due mainly to better exposure, future indirect DM searches will be sensitive enough to probe higher scalar DM masses, if $S$ makes up all of the dark matter. At low scalar masses, the future DM searches are relatively insensitive to the assumed DM density profile (Einasto or a contracted NFW). This is mainly due to the fact that the Fermi log-likelihood dominates in this regime. However, at higher scalar masses, the CTA log-likelihood entering in Eq.~\eqref{future} gives the dominant contribution to the total log-likelihood. Hence, the upcoming CTA experiment will be able to exclude scalar masses up to $\sim$\,176\,GeV (for the Einasto profile) and $\sim$\,9\,TeV (for a contracted NFW profile) at more than $90\%$\,C.L., if $S$ makes up all of the dark matter.

\begin{figure}
    \begin{minipage}[t]{0.48\textwidth}  
      \centering      
      \includegraphics[width=\textwidth]{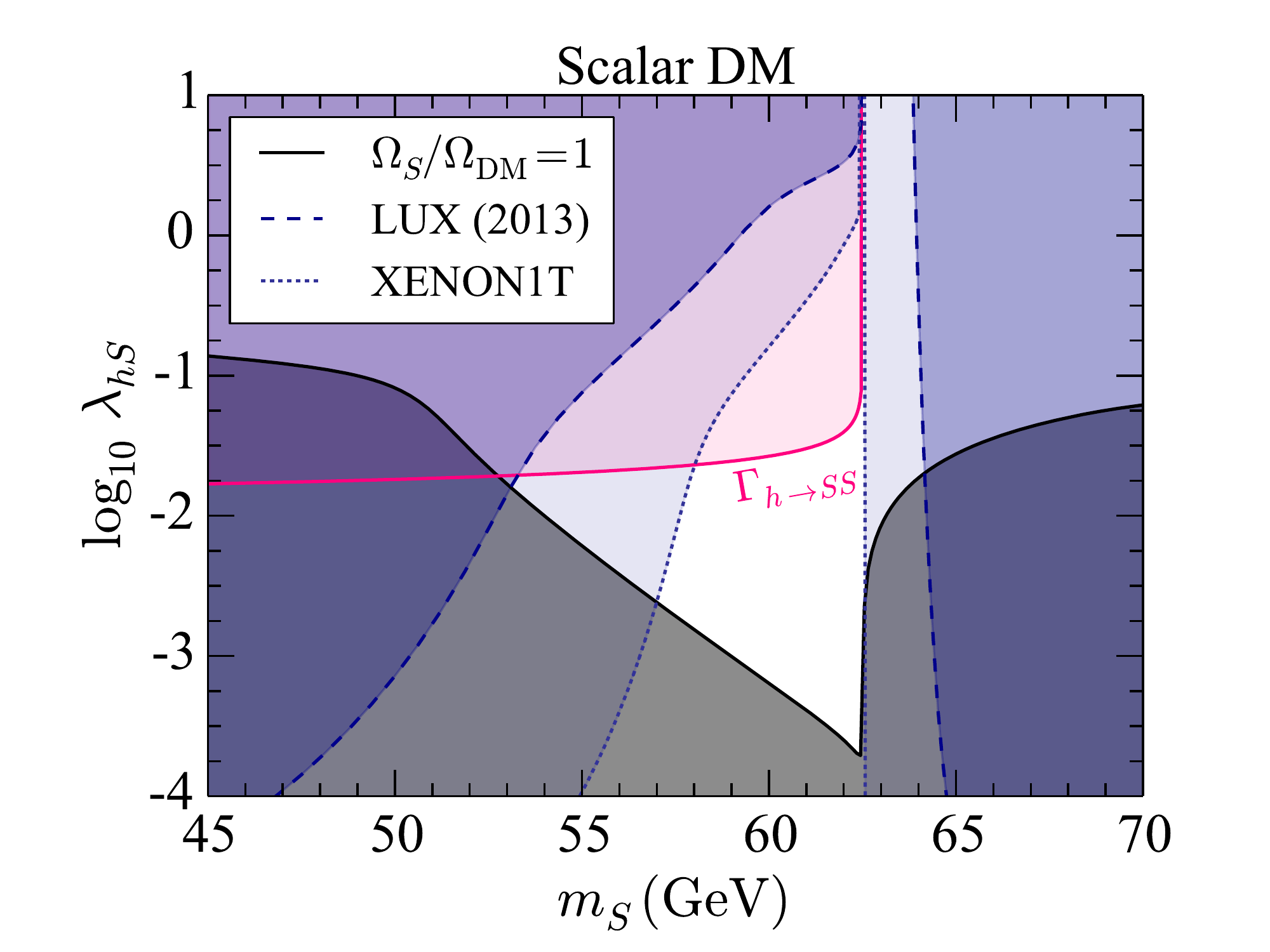}
    \end{minipage}
    \hfil
    \begin{minipage}[t]{0.48\textwidth}
      \centering
      \includegraphics[width=\textwidth]{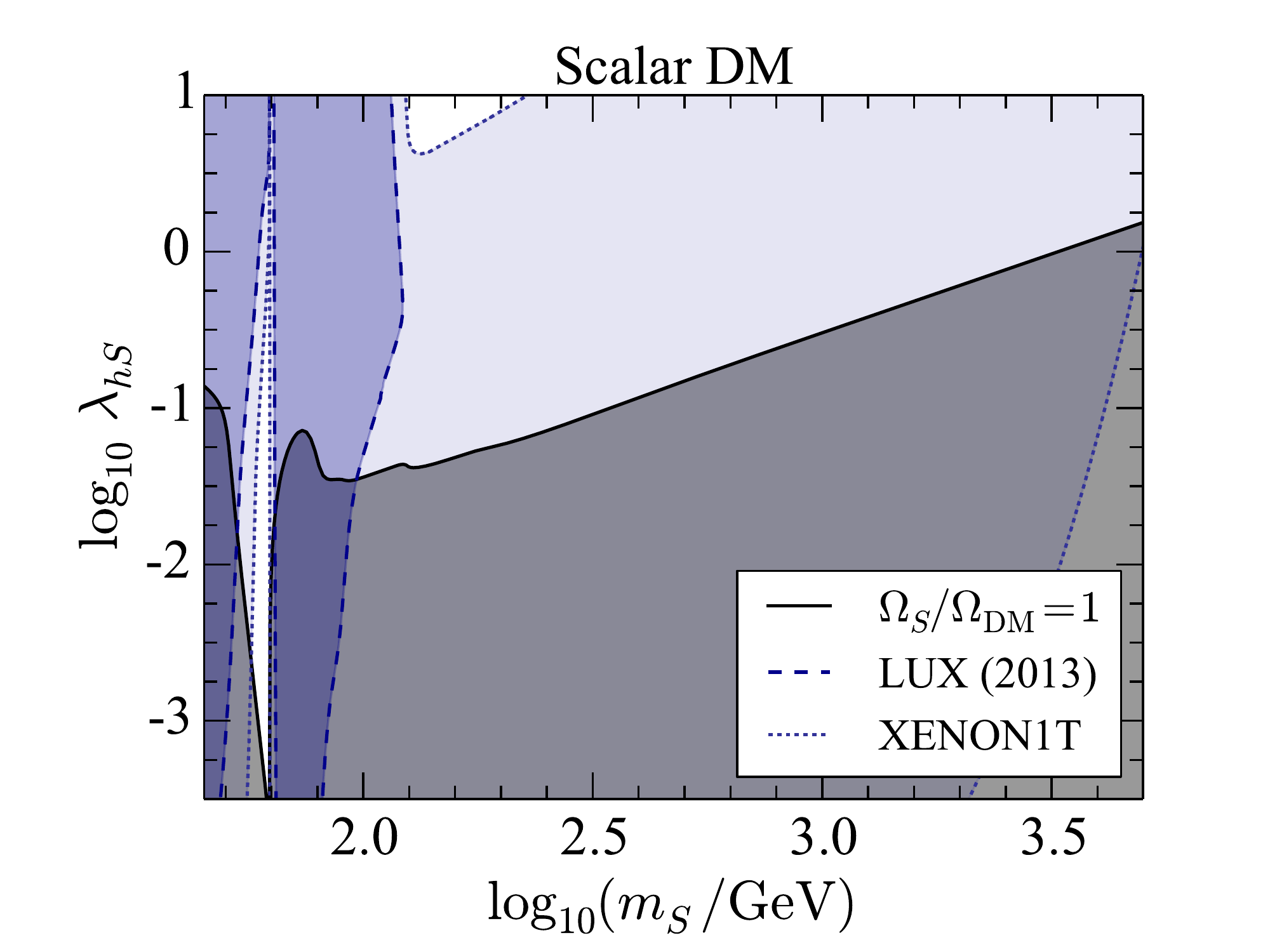} 
    \end{minipage}
    \caption{Direct search limits on the scalar model parameter space. The grey shaded region is ruled out by the observed DM relic density. Regions excluded by the LUX (XENON1T) experiment are delineated with blue dashed (blue dotted) curves and dark (light) shadings. Left: A close-up of the resonantly enhanced annihilation region, $m_S \sim m_h/2$. The pink shaded region is excluded by an upper limit of $19\%$ on $\mathcal{BR}(h \rightarrow SS)$ at 2$\sigma$\,C.L. Right: The full range of $m_S$.}
    \label{scalar-direct}
\end{figure}

The resulting limits in the $(m_S,\lambda_{hS})$ plane from the LUX (blue dashed) and the projected XENON1T (blue dotted) experiment are shown in Fig.~\ref{scalar-direct}. In the left panel near the resonance $m_S \sim m_h/2$, a small triangle will continue to evade detection at the LUX and projected XENON1T experiments. At high scalar masses, the LUX experiment excludes scalar masses up to 120\,GeV for a narrow range of $\lambda_{hS}$ values. Most of the remaining parameter space will be tested and ruled out by the XENON1T experiment for a wide range of couplings $\lambda_{hS}$. In particular, it will be able to exclude scalar masses up to 10\,TeV, if $S$ makes up all of the dark matter. 

\subsection{Vector model}
\begin{figure}
    \begin{minipage}[t]{0.48\textwidth}        
      \centering
      \includegraphics[width=\textwidth]{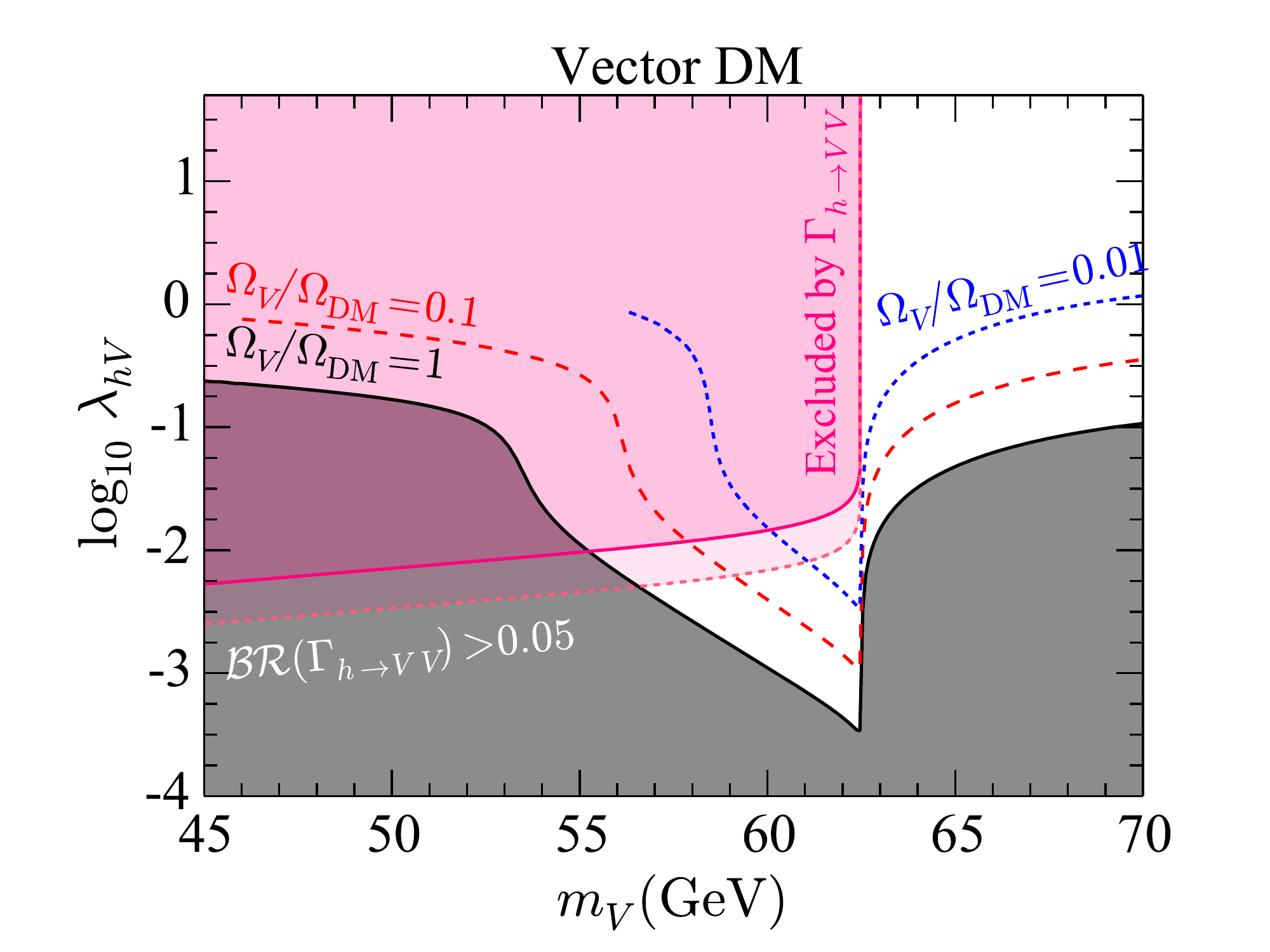}
    \end{minipage}
    \hfil
    \begin{minipage}[t]{0.48\textwidth}
      \centering
      \includegraphics[width=\textwidth]{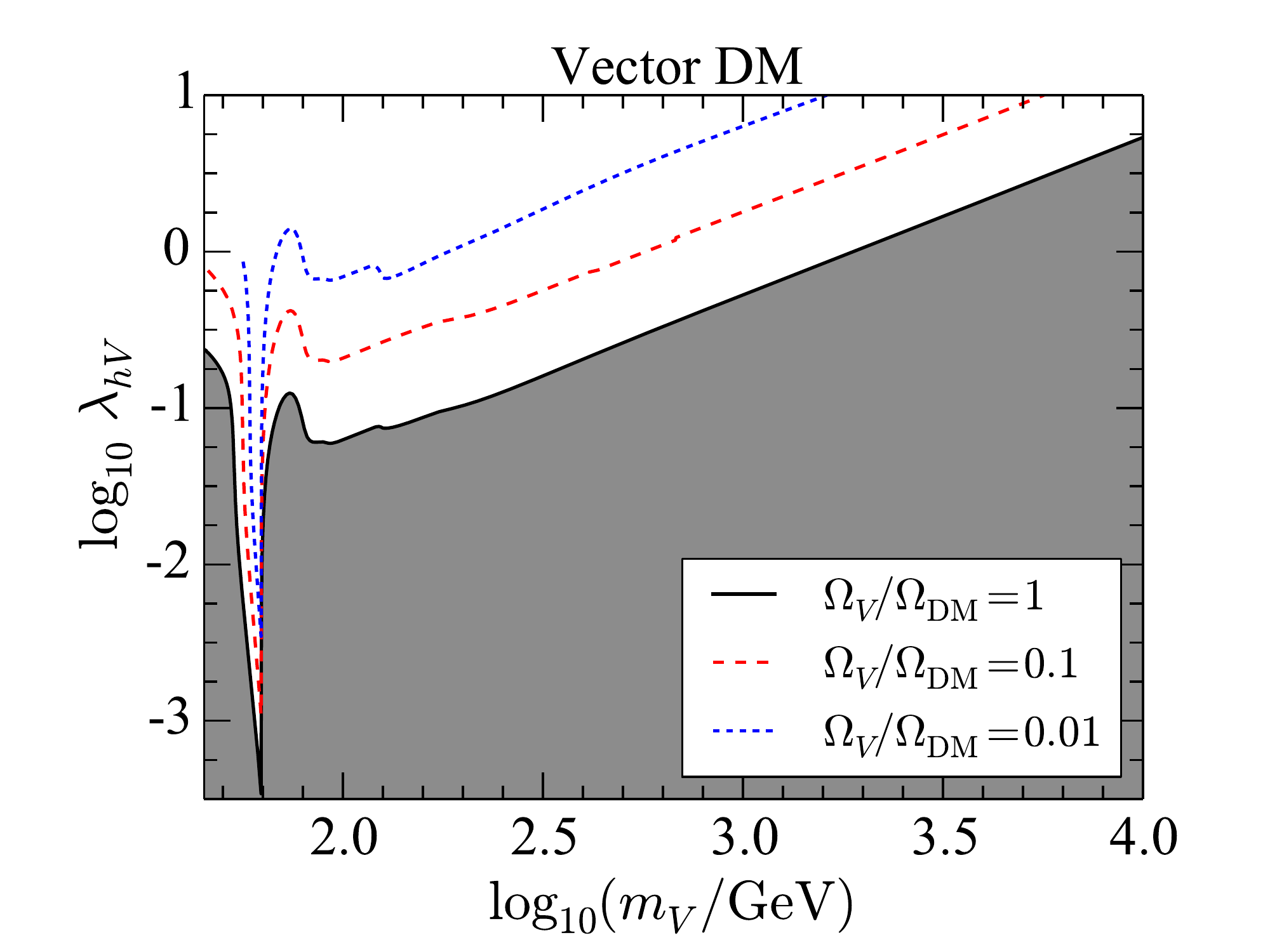}
    \end{minipage}
    \caption{Contours of fixed vector relic density for $f_{\textrm{rel}} = 1$ (black solid), 0.1 (red dashed) and 0.01 (blue dotted). The  grey shaded region is excluded due to an overabundance of dark matter. Left: A close-up of the resonantly enhanced annihilation region, $m_V \sim m_h/2$. Larger values of $\lambda_{hV}$ are excluded by an upper limit of 19\% (pink solid) at 2$\sigma$\,C.L. or 5\% (pink dotted) at 1$\sigma$\,C.L. on $\mathcal{BR}(h \rightarrow VV)$. Right: Relic density contours for the full range of $m_V$.}
    \label{vector-relic}  	    
\end{figure}

The contours of fixed vector relic density $(\Omega_V h^2)$ for $f_{\textrm{rel}} =$ 1 (black solid), 0.1 (red dashed) and 0.01 (blue dotted) are shown in Fig.~\ref{vector-relic}. For $f_\textrm{rel}$ = 0.1 (0.01), no values of $\lambda_{hV}$ satisfy Eq.~\eqref{root-relic} below $46 \,(56)$\,GeV. The minimum relic density below these vector masses stays above the values of the contour being drawn, thereby leaving gaps that are evident in the left panel of Fig.~\ref{vector-relic}.

In the region $m_V < m_h/2$, upper limits of 19\% (pink solid) at 2$\sigma$\,C.L. and 5\% (pink dotted) at 1$\sigma$\,C.L. on the Higgs invisible branching ratio $\mathcal{BR}(h\rightarrow VV)$ exclude couplings larger than $\log_{10} \lambda_{hV} \sim -2.28$ and $\log_{10} \lambda_{hV} \sim -2.6$ respectively. The combined constraints on the vector relic density and the Higgs invisible width exclude most of the low vector masses apart from a small triangle between 56.5\,GeV and $m_h/2$. Above $m_V > m_h/2$, the relic density contours scale linearly with $\log_{10} m_V$ in a similar fashion to the scalar model.
   
\begin{figure} 
	\begin{minipage}[t]{0.48\textwidth}        
      \centering
	  \includegraphics[width=\textwidth]{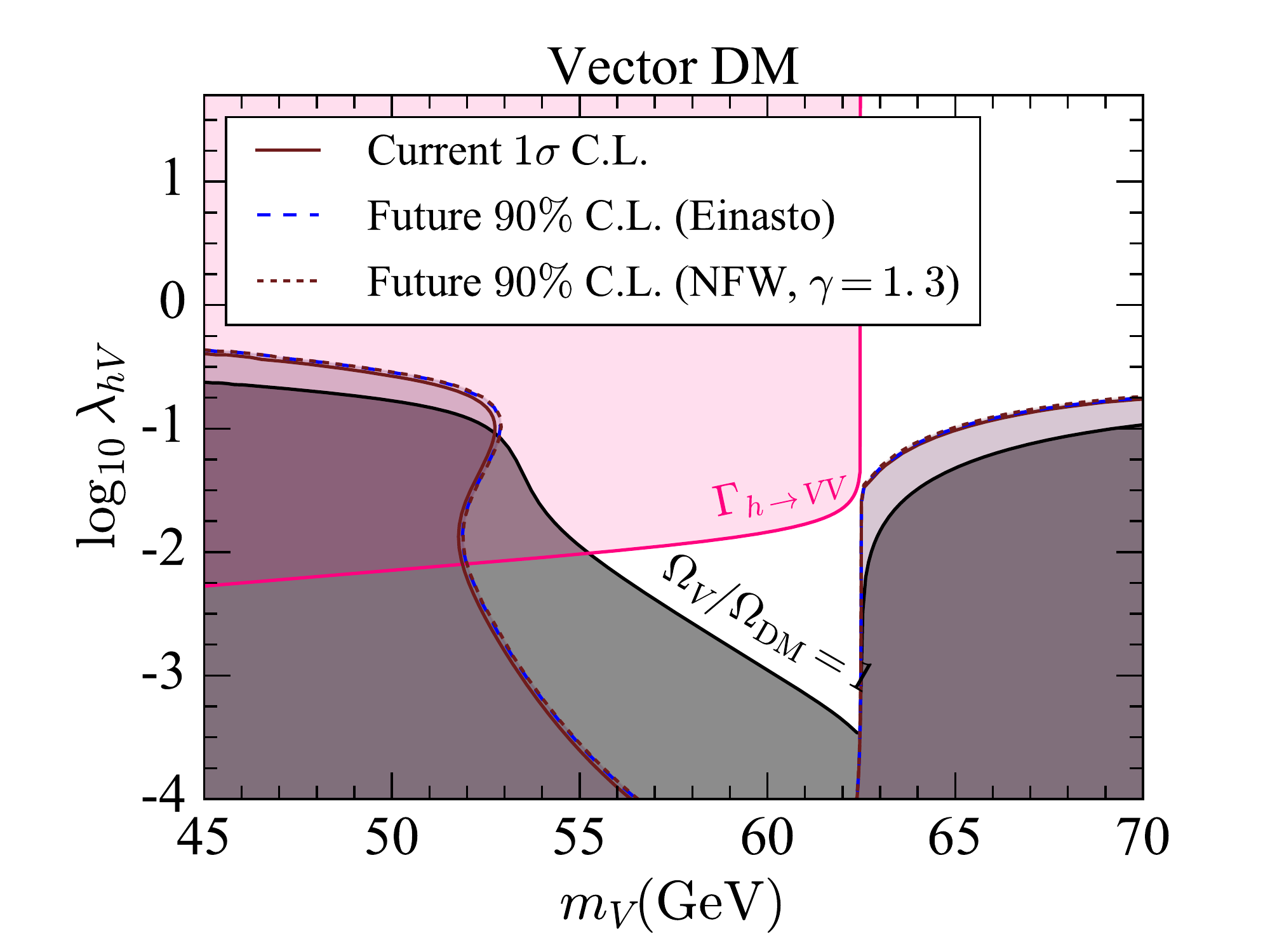}
    \end{minipage}
    \hfil
    \begin{minipage}[t]{0.48\textwidth}
      \centering
      \includegraphics[width=\textwidth]{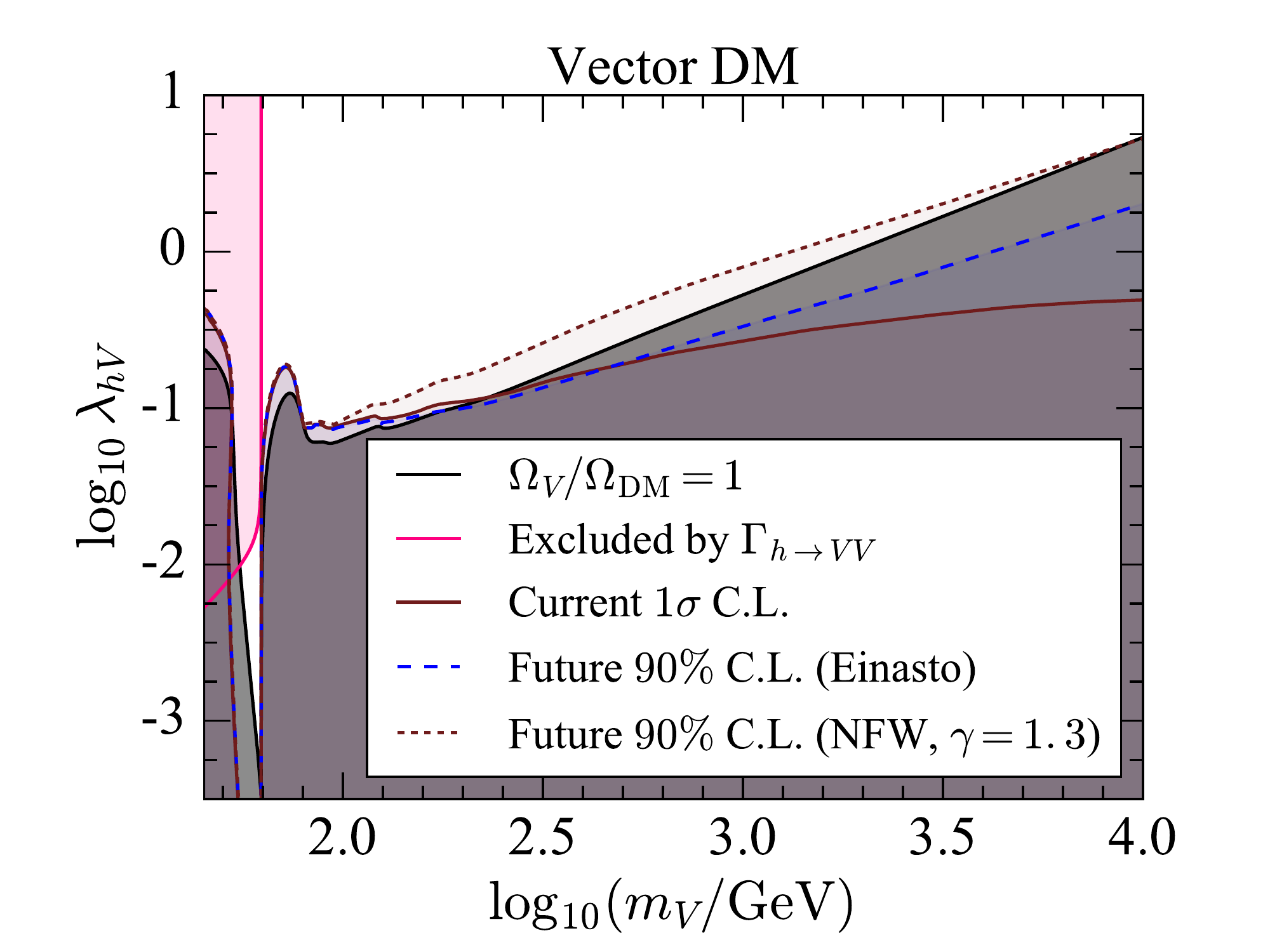}
    \end{minipage}   
    \caption{Indirect search limits on the vector model parameter space. The grey and pink shaded regions are excluded respectively by the observed DM relic density and an upper limit of 19\% on $\mathcal{BR}(h \rightarrow VV)$ at 2$\sigma$\,C.L. Values of $\lambda_{hV}$ below the current 1$\sigma$\,C.L. (brown solid) curve are excluded at more than $1\sigma$\,C.L. Regions below the future 90\%\,C.L. curve with the Einasto (blue dashed) and contracted NFW (brown dotted) profile will be excluded. Left: A close-up of the resonantly enhanced annihilation region, $m_V \sim m_h/2$. Right: The full range of $m_V$.}
    \label{vector-indirect} 
    \vspace{2.5mm}
    \begin{minipage}[t]{0.48\textwidth}        
      \centering
      \includegraphics[width=\textwidth]{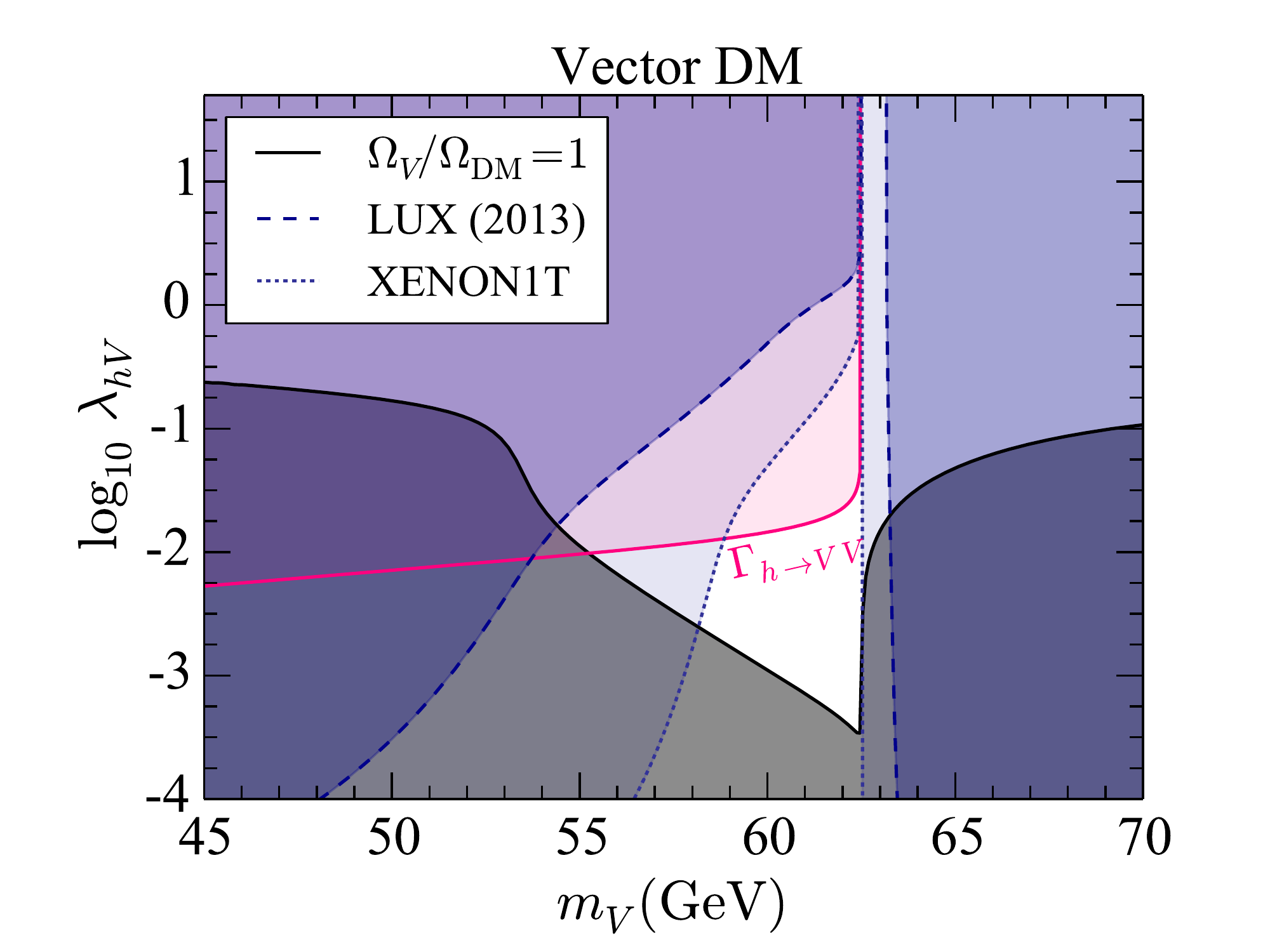}
    \end{minipage}
    \hfil
    \begin{minipage}[t]{0.48\textwidth}
      \centering
      \includegraphics[width=\textwidth]{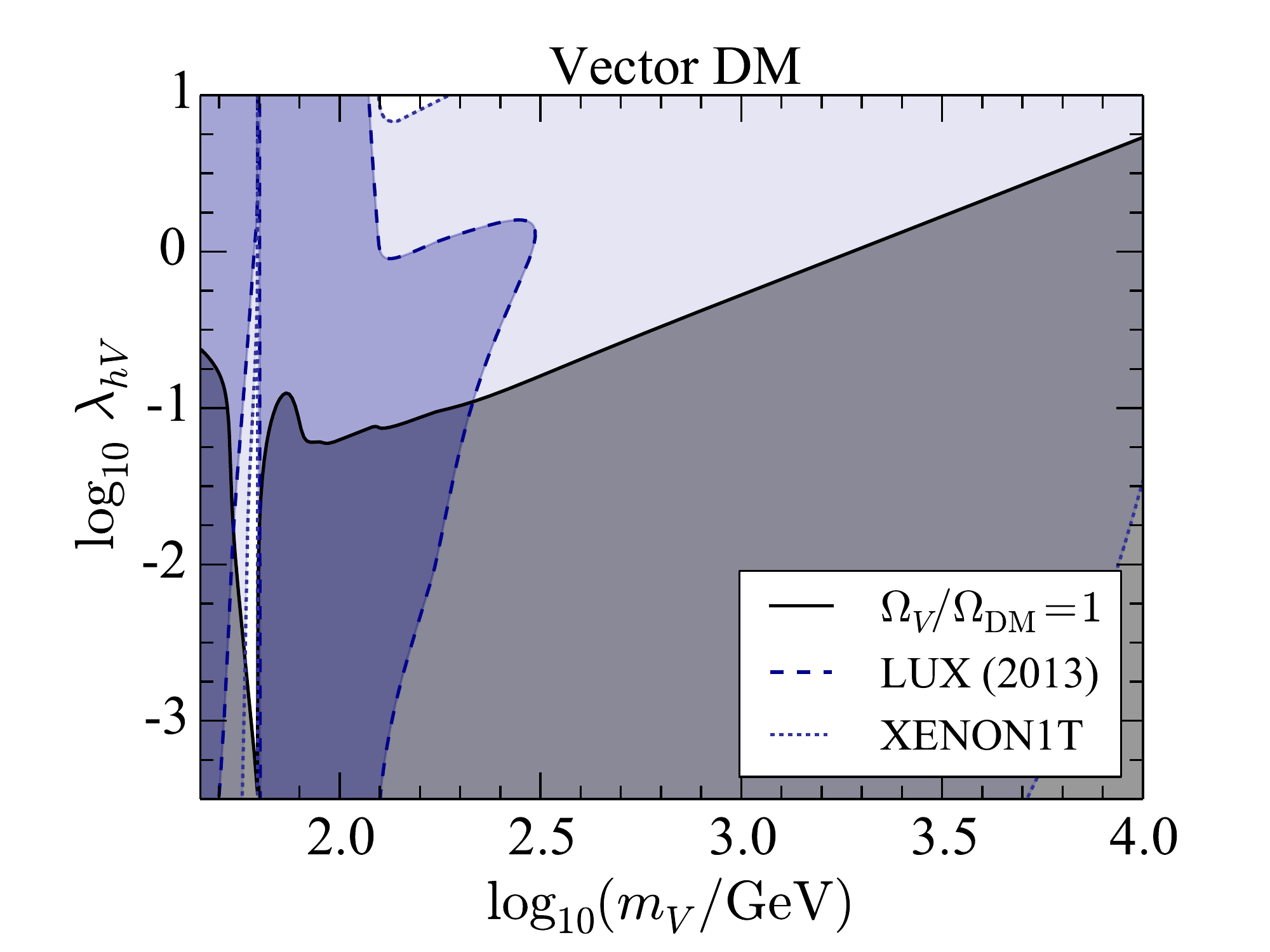} 
    \end{minipage}
    \caption{Direct search limits on the vector model parameter space. The grey shaded region is ruled out by the observed relic density of dark matter. Regions excluded by the LUX (XENON1T) experiment are delineated with blue dashed (blue dotted) curves and dark (light) shadings. Left: A close-up of the resonantly enhanced annihilation region, $m_V \sim m_h/2$. The pink shaded region is excluded by an upper limit of $19\%$ on $\mathcal{BR}(h \rightarrow VV)$ at 2$\sigma$\,C.L. Right: The full range of $m_V$.}
    \label{vector-direct} 
\end{figure}
    
The combined sensitivity of indirect searches to various regions in the $(m_V,\lambda_{hV})$ plane is shown in Fig.~\ref{vector-indirect}. The current limits involve contributions from WMAP7 and the combined analysis of 15 dwarf galaxies by Fermi-LAT. The region $m_h/2 \leq m_V \leq 70$\,GeV with $\log_{10}\lambda_{hV} \in [-2.5,-0.75]$ can be seen to be in tension with the current indirect searches. The same is true at vector masses below $\sim$\,52.5\,GeV; however this region is already excluded by the Higgs invisible width constraint at more than $2\sigma$\,C.L. Vector masses up to $\sim$\,230\,GeV are excluded by the current indirect DM searches at more than $1\sigma$\,C.L., if $V$ makes up all of the dark matter.

Limits from the future indirect DM searches will be able to probe parts of the parameter space that are not already excluded by either the current indirect searches or the relic density constraints. In particular, future searches based on the Einasto and contracted NFW DM profiles will exclude vector masses up to $\sim$\,173\,GeV and $\sim$\,9\,TeV respectively at more than $90\%$\,C.L., if $V$ makes up all of the dark matter.

Limits from the LUX (blue dashed) and the projected XENON1T (blue dotted) experiment in the $(m_V,\lambda_{hV})$ plane are shown in Fig.~\ref{vector-direct}. The LUX experiment excludes vector masses up to $\sim$\,300\,GeV for a moderate range of $\lambda_{hV}$ values, whereas the projected XENON1T experiment will exclude most parts of the parameter space that are not presently ruled out by the DM relic density and the Higgs invisible width constraints. In particular, it will be able to exclude vector masses up to $\sim$\,30\,TeV, if $V$ makes up all of the dark matter.

\subsection{Majorana fermion model}
\begin{figure}
    \begin{minipage}[t]{0.48\textwidth}        
      \centering
      \includegraphics[width=\textwidth]{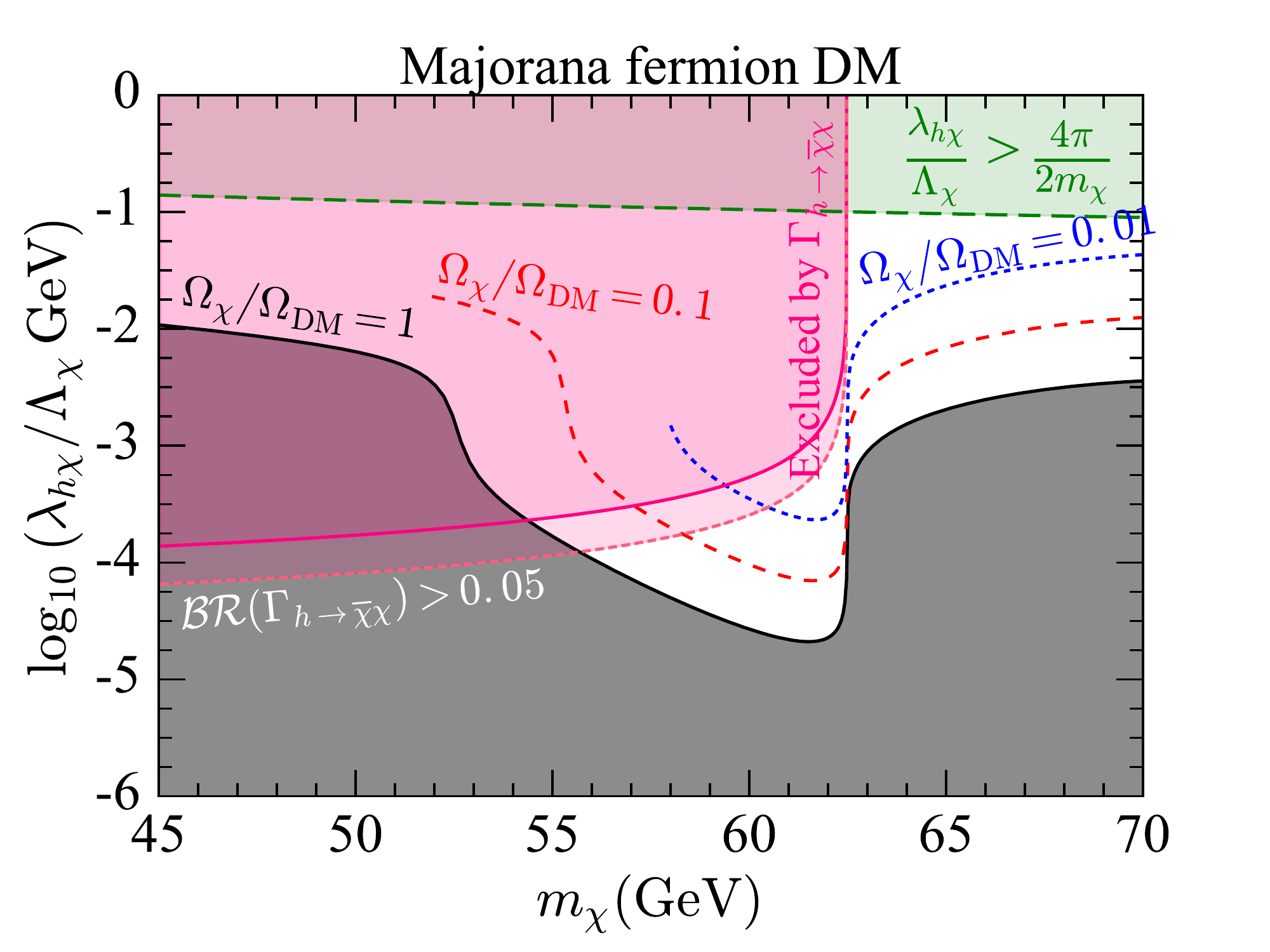}
    \end{minipage}
    \hfil
    \begin{minipage}[t]{0.48\textwidth}
      \centering
      \includegraphics[width=\textwidth]{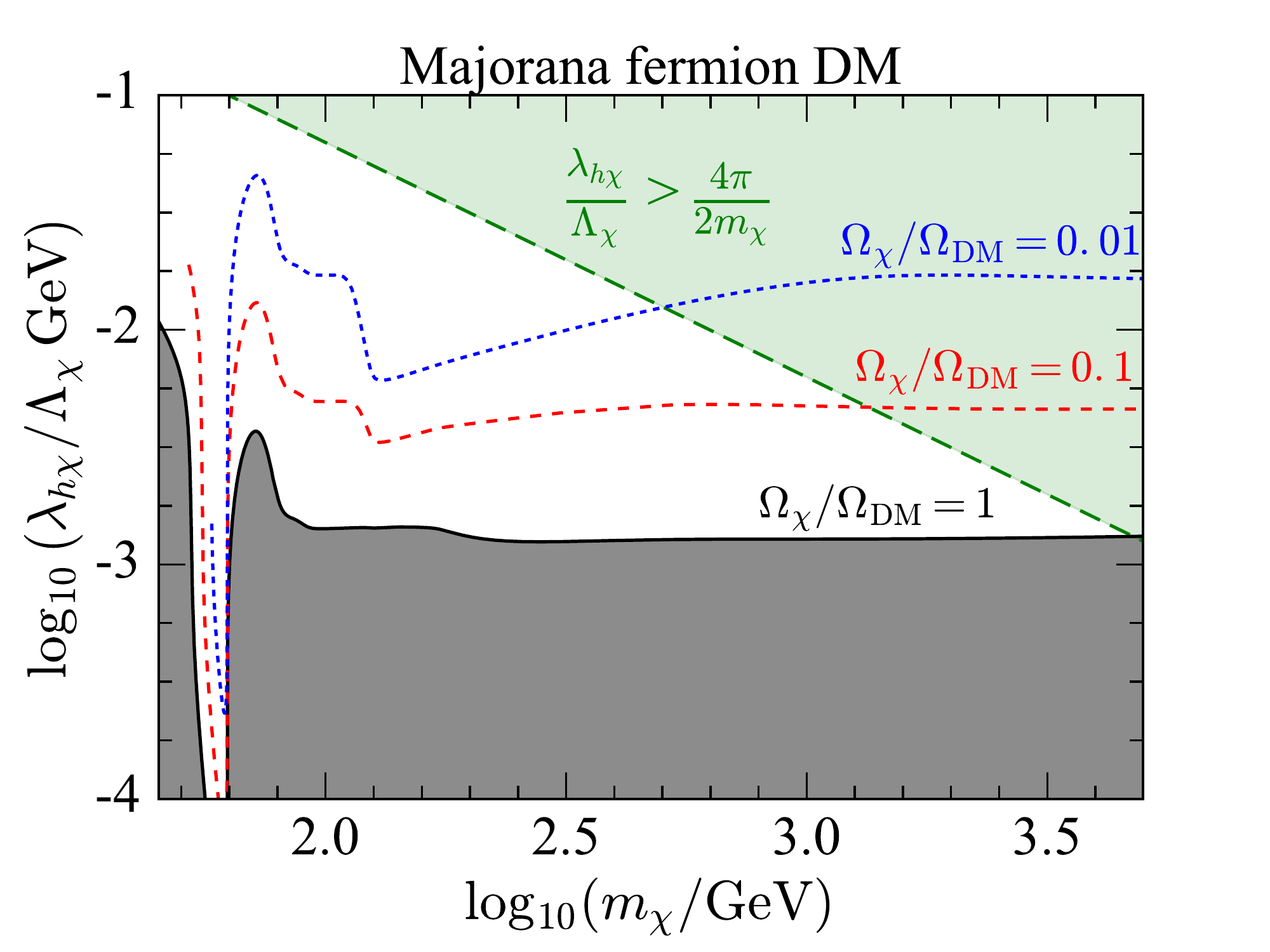}  
    \end{minipage} 
	
	\small{\emph{a}). Pure scalar interaction, $\cos \xi = 1$.} 
	
	\begin{minipage}[t]{0.48\textwidth}        
      \centering
      \includegraphics[width=\textwidth]{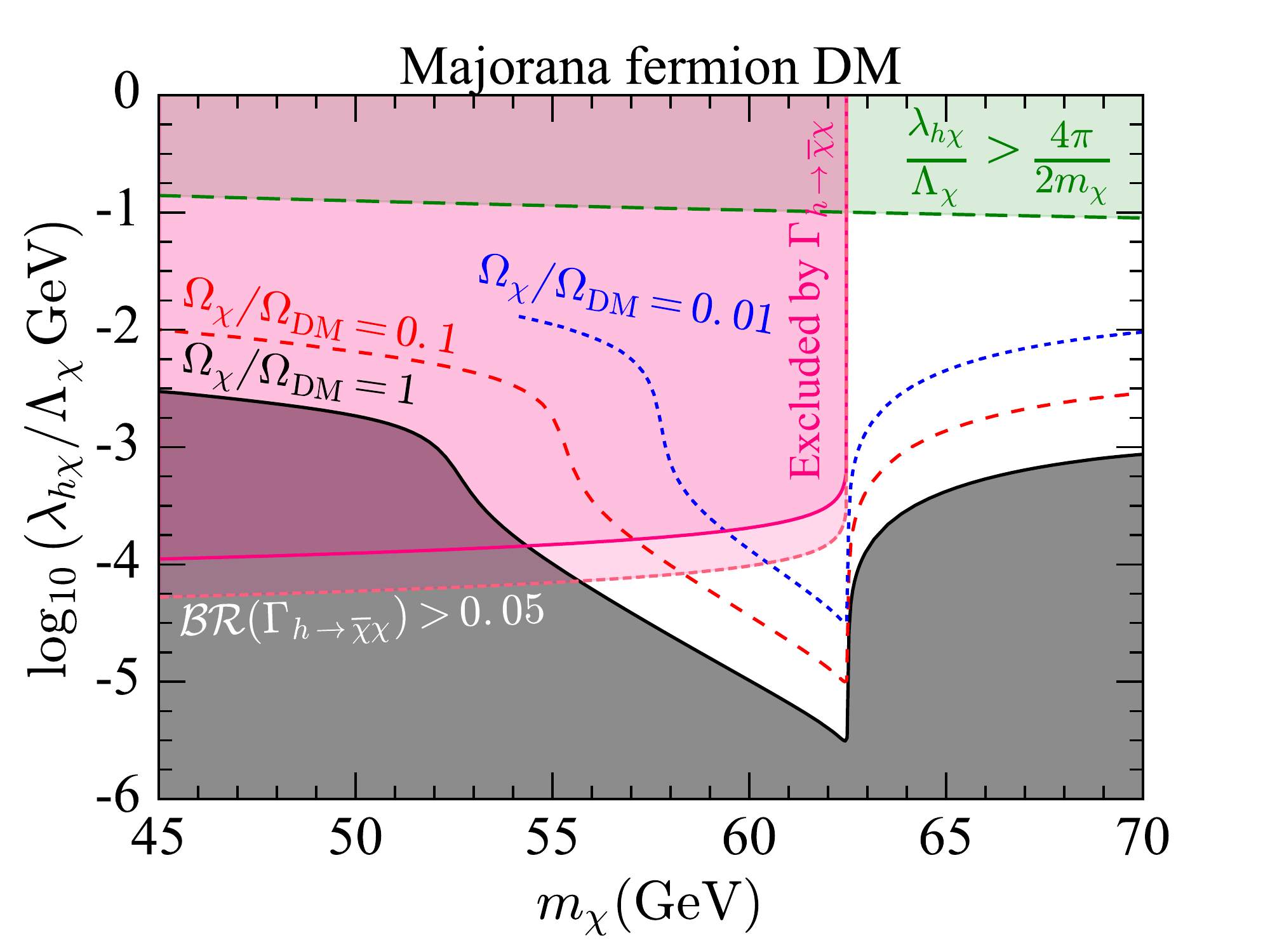}
    \end{minipage}
    \hfil
    \begin{minipage}[t]{0.48\textwidth}
      \centering
      \includegraphics[width=\textwidth]{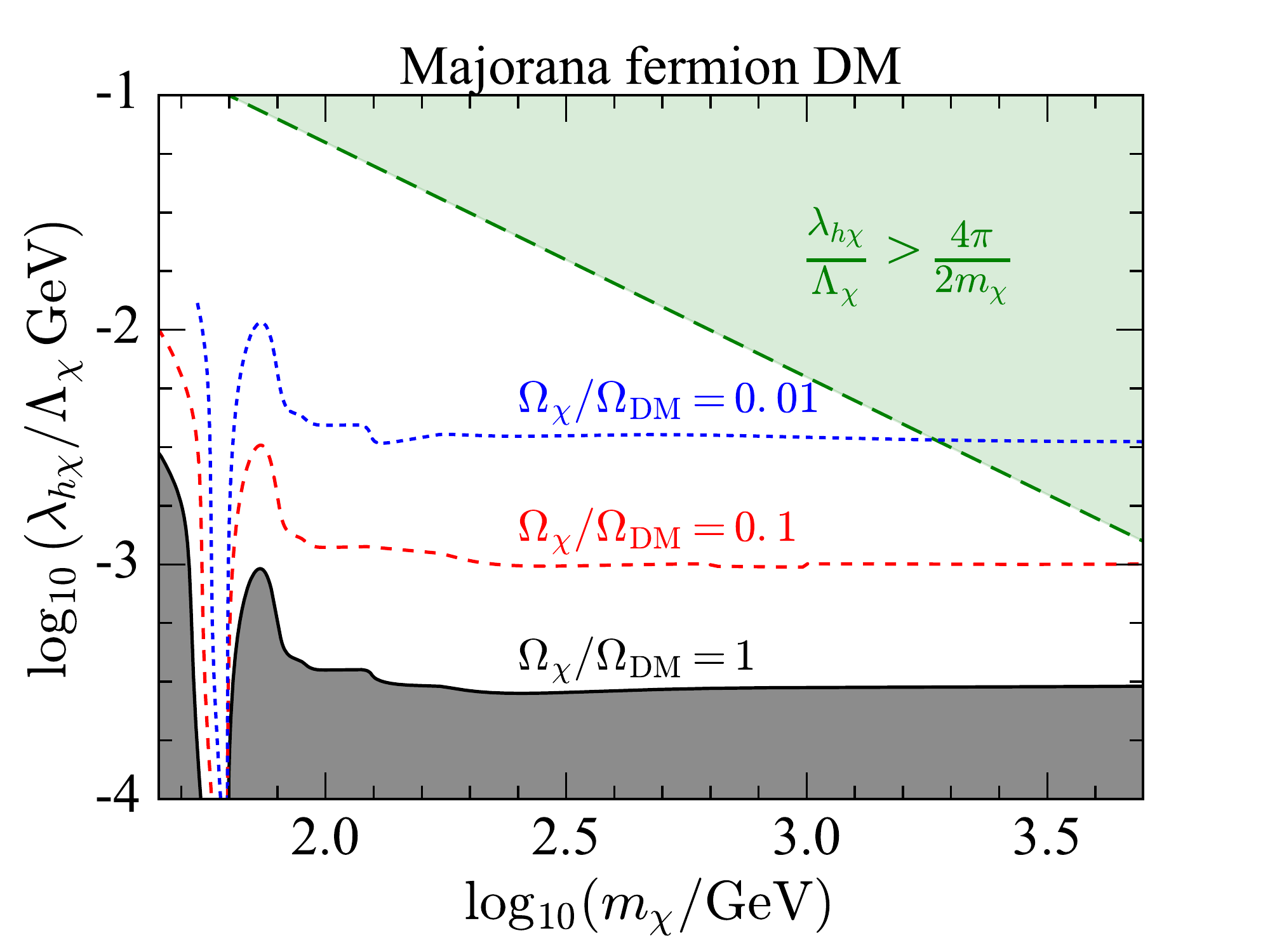}  
    \end{minipage} 
    
	\small{\emph{b}). Equally mixed scalar-pseudoscalar interaction, $\cos \xi = 1/\sqrt{2}$.} 
	
    \begin{minipage}[t]{0.48\textwidth}        
      \centering
      \includegraphics[width=\textwidth]{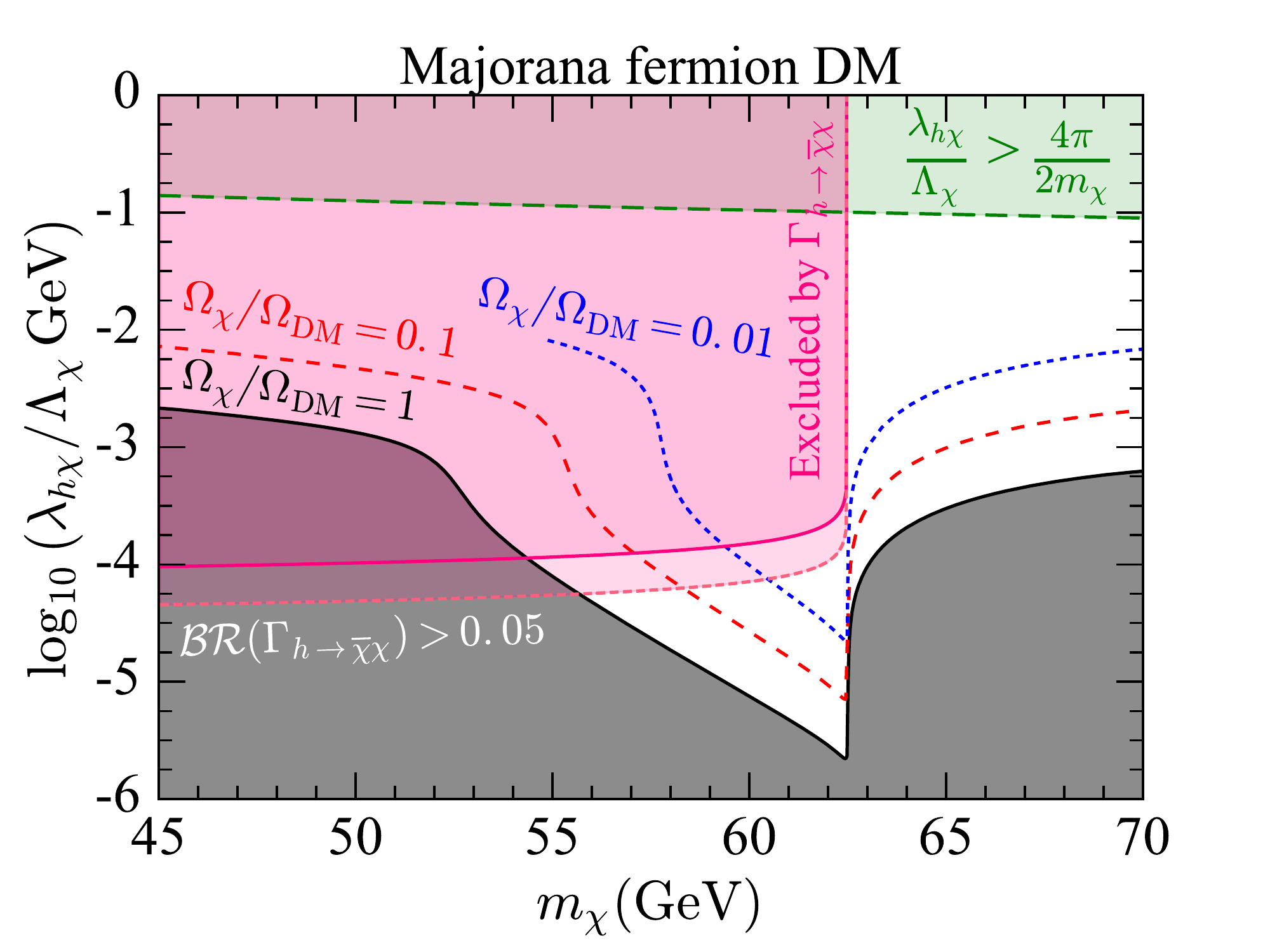}
    \end{minipage}
    \hfil
    \begin{minipage}[t]{0.48\textwidth}
      \centering
      \includegraphics[width=\textwidth]{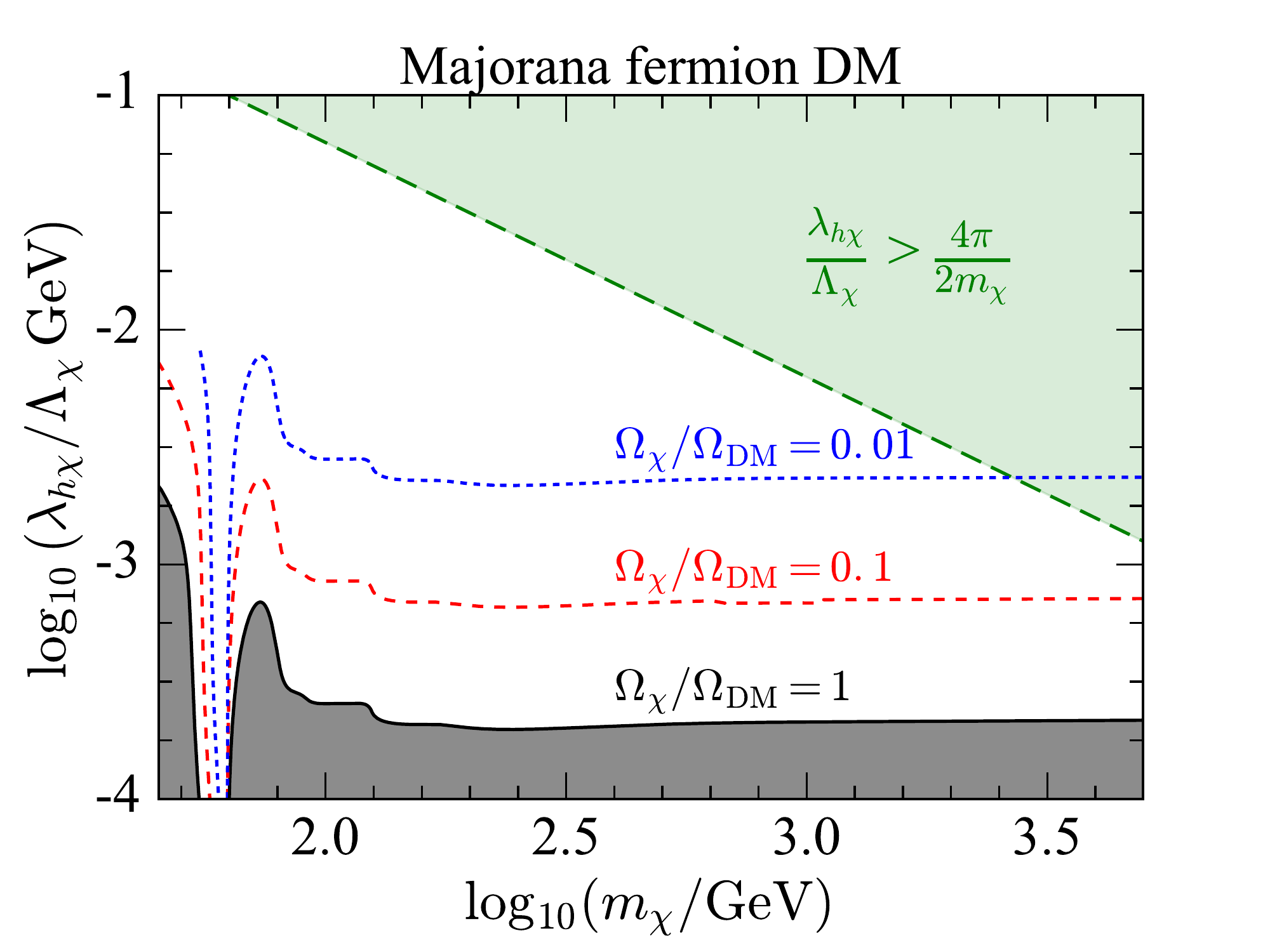}  
    \end{minipage}
    
	\small{\emph{c}). Pure pseudoscalar interaction, $\cos \xi = 0$.}
	
    \caption{Contours of fixed Majorana relic density for $f_{\textrm{rel}} = 1$ (black solid), 0.1 (red dashed) and 0.01 (blue dotted). The grey shaded region is ruled out due to an overabundance of dark matter. The green shaded region is where the EFT approximation of the full theory breaks down for $\lambda_{h\chi}/\Lambda_\chi > 4\pi/2m_\chi$. Left: A close-up of the resonantly enhanced annihilation region, $m_\chi \sim m_h/2$. Larger values of $\lambda_{h\chi}/\Lambda_\chi$ are excluded by an upper limit of 19\% (pink solid) at 2$\sigma$\,C.L. or 5\% (pink dotted) at 1$\sigma$\,C.L. on $\mathcal{BR}(h \rightarrow \ovr{\chi}\chi)$. Right: Relic density contours for the full range of $m_\chi$. }   
    \label{majorana-relic}
\end{figure}

The contours of fixed Majorana relic density in the $(m_\chi,\lambda_{h\chi}/\Lambda_\chi)$ plane for $f_{\textrm{rel}} =$ 1 (black solid), 0.1 (red dashed) and 0.01 (blue dotted) are shown in Fig.~\ref{majorana-relic}. Contours in each row are generated at fixed values of $\cos\xi$: $\cos\xi = 1$ for a pure scalar interaction (top row), $\cos\xi = 1/\sqrt{2}$ for an equal mix between the scalar and pseudoscalar terms (middle row), and $\cos\xi = 0$ for a pure pseudoscalar interaction (bottom row). As the interactions between the SM Higgs boson and the Majorana fermion DM change from pure scalar to pure pseudoscalar, the Majorana relic density contours move to lower values of $\lambda_{h\chi}/\Lambda_\chi$. 

In the region $m_\chi < m_h/2$, most values of $\lambda_{h\chi}/\Lambda_\chi$ for $\cos\xi = 1, \,1/\sqrt{2}$ and 0 are excluded by the combined constraints on the Majorana relic density and the Higgs invisible width apart from a small triangle that continues to evade these limits in a similar fashion to the scalar and vector models. Above Majorana masses of $\sim$\,300\,GeV, the relic density contours for $\cos\xi = 1/\sqrt{2}$ and $0$ essentially remain constant and independent of the coupling $\lambda_{h\chi}/\Lambda_\chi$. When $f_{\textrm{rel}}$ = 0.1 or 0.01, the relic density contours at higher DM masses lie in the green shaded region where $\lambda_{h\chi}/\Lambda_\chi > 4\pi/2m_\chi$. Hence, the validity of the results in these regions cannot be guaranteed within the EFT framework.

\begin{figure}
    \begin{minipage}[t]{0.48\textwidth}        
      \centering
      \includegraphics[width=\textwidth]{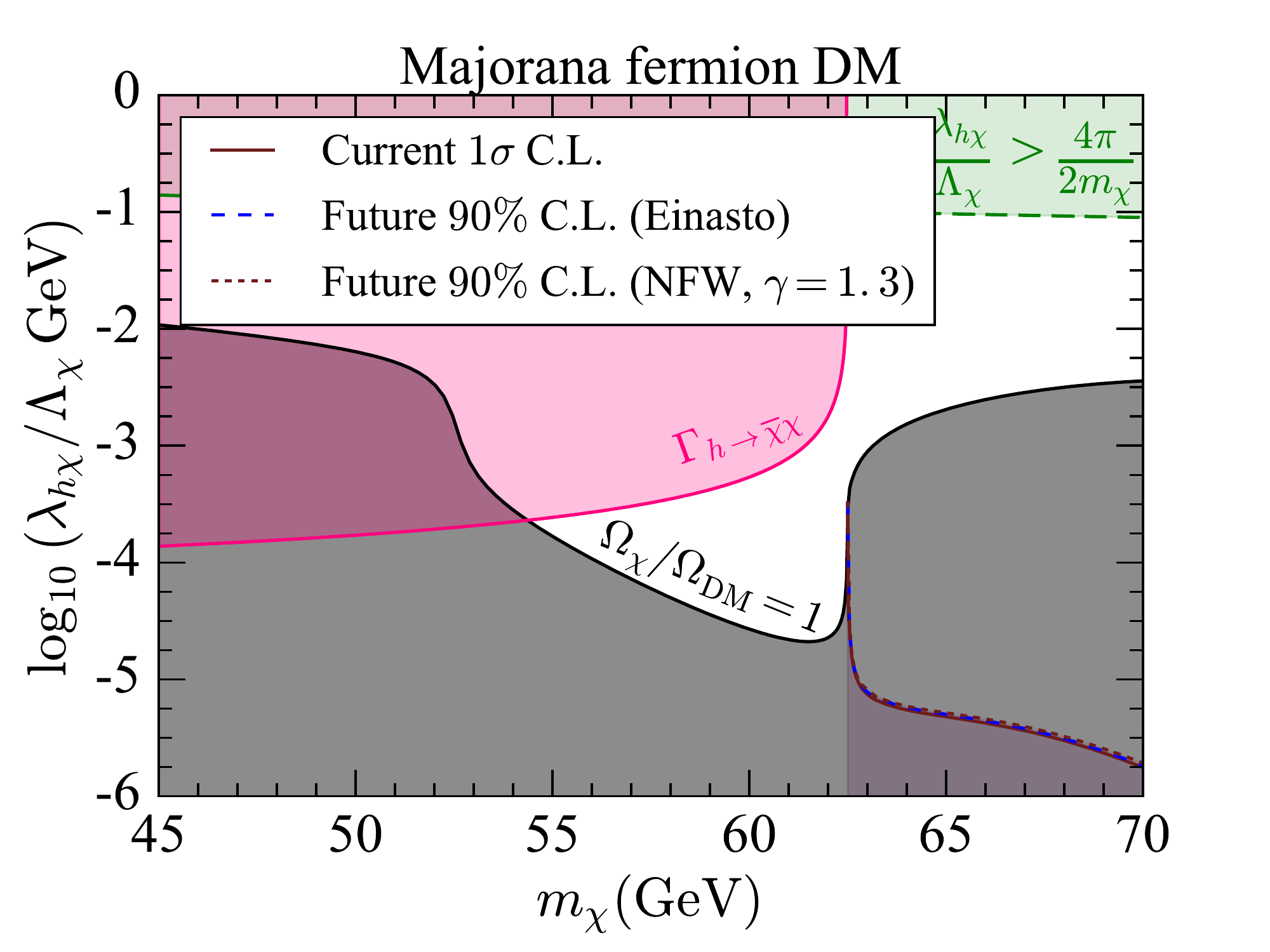}
    \end{minipage}
    \hfil
    \begin{minipage}[t]{0.48\textwidth}
      \centering
      \includegraphics[width=\textwidth]{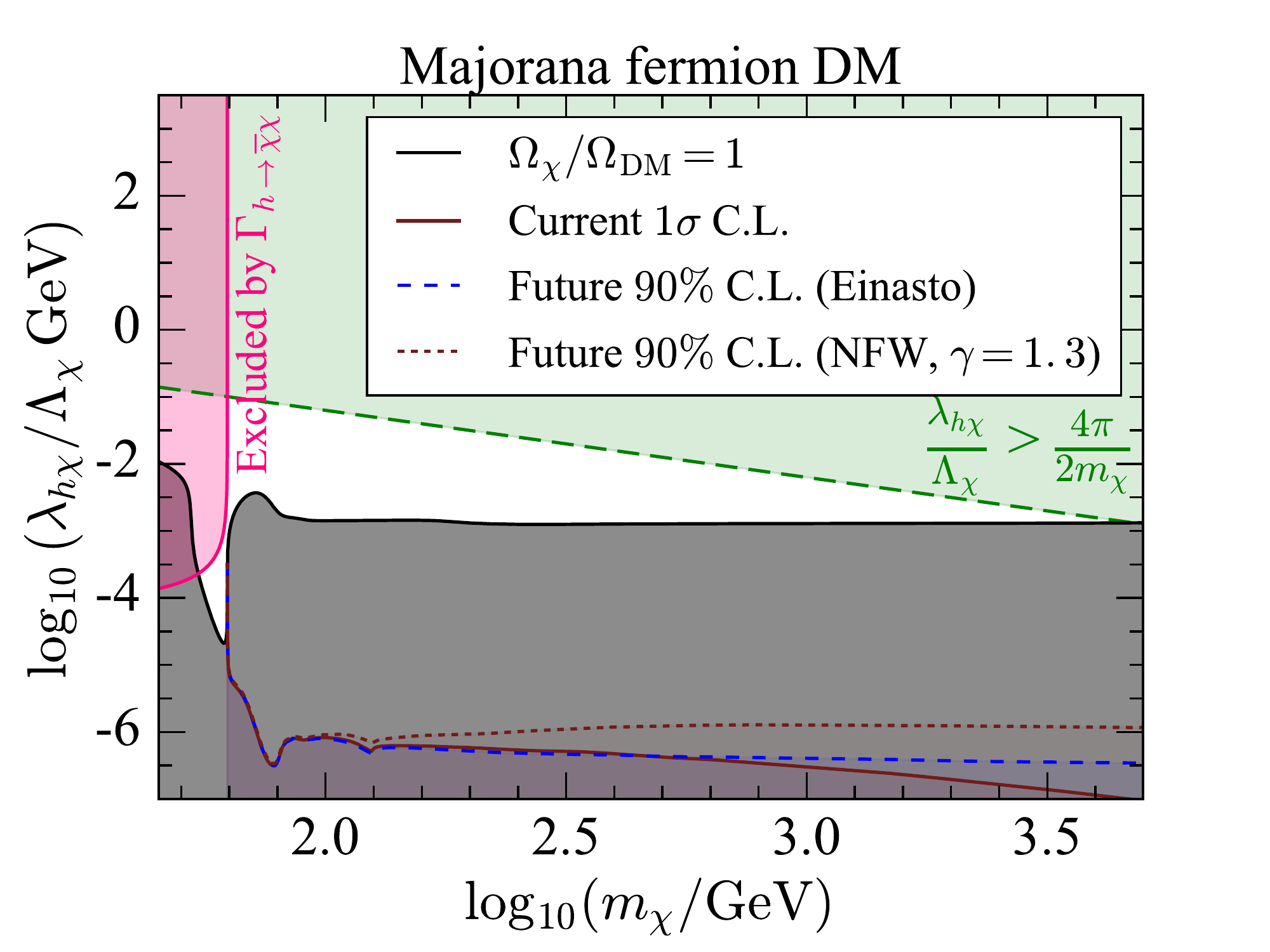}
    \end{minipage} 
	
	\small{\emph{a}). Pure scalar interaction, $\cos \xi = 1$.}
	
    \begin{minipage}[t]{0.48\textwidth}        
      \centering
      \includegraphics[width=\textwidth]{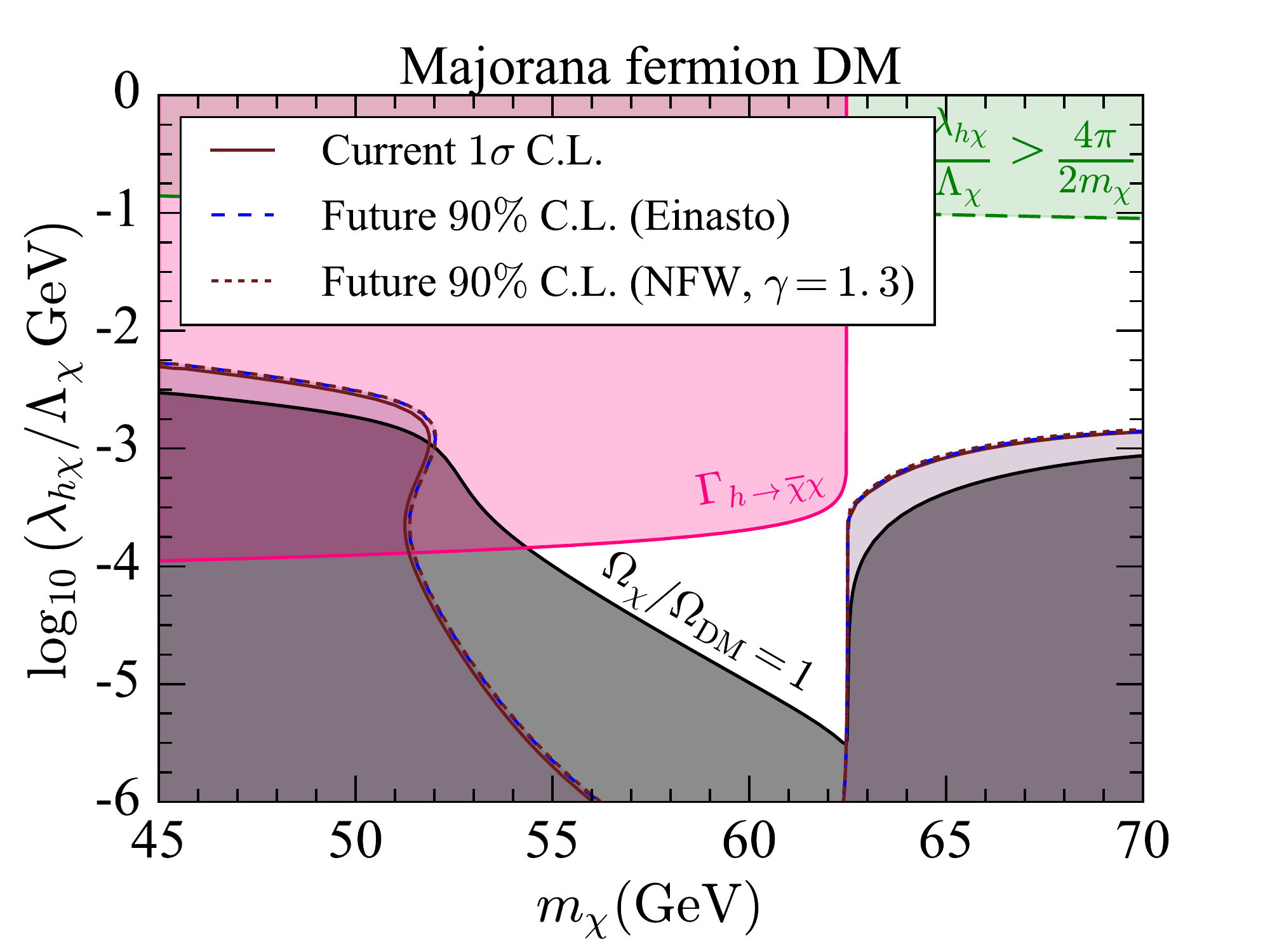}
    \end{minipage}
    \hfil
    \begin{minipage}[t]{0.48\textwidth}
      \centering
      \includegraphics[width=\textwidth]{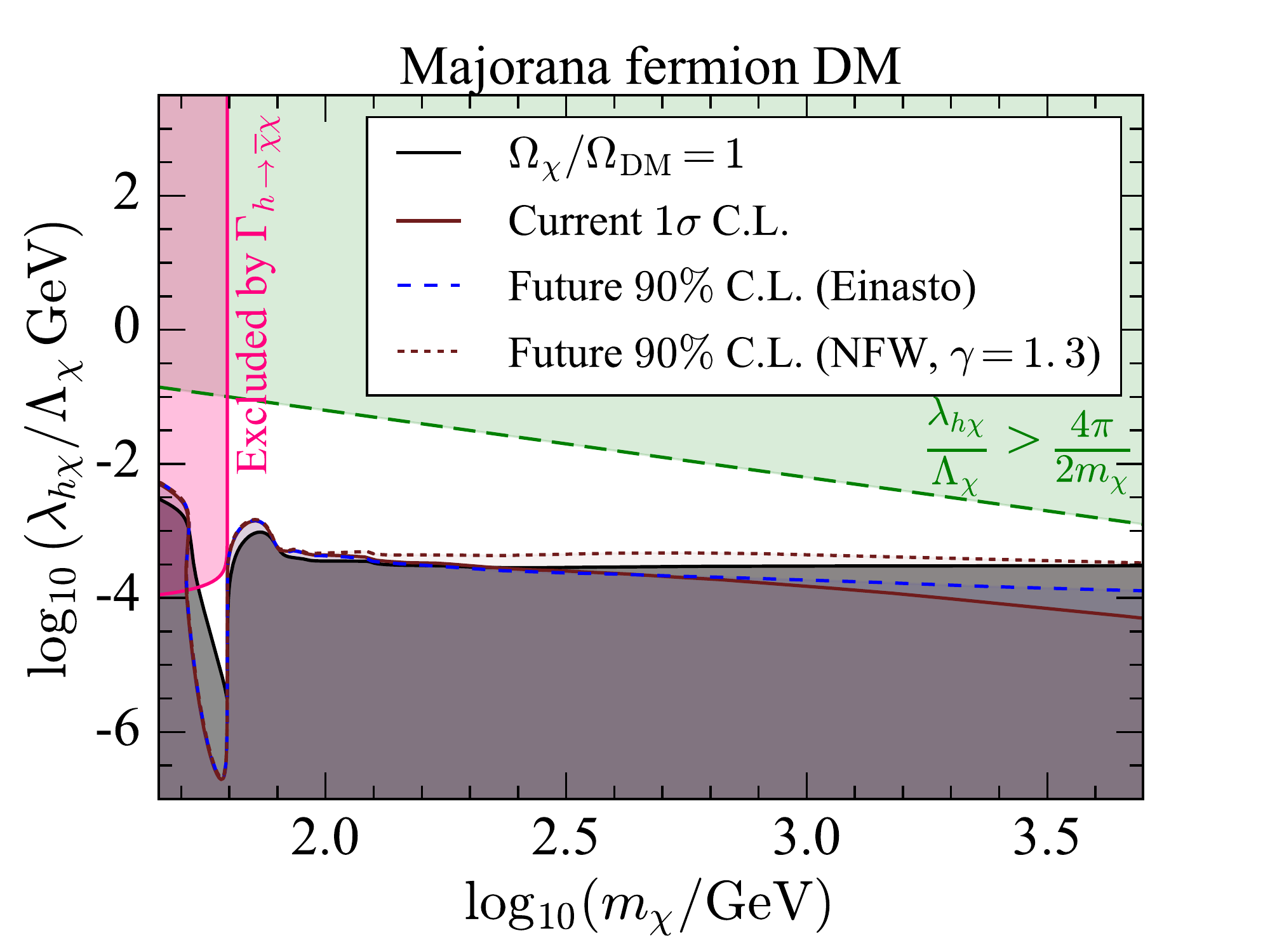}
    \end{minipage}
    
    \small{\emph{b}). Equally mixed scalar-pseudoscalar interaction, $\cos \xi = 1/\sqrt{2}$.}            
    
    \begin{minipage}[t]{0.48\textwidth}        
      \centering
      \includegraphics[width=\textwidth]{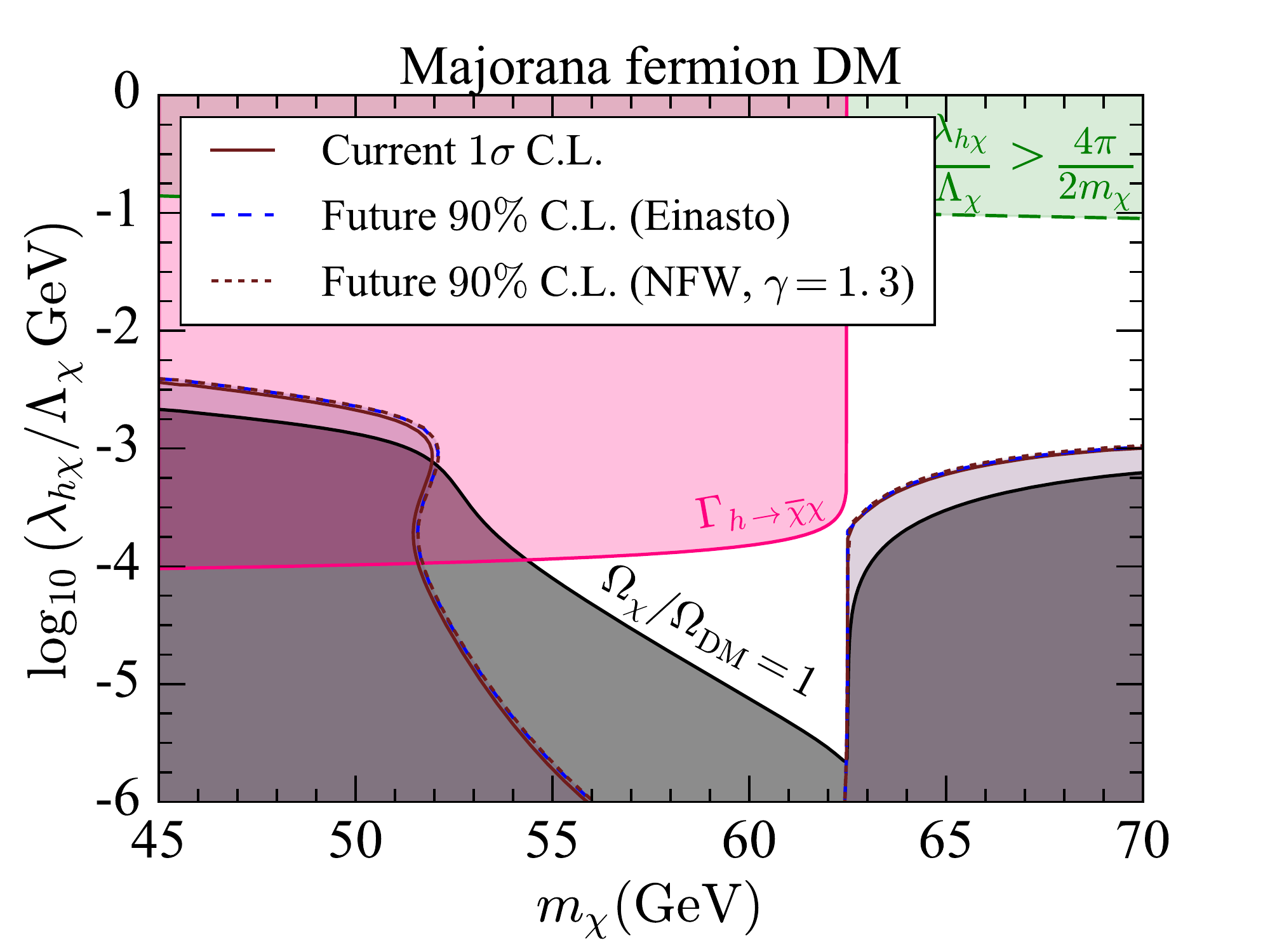}
    \end{minipage}
    \hfil
    \begin{minipage}[t]{0.48\textwidth}
      \centering
      \includegraphics[width=\textwidth]{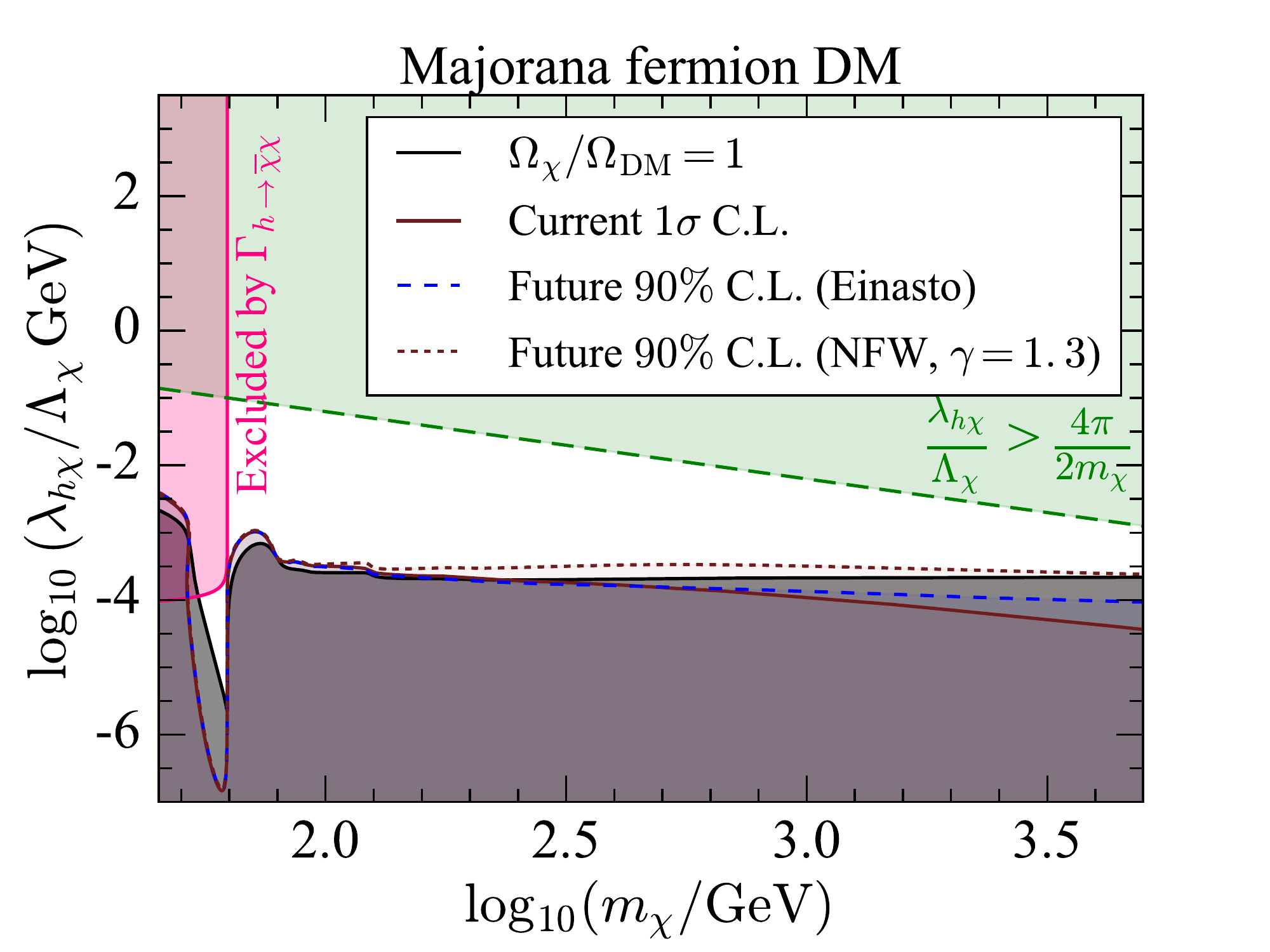}
    \end{minipage}
     
	\small{\emph{c}). Pure pseudoscalar interaction, $\cos\xi = 0$.}
     
    \caption{Indirect search limits on the Majorana model parameter space. The grey and pink shaded regions are excluded respectively by the relic density and the Higgs invisible width constraints. The green shaded region is where the EFT approximation of the full theory breaks down for $\lambda_{h\chi}/\Lambda_\chi > 4\pi/2m_\chi$. Values of $\lambda_{h\chi}/\Lambda_\chi$ below the current 1$\sigma$\,C.L. (brown solid) curve are excluded at more than $1\sigma$\,C.L. Regions below the future 90\%\,C.L. curve with the Einasto (blue dashed) and contracted NFW (brown dotted) profiles will be excluded. Left: A close-up of the resonantly enhanced annihilation region, $m_\chi \sim m_h/2$. Right: The full range of $m_\chi$.}
    \label{majorana-indirect} 
\end{figure}

The combined sensitivity of indirect searches to various regions in the $(m_\chi,\lambda_{h\chi}/\Lambda_\chi)$ plane is shown in Fig.~\ref{majorana-indirect} for $\cos\xi = 1$ (top row), $1/\sqrt{2}$ (middle row) and $0$ (bottom row). In the case of $\cos\xi = 1$, the annihilation cross section $\sigma v_{\textrm{rel}}$ entering in the calculation of indirect detection rates receive a $v^2$ suppression, where $v\sim 10^{-3}$\,$c$ is the typical speed of dark matter in a local halo. Consequently, the resulting indirect search limits are weak and no exclusion is possible.

\begin{figure}
	\begin{minipage}[t]{0.48\textwidth}
		\centering
		\includegraphics[width=\textwidth]{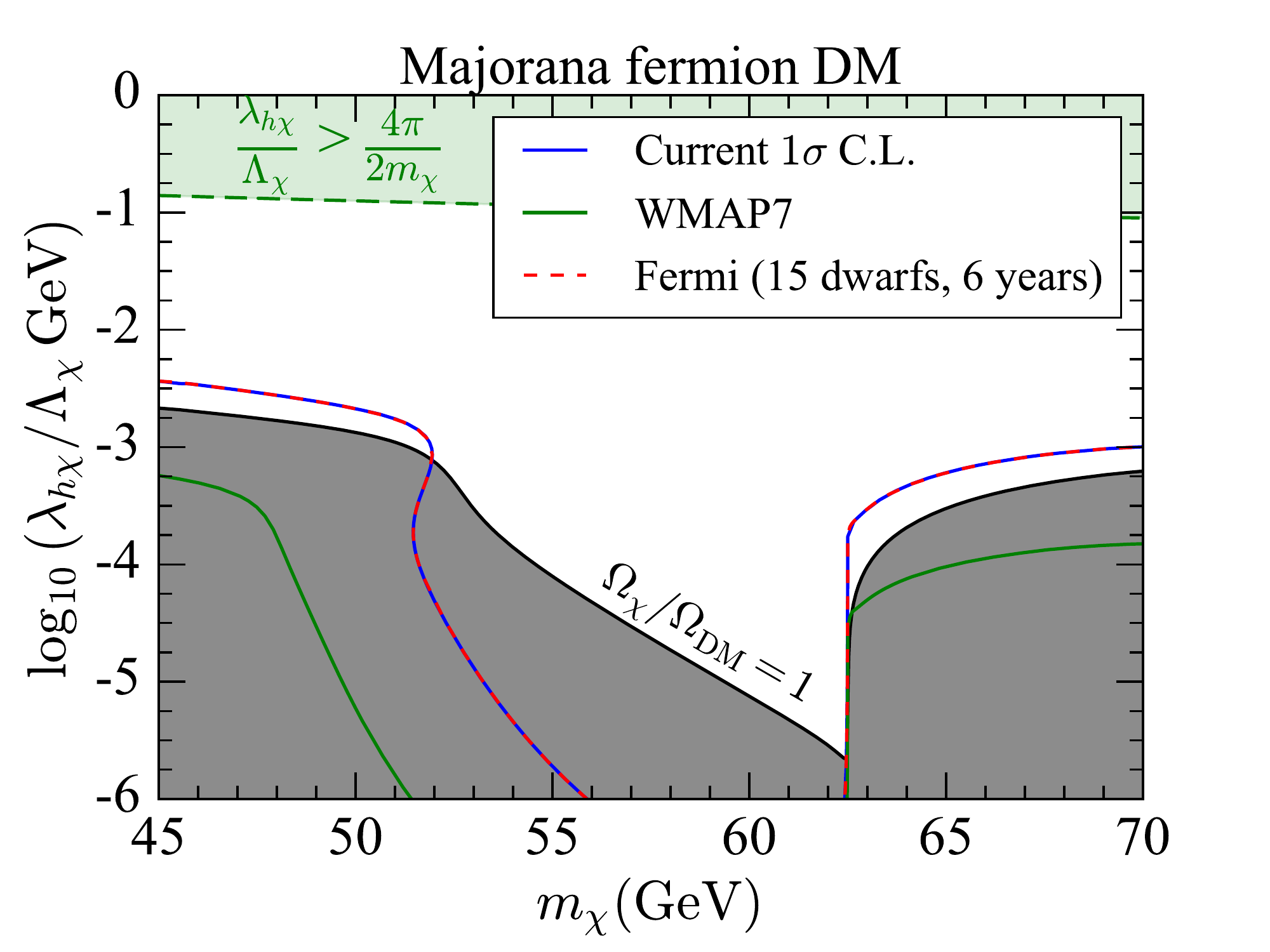}	
	\end{minipage}
	\hfil
	\begin{minipage}[t]{0.48\textwidth}
		\centering
		\includegraphics[width=\textwidth]{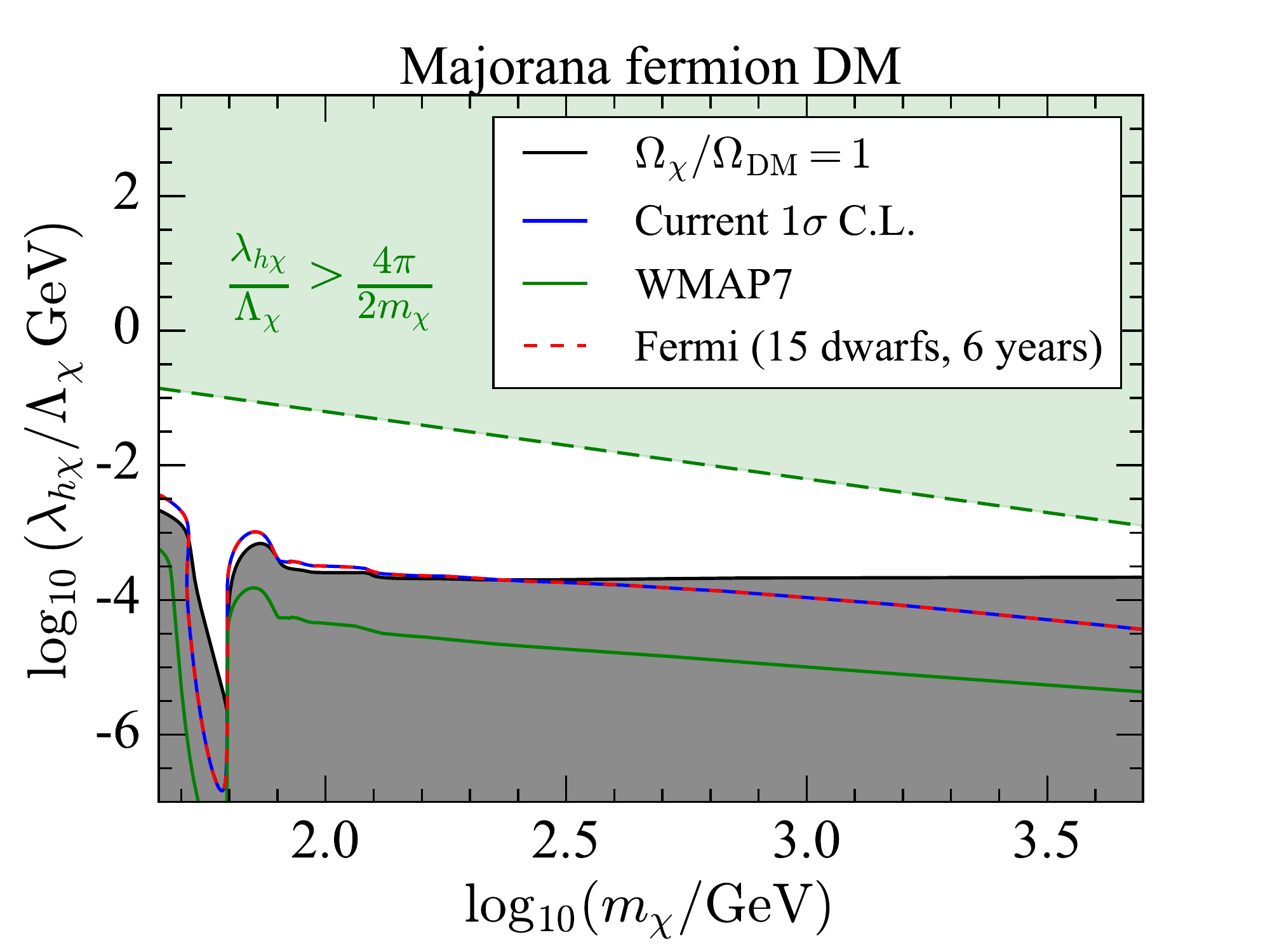}
	\end{minipage}	
	\caption{Breakdown of the current 1$\sigma$\,C.L. (blue solid) in the Majorana fermion parameter space when $\cos\xi= 0$. The grey shaded region is excluded by the relic density constraint. The green shaded region is where the EFT approximation of the full theory breaks down for $\lambda_{h\chi}/\Lambda_\chi > 4\pi/2m_\chi$. Contributions to the combined current 1$\sigma$\,C.L. come from WMAP 7-year observations of the CMB (green solid) and a combined analysis of 15 dwarf galaxies using 6 years of the Fermi-LAT data (red dashed). Left: A close-up of the resonantly enhanced annihilation region, $m_\chi \sim m_h/2$. Right: The full range of $m_\chi$.}
	\label{1sigCurrentBreak}
	\vspace{5mm}
	\begin{minipage}[t]{0.48\textwidth}
		\centering
		\includegraphics[width=\textwidth]{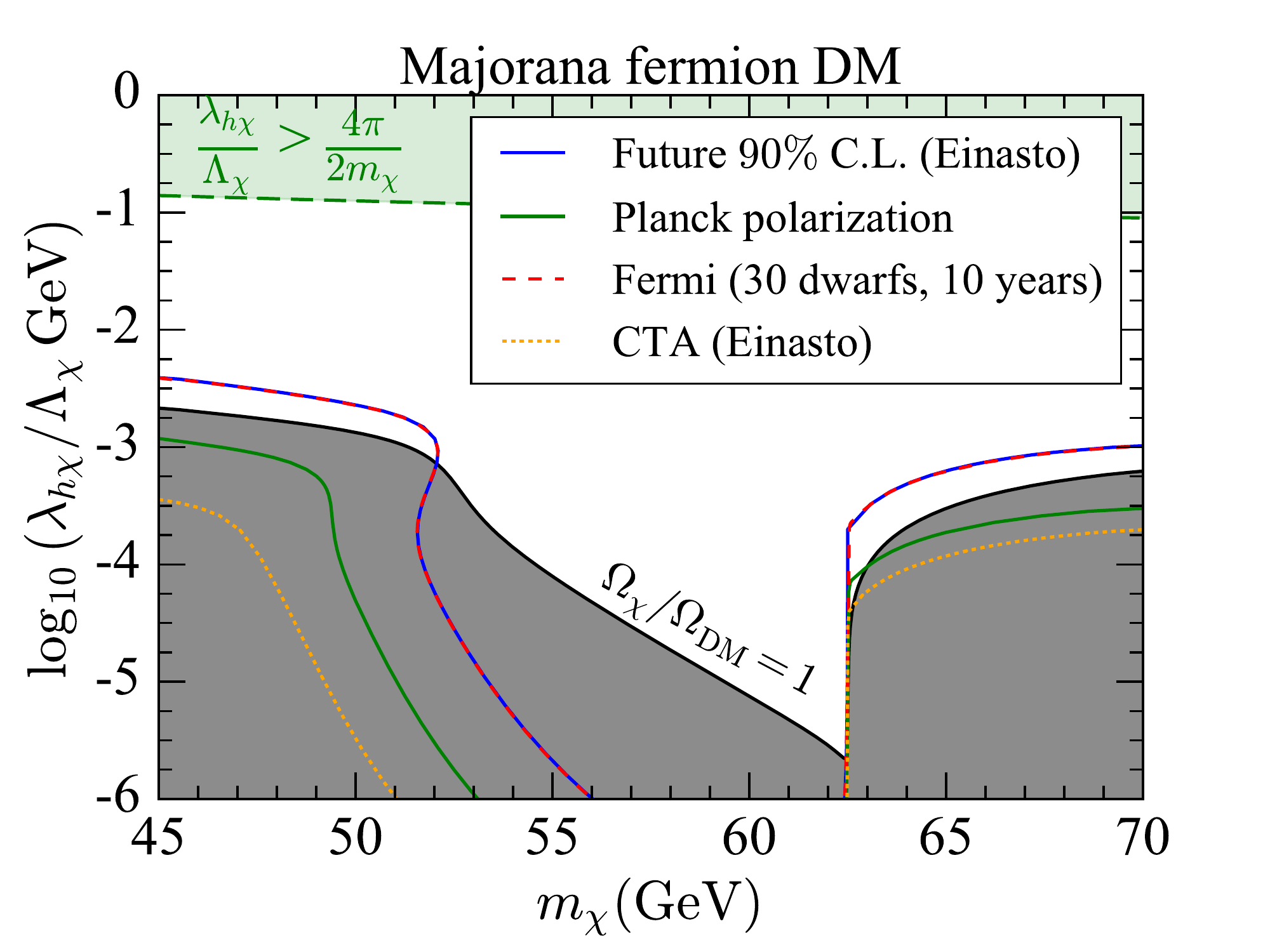}	
	\end{minipage}
	\hfil
	\begin{minipage}[t]{0.48\textwidth}
		\centering
		\includegraphics[width=\textwidth]{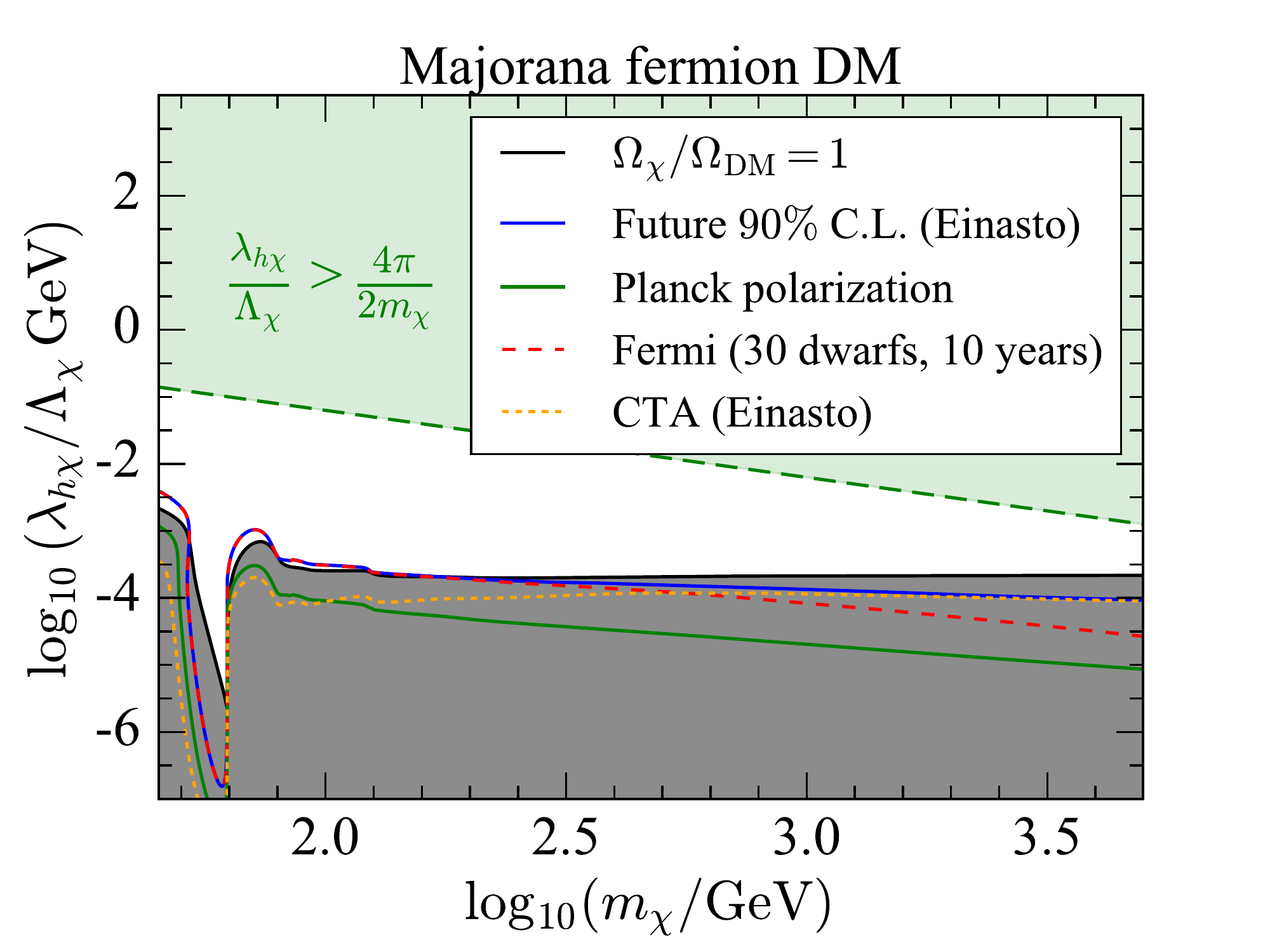}
	\end{minipage}
	\caption{Breakdown of the future 90\%\,C.L. (Einasto) in the Majorana fermion parameter space when $\cos\xi = 0$ (blue solid). The grey shaded region is excluded by the relic density constraint. The green shaded region is where the EFT approximation of the full theory breaks down for $\lambda_{h\chi}/\Lambda_\chi > 4\pi/2m_\chi$. Contributions to the combined future 90\%\,C.L. (Einasto) come from the Planck polarization data (green solid), projected improvements in Fermi-LAT sensitivity towards observation of a further 15 southern dwarf galaxies over 10 years (red dashed) and projected limits from the CTA using the Einasto profile (orange dotted). Left: A close-up of the resonantly enhanced annihilation region, $m_\chi \sim m_h/2$. Right: The full range of $m_\chi$.}
	\label{90PfutureEinasto}					
\end{figure}

When interactions are pure pseudoscalar, the velocity suppression of $\sigma v_{\textrm{rel}}$ is lifted. Consequently, the indirect search limits are nontrivial. In the case of $\cos\xi = 1/\sqrt{2}$ and $0$, Majorana masses between $m_h/2$ and 70\,GeV with $\log_{10} (\lambda_{h\chi}/\Lambda_{\chi} \textrm{\,GeV}) \in [-4.1,-2.8]$ can be seen to be in tension with the current indirect searches at more than $1\sigma$\,C.L. Analogous to the indirect search limits in the scalar and vector models, a small region around $m_\chi \lesssim m_h/2$ will continue to evade the current and future indirect searches. 	

\begin{figure}	
	\begin{minipage}[t]{0.48\textwidth}
		\centering
		\includegraphics[width=\textwidth]{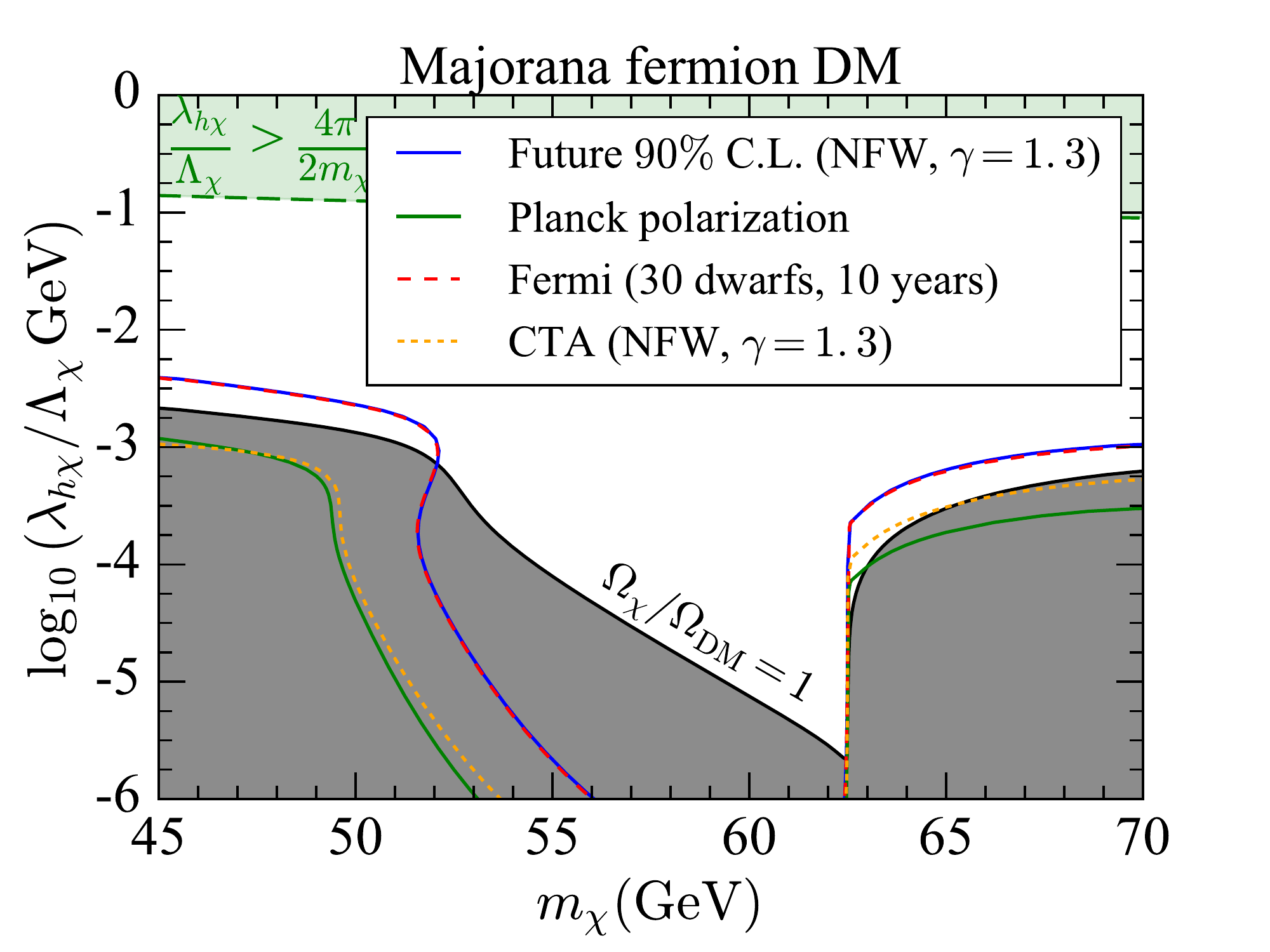}
	\end{minipage}
	\hfil
	\begin{minipage}[t]{0.48\textwidth}
		\centering
		\includegraphics[width=\textwidth]{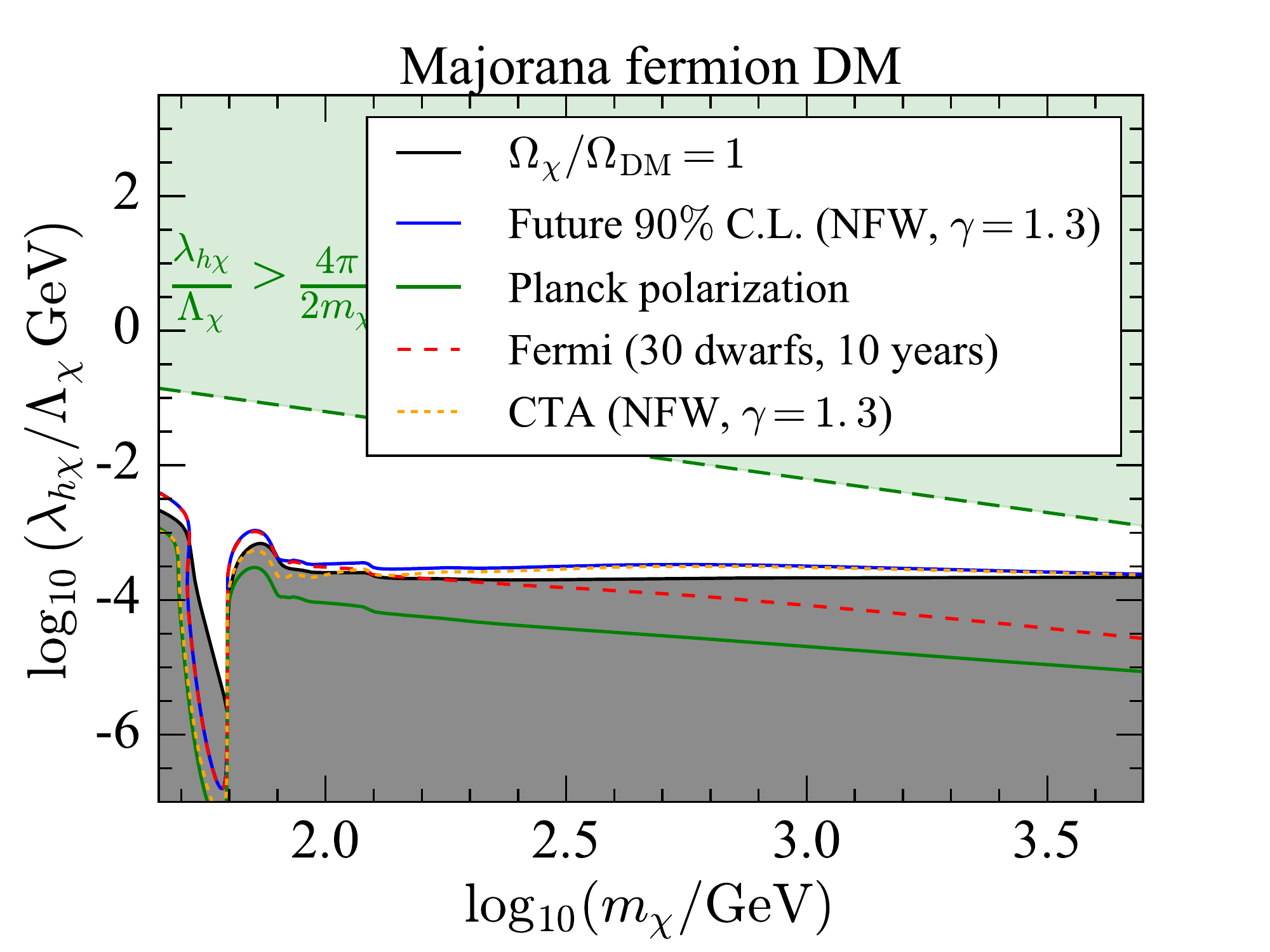}
	\end{minipage}	
	\caption{Breakdown of the future $90\%$\,C.L. (NFW, $\gamma=1.3$) in the Majorana fermion parameter space when $\cos\xi= 0$  (blue solid). The grey shaded region is excluded by the relic density constraint. The green shaded region is where the EFT approximation of the full theory breaks down for $\lambda_{h\chi}/\Lambda_\chi > 4\pi/2m_\chi$. Contributions to the combined future $90\%$\,C.L. (NFW, $\gamma=1.3$) come from the Planck polarization data (green solid), projected improvements in Fermi-LAT sensitivity towards observation of a further 15 southern dwarf galaxies over 10 years (red dashed) and projected limits from the CTA experiment using the contracted NFW ($\gamma=1.3$) profile (orange dotted). Left: A close-up of the resonantly enhanced annihilation region, $m_\chi \sim m_h/2$. Right: The full range of $m_\chi$. }
	\label{90PfutureNFW}
\end{figure}

To illustrate the degree to which different indirect searches contribute to the limits, in Figs.~\ref{1sigCurrentBreak}--\ref{90PfutureNFW} we show the breakdown of the current 1$\sigma$\,C.L., future 90\%\,C.L. (Einasto) and future 90\%\,C.L. (NFW, $\gamma=1.3$) for a pure pseudoscalar coupling ($\cos\xi = 0$). Current indirect search limits incorporate the WMAP 7-year observations of the CMB and a combined analysis of 15 dwarf galaxies using 6 years of the Fermi-LAT data. Projected future limits include contributions from the Planck polarization data, projected Fermi-LAT results based on the discovery of a further 15 southern dwarf galaxies over 10 years, and projected CTA results from the GC. At low DM masses, limits from the Fermi-LAT are strongest, whereas at higher DM masses, limits from the upcoming CTA experiment are strongest. Constraints from the CMB are weak in all parts of the parameter space. Future indirect searches will be sensitive enough to exclude Majorana masses up to $\sim$\,170\,GeV (Einasto) and $\sim$\,12\,TeV (NFW, $\gamma=1.3$) at more than $90\%$\,C.L., if $\chi$ makes up all of the dark matter.

\begin{figure}
    \begin{minipage}[t]{0.48\textwidth}        
      \centering
      \includegraphics[width=\textwidth]{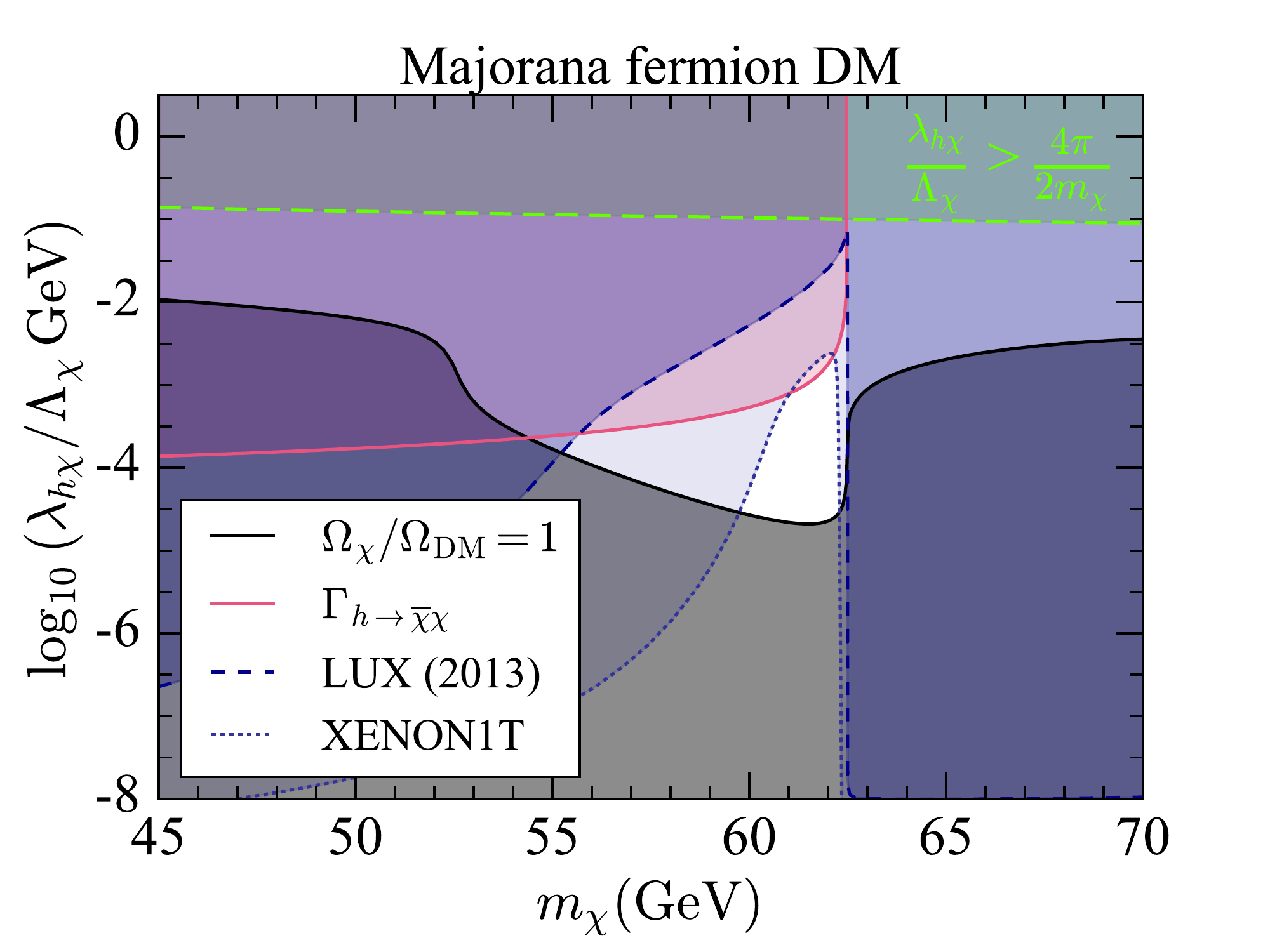}
    \end{minipage}
    \hfil
    \begin{minipage}[t]{0.48\textwidth}
      \centering
      \includegraphics[width=\textwidth]{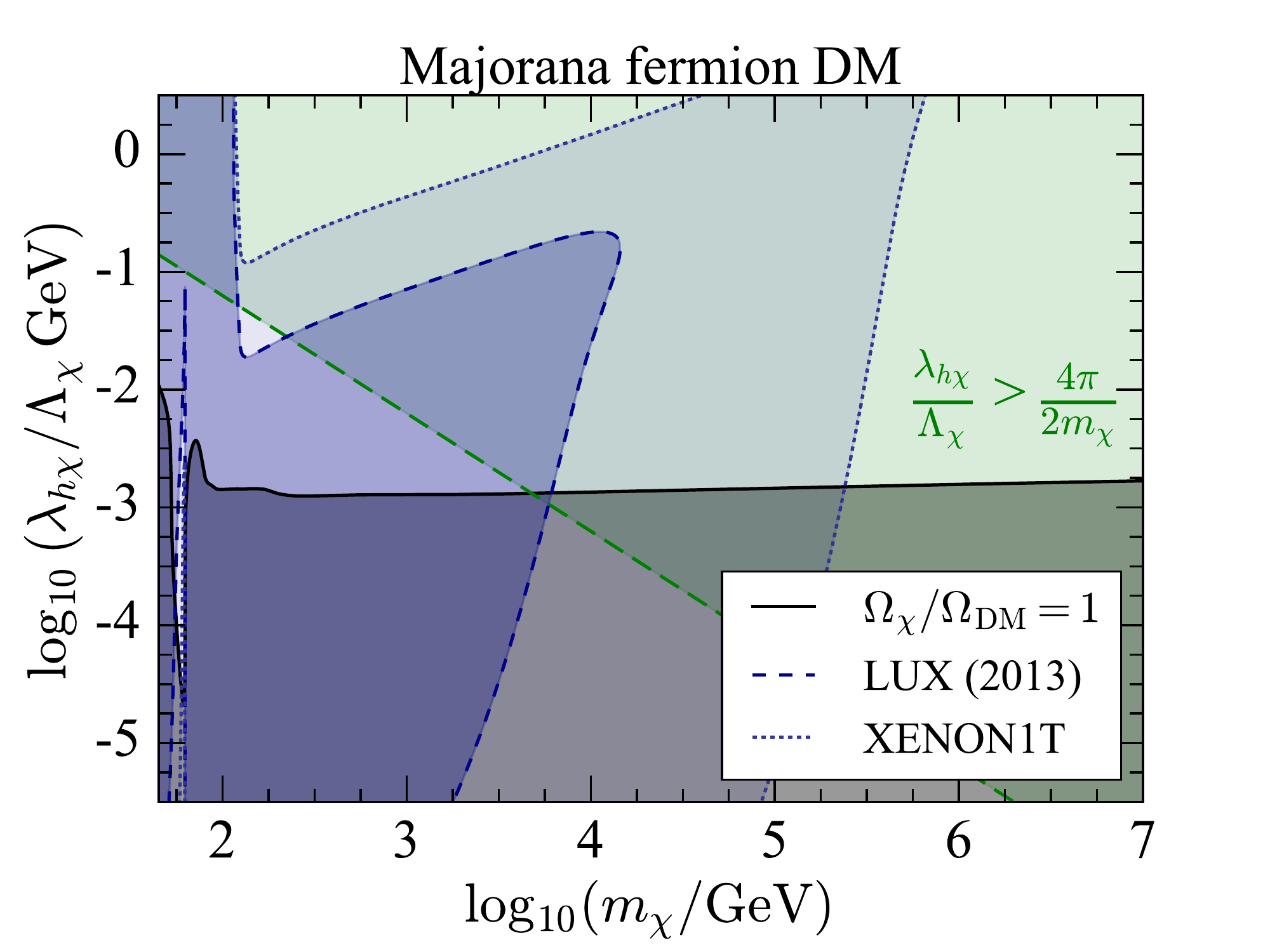} 
    \end{minipage}       
    
	\small{\emph{a}). Pure scalar interaction, $\cos \xi = 1$.}
	
    \begin{minipage}[t]{0.48\textwidth}        
      \centering
      \includegraphics[width=\textwidth]{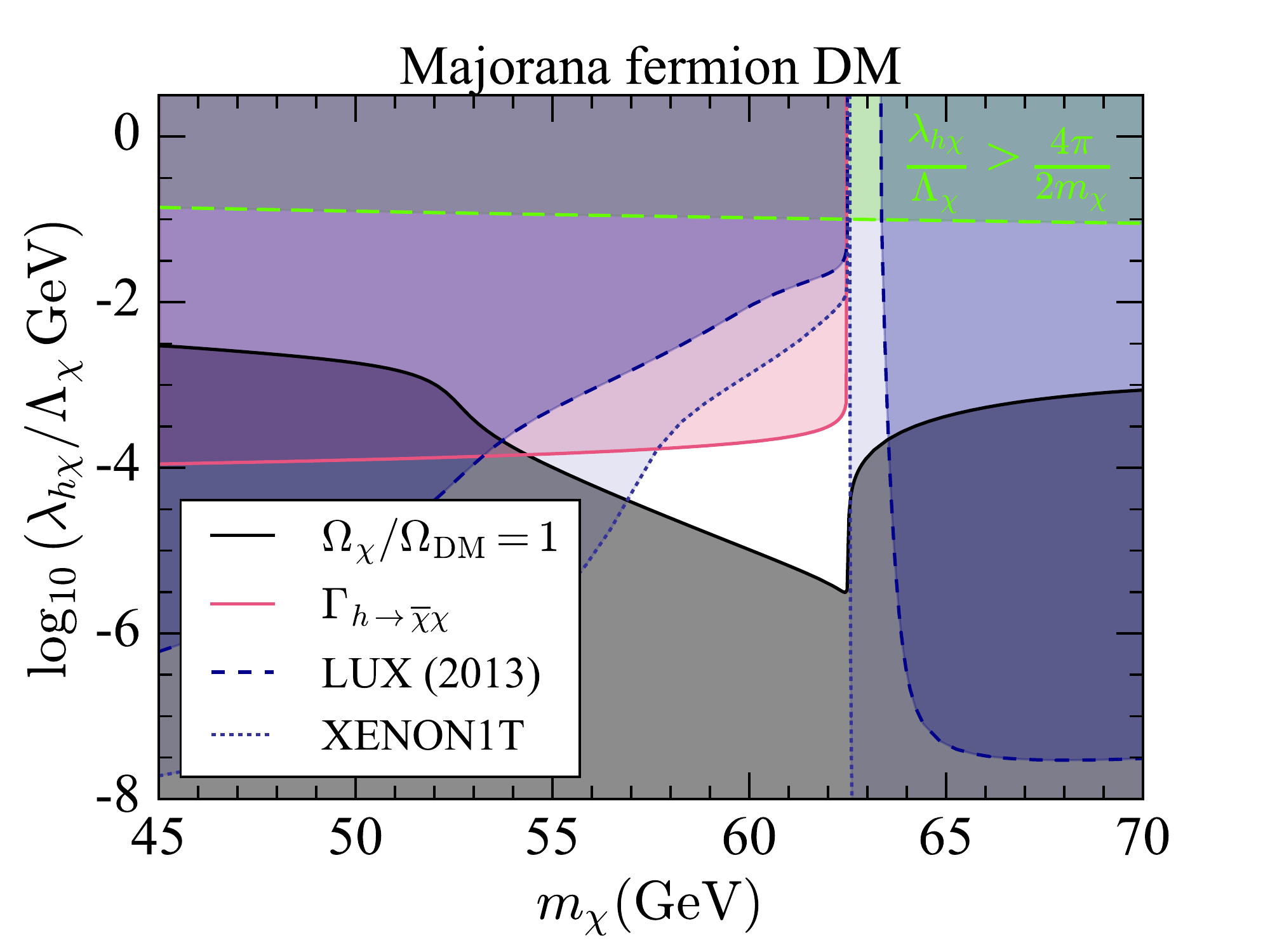}
    \end{minipage}
    \hfil
    \begin{minipage}[t]{0.48\textwidth}
      \centering
      \includegraphics[width=\textwidth]{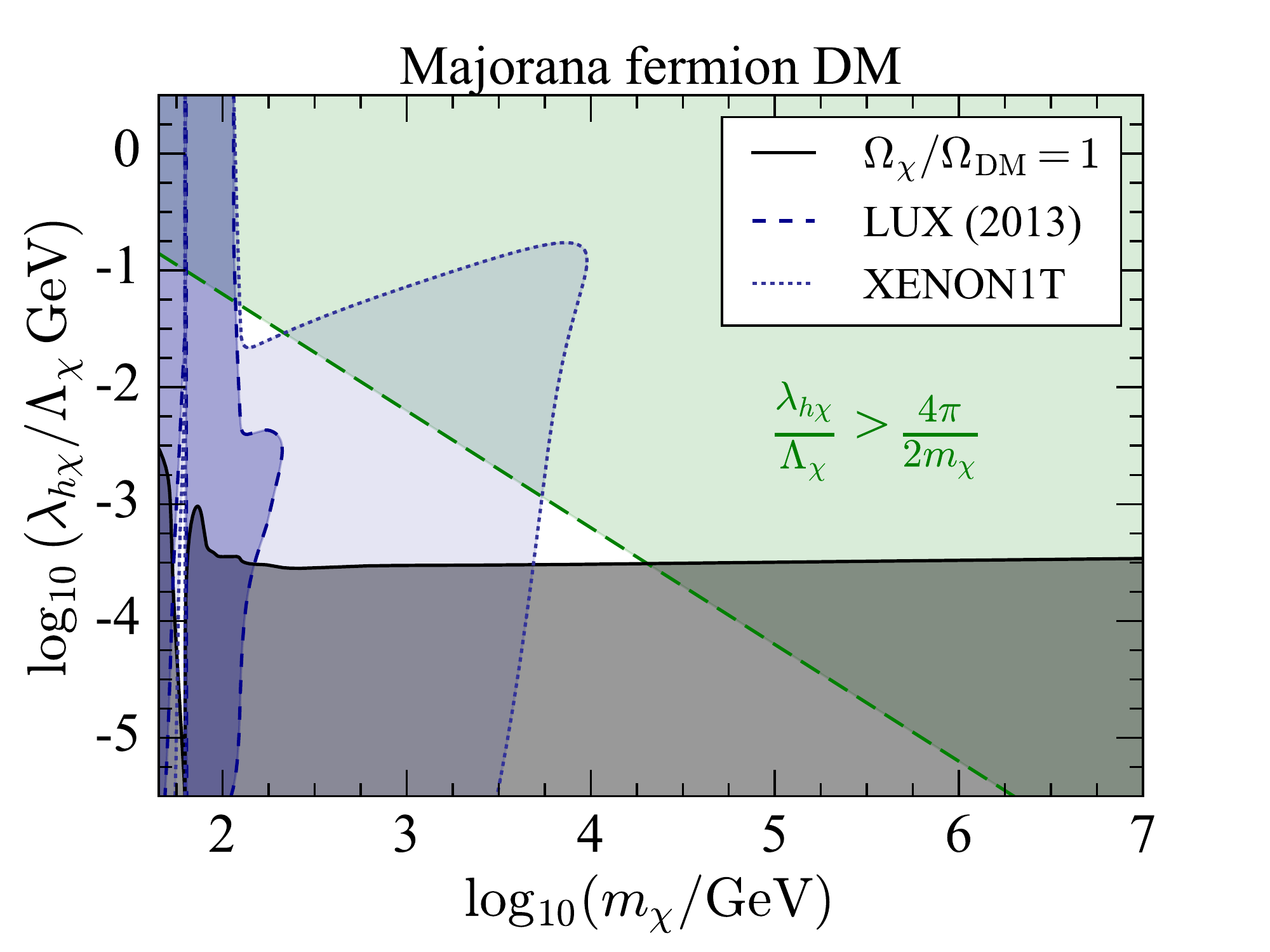} 
    \end{minipage} 
	
	\small{\emph{b}). Equally mixed scalar-pseudoscalar interaction, $\cos \xi = 1/\sqrt{2}$.}
	
    \begin{minipage}[t]{0.48\textwidth}        
      \centering
      \includegraphics[width=\textwidth]{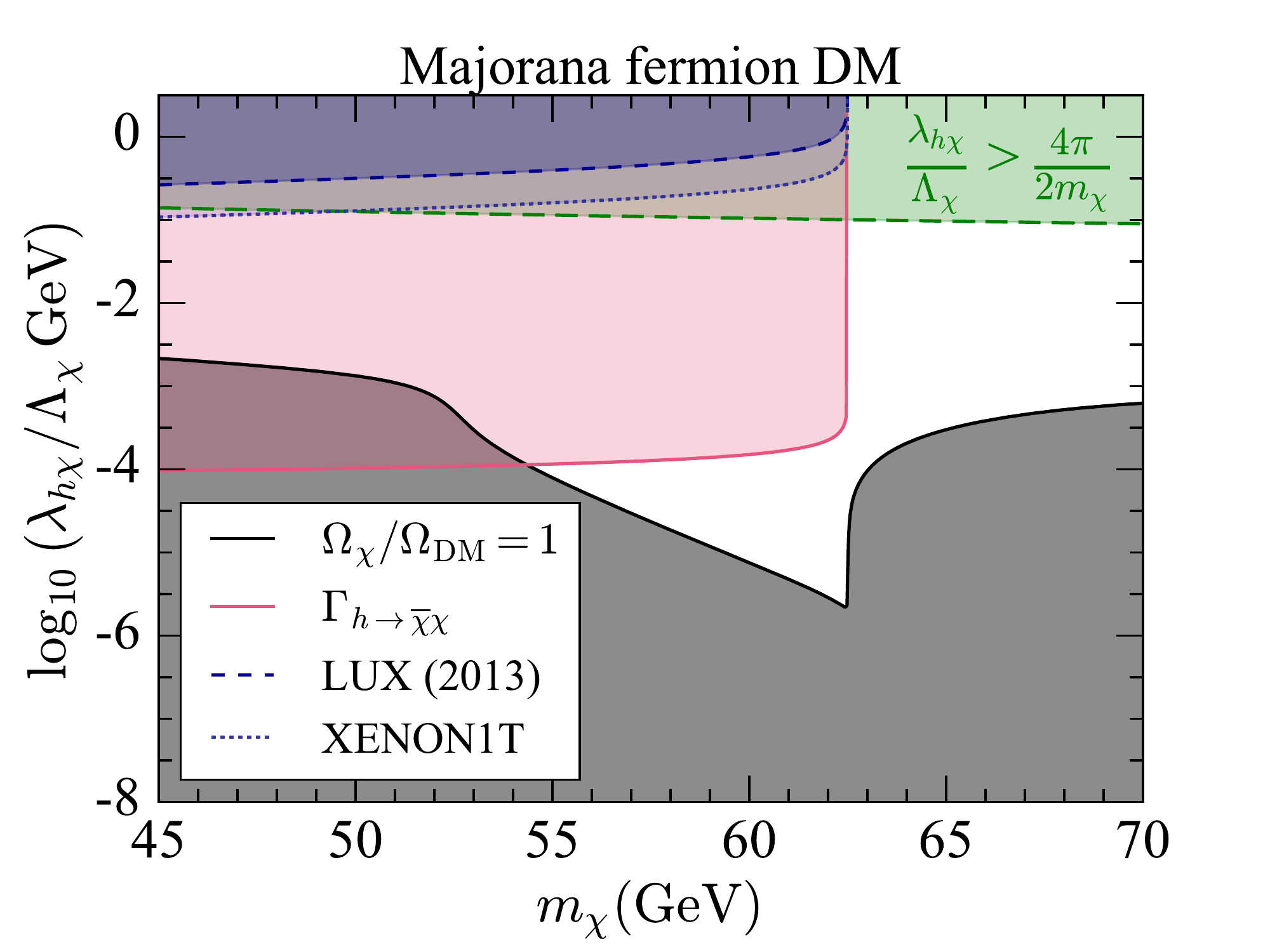}
    \end{minipage}
    \hfil
    \begin{minipage}[t]{0.48\textwidth}
      \centering
      \includegraphics[width=\textwidth]{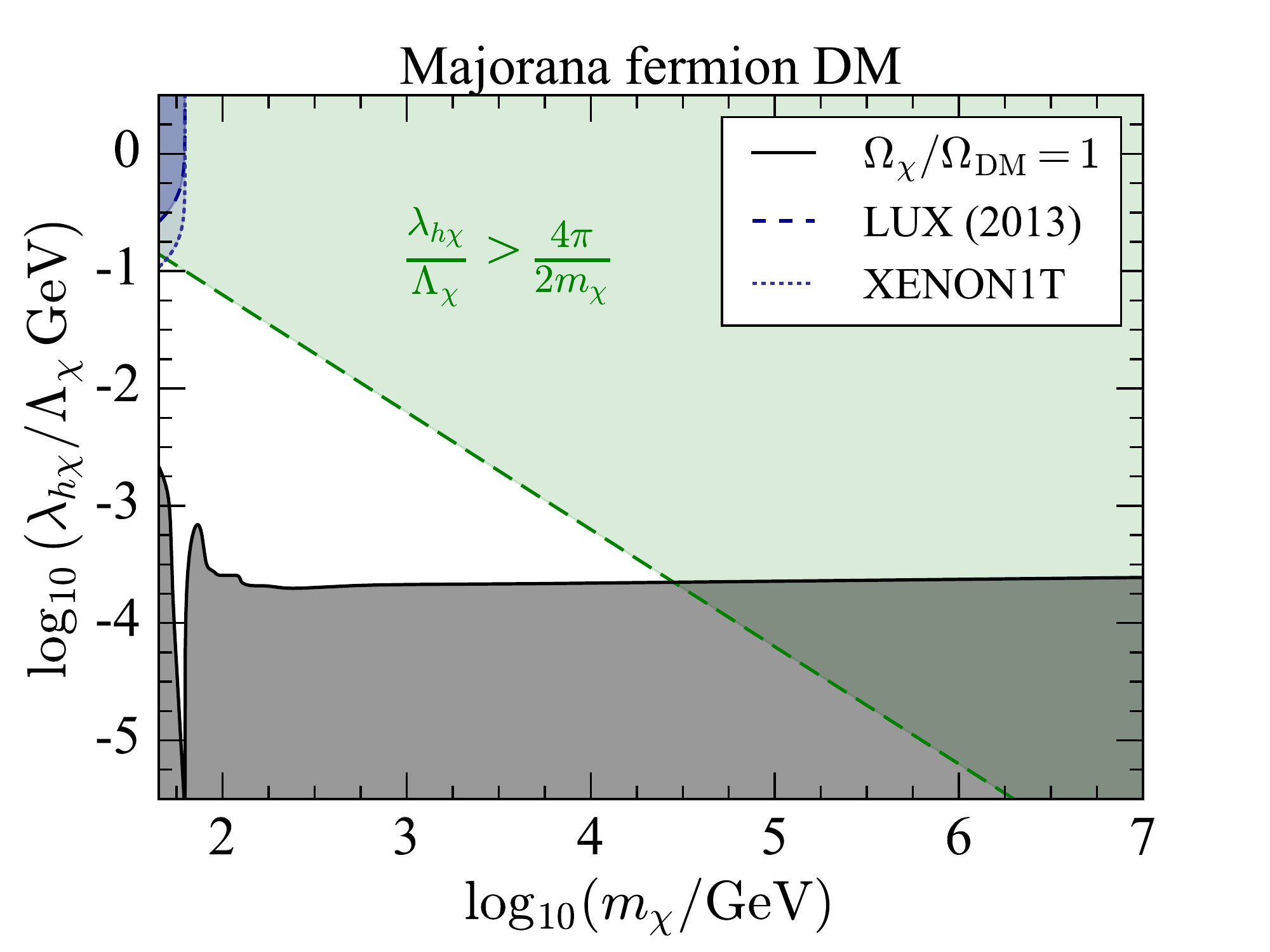} 
    \end{minipage}
    
    \small{\emph{c}). Pure pseudoscalar interaction, $\cos \xi = 0$.}
    
    \caption{Direct search limits on the Majorana model parameter space. The grey shaded region is ruled out by the relic density constraint. Regions excluded by the LUX (XENON1T) experiment are delineated with blue dashed (blue dotted) curves and dark (light) shadings. Although EFTs are valid at direct search experiments, our scaling of the LUX/XENON1T limits by the relic abundance parameter $f_{\textrm{rel}} = \Omega_\chi/\Omega_{\textrm{DM}}$ introduces a sensitivity to UV corrections when the EFT approximation in DM annihilations breaks down for $\lambda_{h\chi}/\Lambda_\chi > 4\pi/2m_\chi$. Left: A close-up of the resonantly enhanced annihilation region, $m_\chi \sim m_h/2$. The pink shaded region is excluded by an upper limit of $19\%$ on $\mathcal{BR}(h \rightarrow \ovr{\chi}\chi)$ at 2$\sigma$\,C.L. Right: The full range of $m_\chi$.}     
    \label{majorana-direct} 
\end{figure}

In Fig.~\ref{majorana-direct}, we present limits from the LUX (blue dashed) and the projected XENON1T (blue dotted) experiment in the $(m_\chi, \lambda_{h\chi}/\Lambda_\chi)$ plane for $\cos\xi = 1$ (top row), $1/\sqrt{2}$ (middle row) and $0$ (bottom row). Although the use of an EFT at direct search experiments is perfectly valid, our LUX and projected XENON1T limits within the green shaded regions are still expected to be subject to unaccounted-for UV corrections. This is due to the scaling of our limits by the relic abundance parameter $f_{\textrm{rel}} = \Omega_\chi/\Omega_{\textrm{DM}}$. Consequently, values of $\lambda_{h\chi}/\Lambda_\chi$ in the green shaded region for the Majorana fermion masses above $\sim$\,4.76\,TeV ($\cos\xi = 1$), $\sim$\,20.3\,TeV ($\cos\xi = 1/\sqrt{2}$) and $\sim$\,28\,TeV ($\cos\xi = 0)$ cannot be guaranteed to keep the EFT approximation for DM annihilations valid. 

When $\cos\xi = 0$, the SI cross section is momentum suppressed by a factor of $q^2/4m_\chi^2$ and hence the expected number of signal events $(s)$ is small. In fact, the imposed direct search limits are significantly weaker than the Higgs invisible width constraint at $2\sigma$\,C.L. Hence, the higher Majorana mass range with a pure pseudoscalar coupling (i.e., $\cos\xi = 0$) will be inaccessible at the XENON1T experiment. This coincides with better prospects from indirect detection (bottom row in Fig.~\ref{majorana-indirect}), making the latter class of observation the \emph{only} type of experiment capable of probing the higher DM mass range should nature have chosen to have DM interact with the SM Higgs boson only by a pure pseudoscalar coupling.

When interactions are pure scalar, the momentum suppression of the SI cross section is lifted. In the case of an equal mixing between the scalar and pseudoscalar terms, although the DM-nucleon effective couplings $(G_{N}^{\textrm{SI}},\,\widetilde{G}_{N}^{\textrm{SI}})$ are equal, the pseudoscalar effective couplings ($\widetilde{G}_{N}^{\textrm{SI}}$) carry a momentum dependence of $q^2/4m_\chi^2$ as evident in Eq.~\eqref{sigSI}. Consequently, the direct search limits are strongest in the case of a pure scalar interaction and moderate for an equal mix between the scalar and pseudoscalar terms .

At Majorana masses above 70\,GeV, direct search experiments will exclude large portions of the model parameter space as long as the EFT approximation remains valid for the model to make up all of the dark matter. The LUX experiment excludes Majorana masses up to $\sim$\,4.7\,TeV when $\cos \xi = 1$ and $\sim$\,200\,GeV when $\cos\xi = 1/\sqrt{2}$. Further exclusion will also be possible with the projected XENON1T experiment.

\subsection{Dirac fermion model}
The Dirac fermion model is similar to the Majorana fermion model. One aspect that separates Majorana fermion from Dirac fermion dark matter is the conventional factor of $1/2$ in front of each fermion bilinear $\ovr{\chi}\chi$ in defining a Majorana fermion field $\chi$. This factor of $1/2$  accounts for the field normalization and self-conjugation. The Majorana model results carry over to the Dirac model in a relatively straightforward way after this factor of $1/2$ is accounted for. 

\begin{figure}
    \begin{minipage}[t]{0.48\textwidth}        
      \centering
      \includegraphics[width=\textwidth]{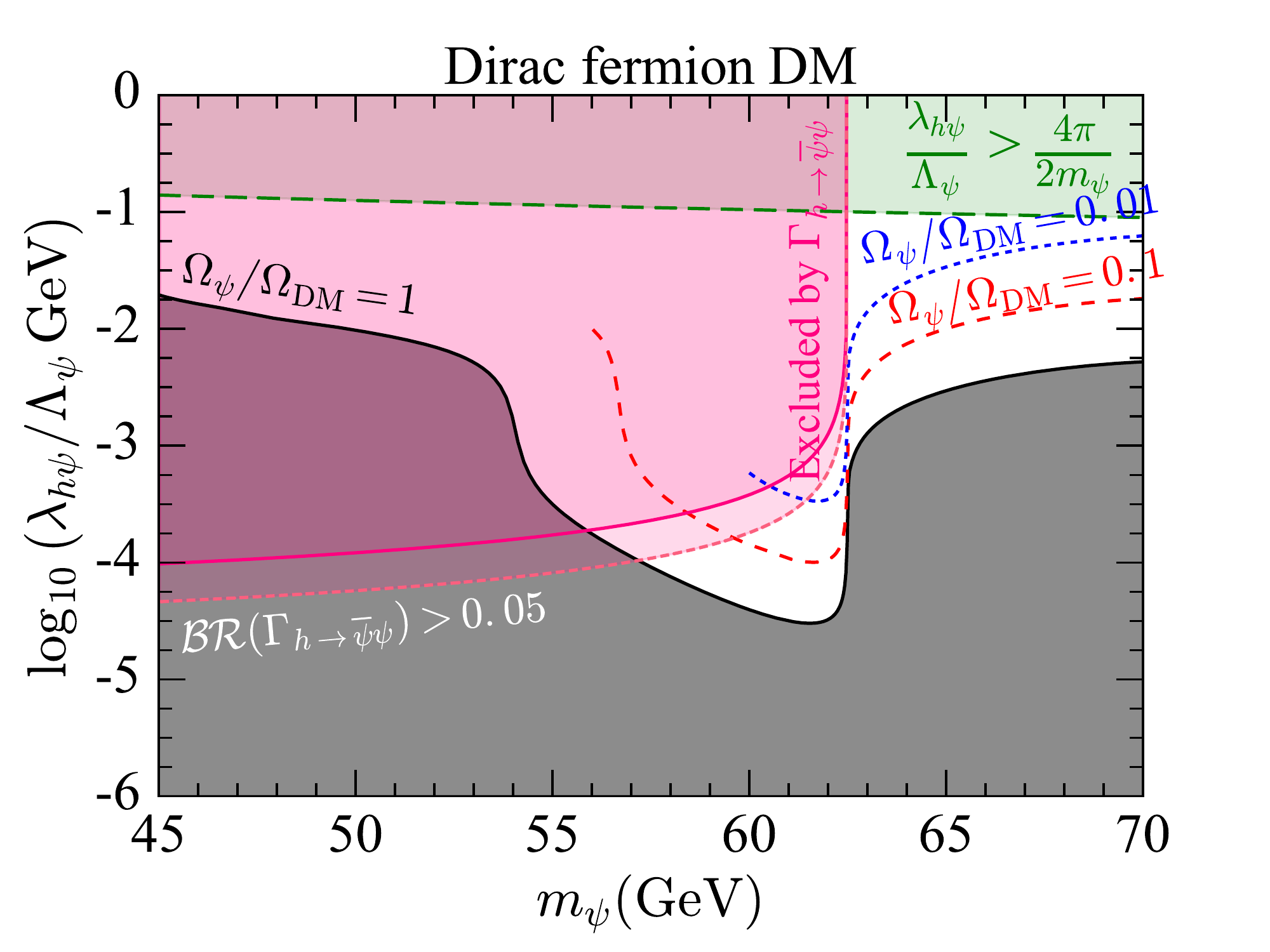}
    \end{minipage}
    \hfil
    \begin{minipage}[t]{0.48\textwidth}
      \centering
      \includegraphics[width=\textwidth]{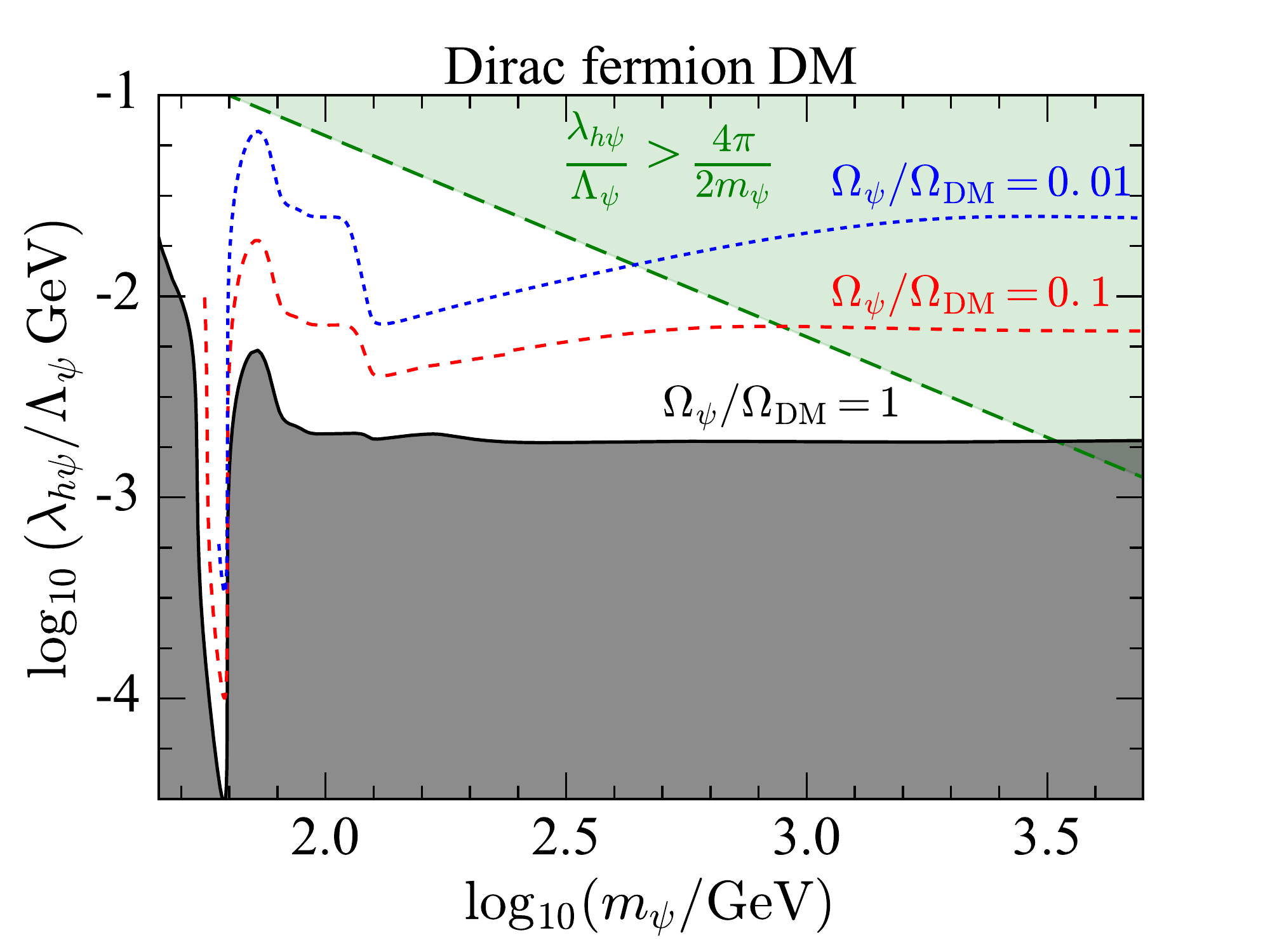} 
    \end{minipage}   
	
	\small{\emph{a}). Pure scalar interaction, $\cos \xi = 1$.}
	
    \begin{minipage}[t]{0.48\textwidth}        
      \centering
      \includegraphics[width=\textwidth]{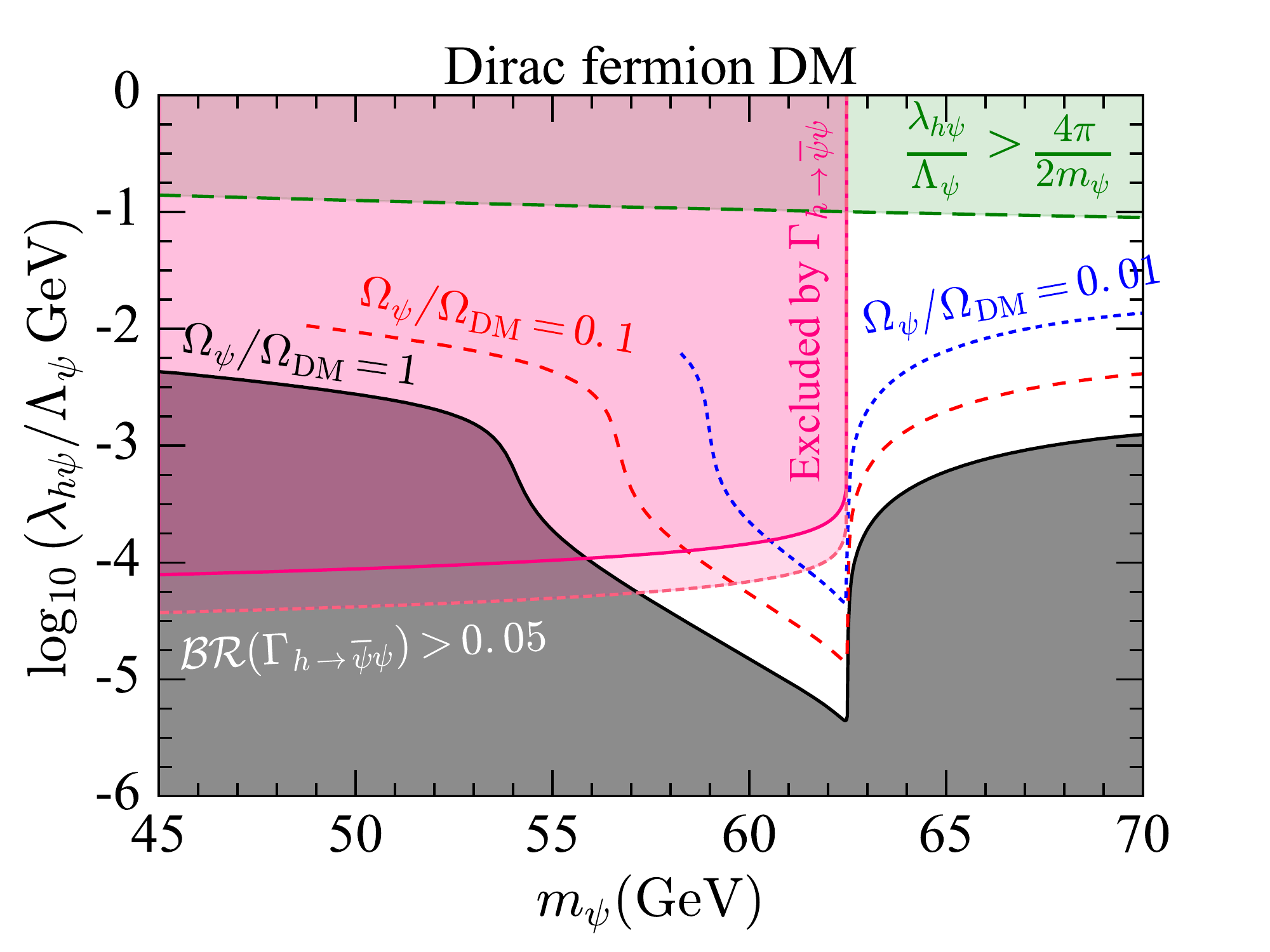}
    \end{minipage}
    \hfil
    \begin{minipage}[t]{0.48\textwidth}
      \centering
      \includegraphics[width=\textwidth]{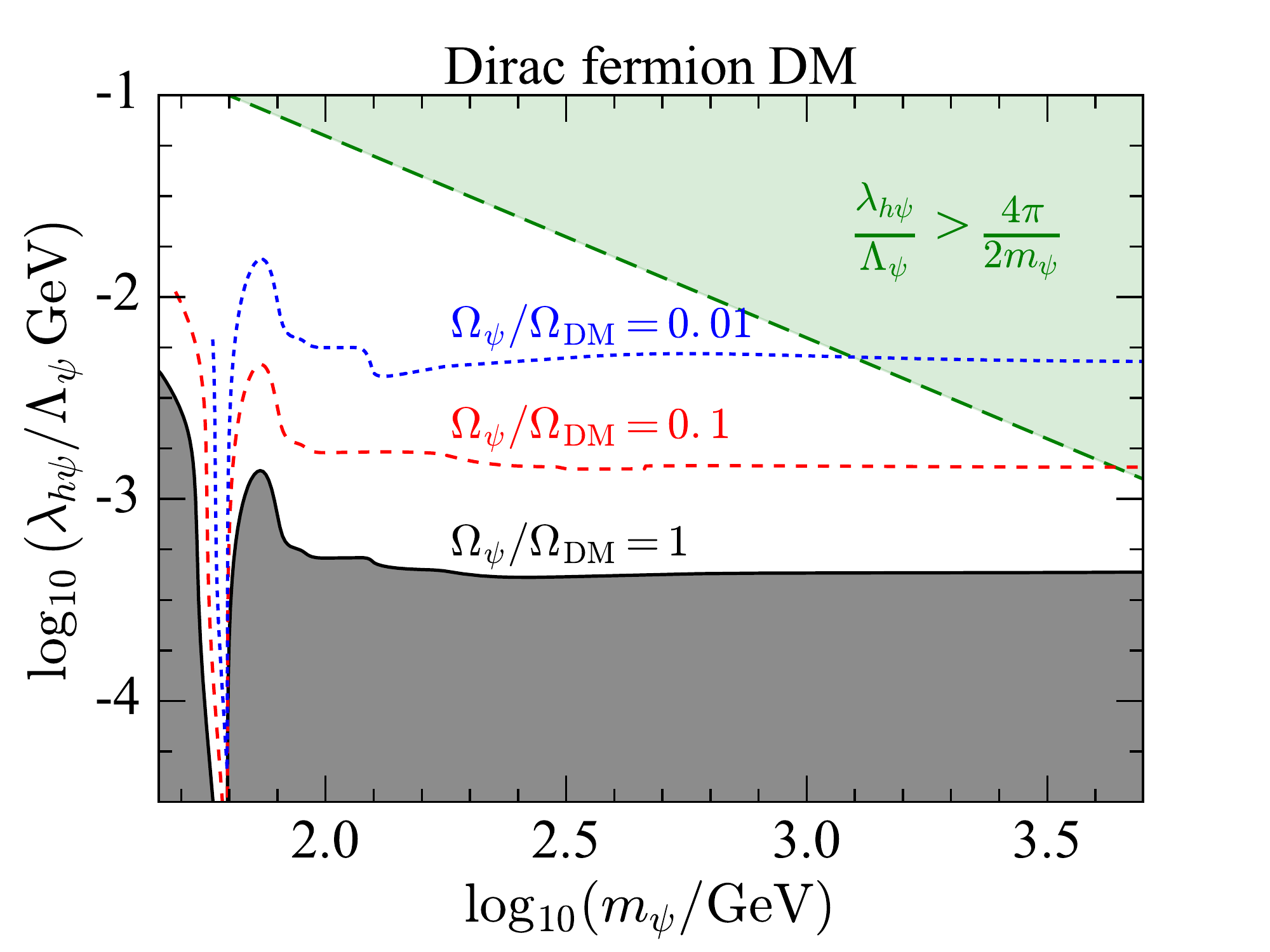}
    \end{minipage} 
	
	\small{\emph{b}). Equally mixed scalar-pseudoscalar interaction, $\cos \xi = 1/\sqrt{2}$.}
	     
    \begin{minipage}[t]{0.48\textwidth}        
      \centering
      \includegraphics[width=\textwidth]{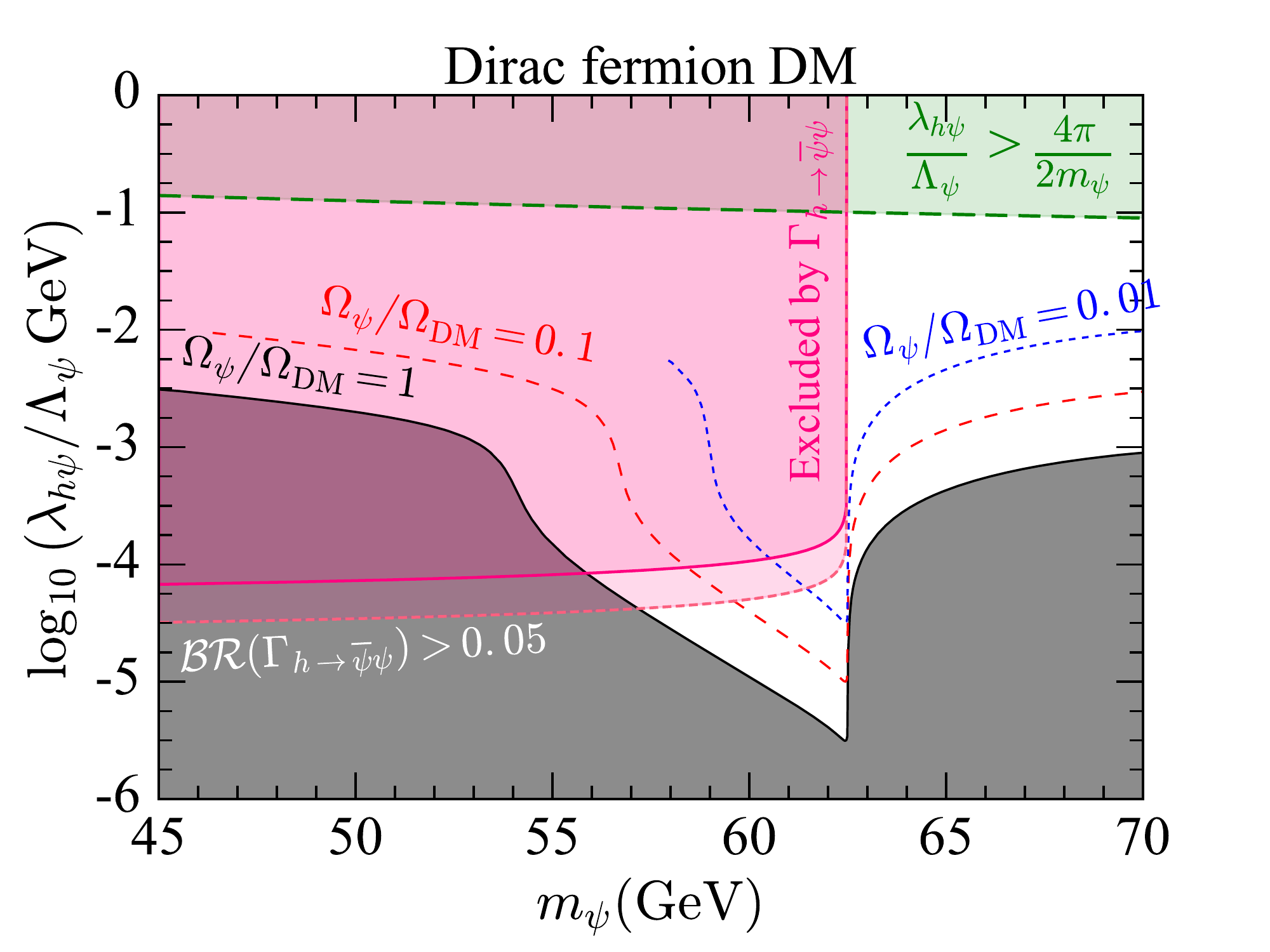}
    \end{minipage}
    \hfil
    \begin{minipage}[t]{0.48\textwidth}
      \centering
      \includegraphics[width=\textwidth]{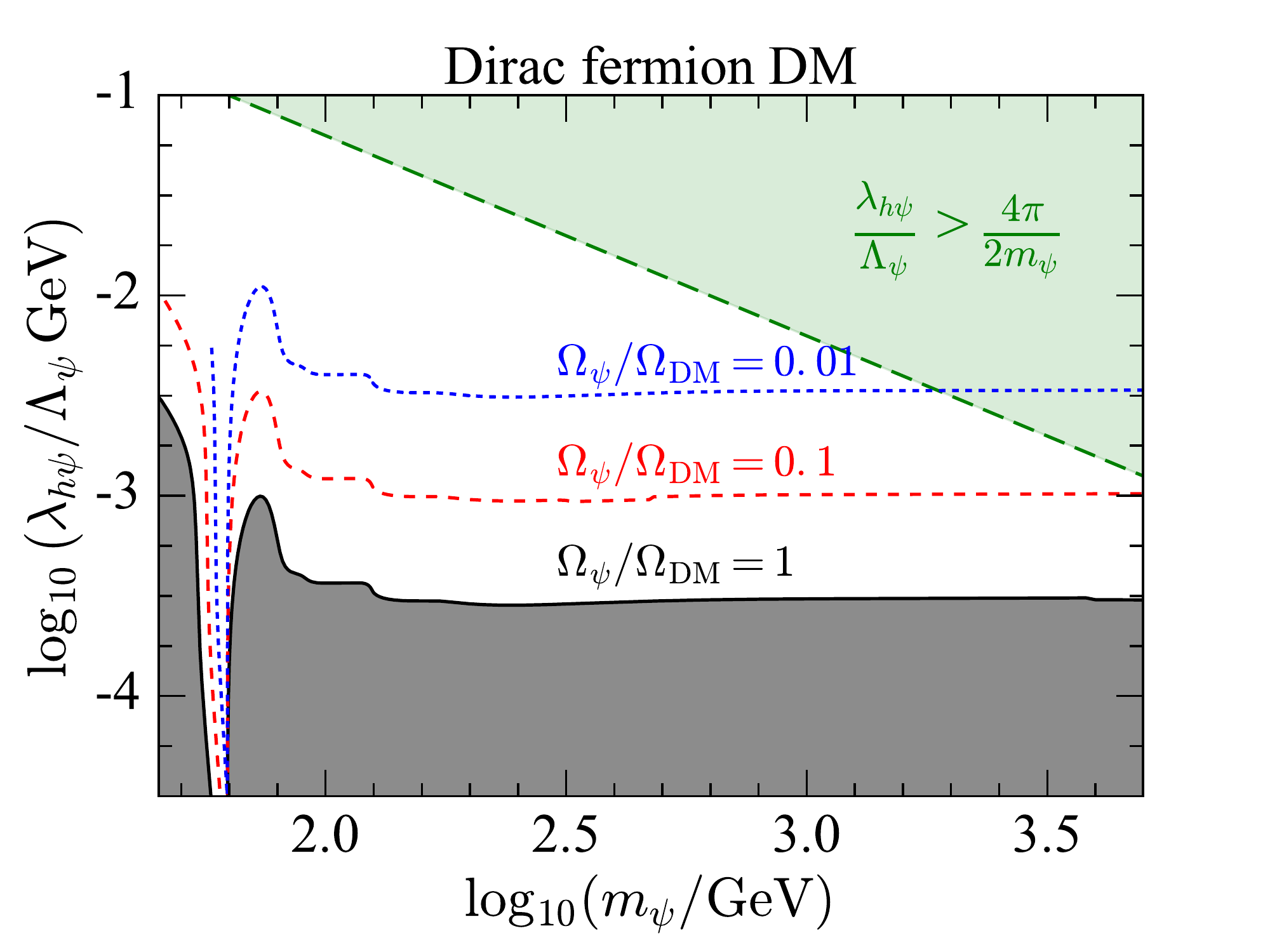} 
    \end{minipage}
    
    \small{\emph{c}). Pure pseudoscalar interaction, $\cos \xi = 0$.}
    
    \caption{Contours of fixed Dirac relic density for $f_{\textrm{rel}} = 1$ (black solid), 0.1 (red dashed) and 0.01 (blue dotted). The grey shaded region is ruled out due to an overabundance of dark matter. The green shaded region is where the EFT approximation of the full theory breaks down for $\lambda_{h\psi}/\Lambda_\psi > 4\pi/2m_\psi$. Left: A close-up of the resonantly enhanced annihilation region, $m_\psi \sim m_h/2$. Larger values of $\lambda_{h\psi}/\Lambda_\psi$ are excluded by an upper limit of 19\% (pink solid) at 2$\sigma$ C.L. or 5\% (pink dotted) at 1$\sigma$\,C.L. on $\mathcal{BR}(h \rightarrow \ovr{\psi}\psi)$. Right: Relic density contours for the full range of $m_\psi$.}
    \label{dirac-relic}
\end{figure}

In Fig.~\ref{dirac-relic}, we show the contours of fixed Dirac relic density in $(m_\psi,\lambda_{h\psi}/\Lambda_\psi)$ plane for $f_{\textrm{rel}} =$ 1 (black solid), 0.1 (red dashed) and 0.01 (blue dotted). Analogous to the Majorana model, contours in each row are generated at fixed values of $\cos\xi$: $\cos\xi = 1$ for a pure scalar interaction (top row), $\cos\xi = 1/\sqrt{2}$ for an equal mix between the scalar and pseudoscalar terms (middle row), and $\cos\xi = 0$ for a pure pseudoscalar interaction (bottom row). For $m_\psi < m_h/2$, most of the model parameter space except a small triangular region between $\sim$\,57.5\,GeV and $m_h/2$ is excluded by the combined constraints on the Dirac relic density and the Higgs invisible width. Similar to the Majorana model, roots of Eq.~\eqref{root-relic} for $f_{\textrm{rel}} = 0.1$ and 0.01 do not exist at Dirac masses below $m_h/2$. At higher Dirac fermion masses, the relic density contours continue to increase and ultimately become independent of the coupling $\lambda_{h\psi}/\Lambda_\psi$.  

\begin{figure}
    \begin{minipage}[t]{0.48\textwidth}        
      \centering
      \includegraphics[width=\textwidth]{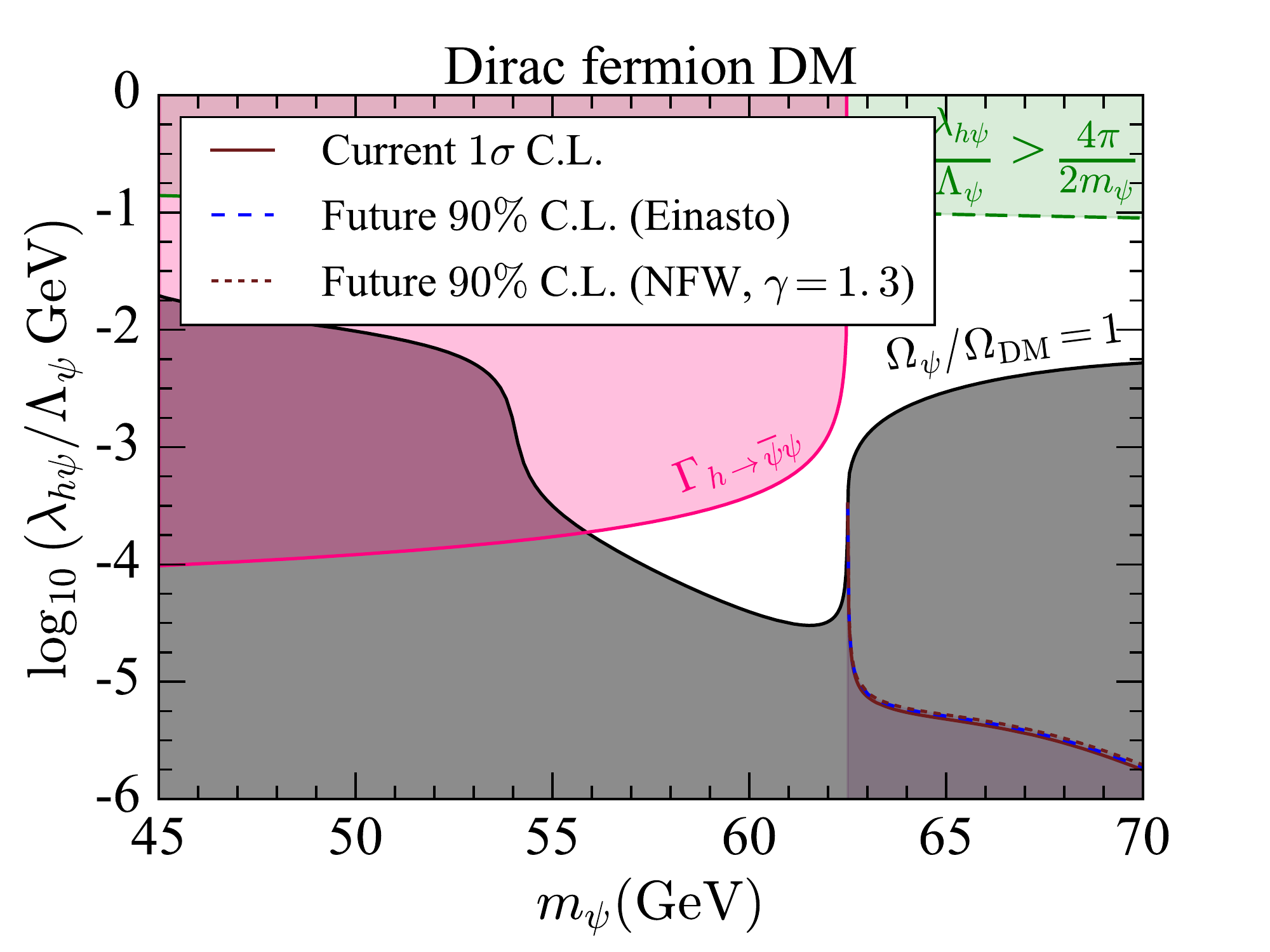}
    \end{minipage}
    \hfil
    \begin{minipage}[t]{0.48\textwidth}
      \centering
      \includegraphics[width=\textwidth]{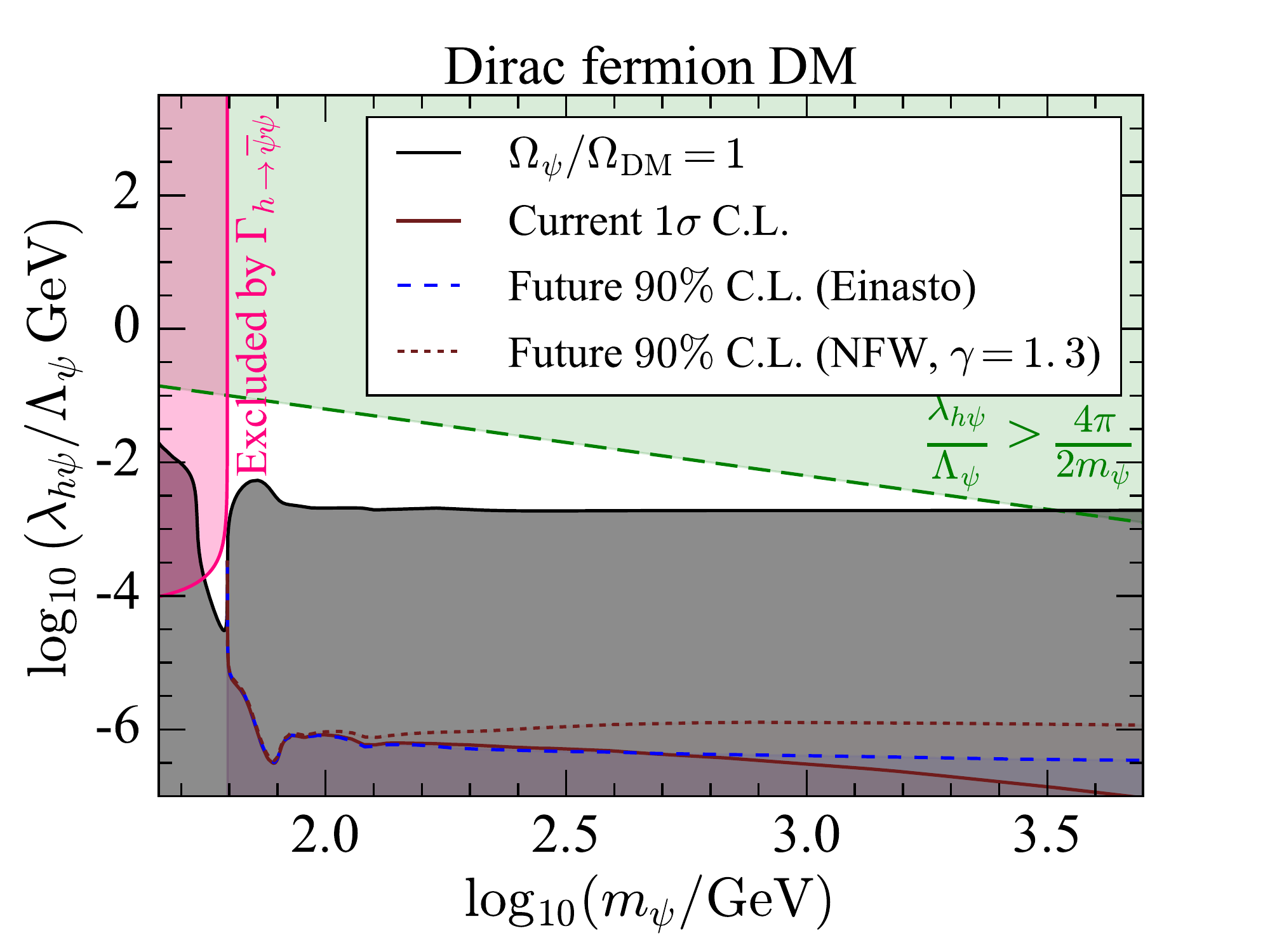}
    \end{minipage}
	
	\small{\emph{a}). Pure scalar interaction, $\cos \xi = 1$.}
	
    \begin{minipage}[t]{0.48\textwidth}        
      \centering
      \includegraphics[width=\textwidth]{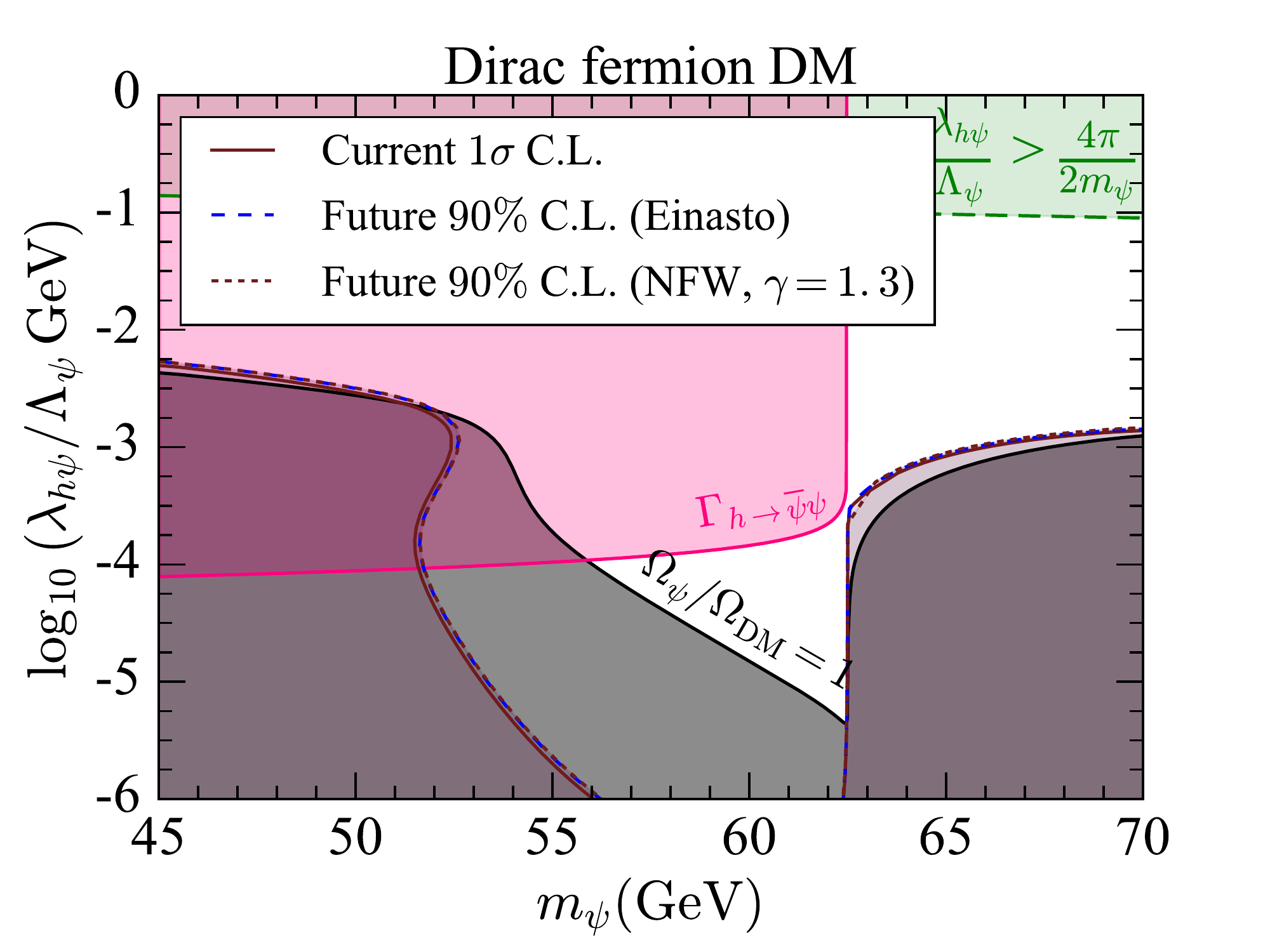}
    \end{minipage}
    \hfil
    \begin{minipage}[t]{0.48\textwidth}
      \centering
      \includegraphics[width=\textwidth]{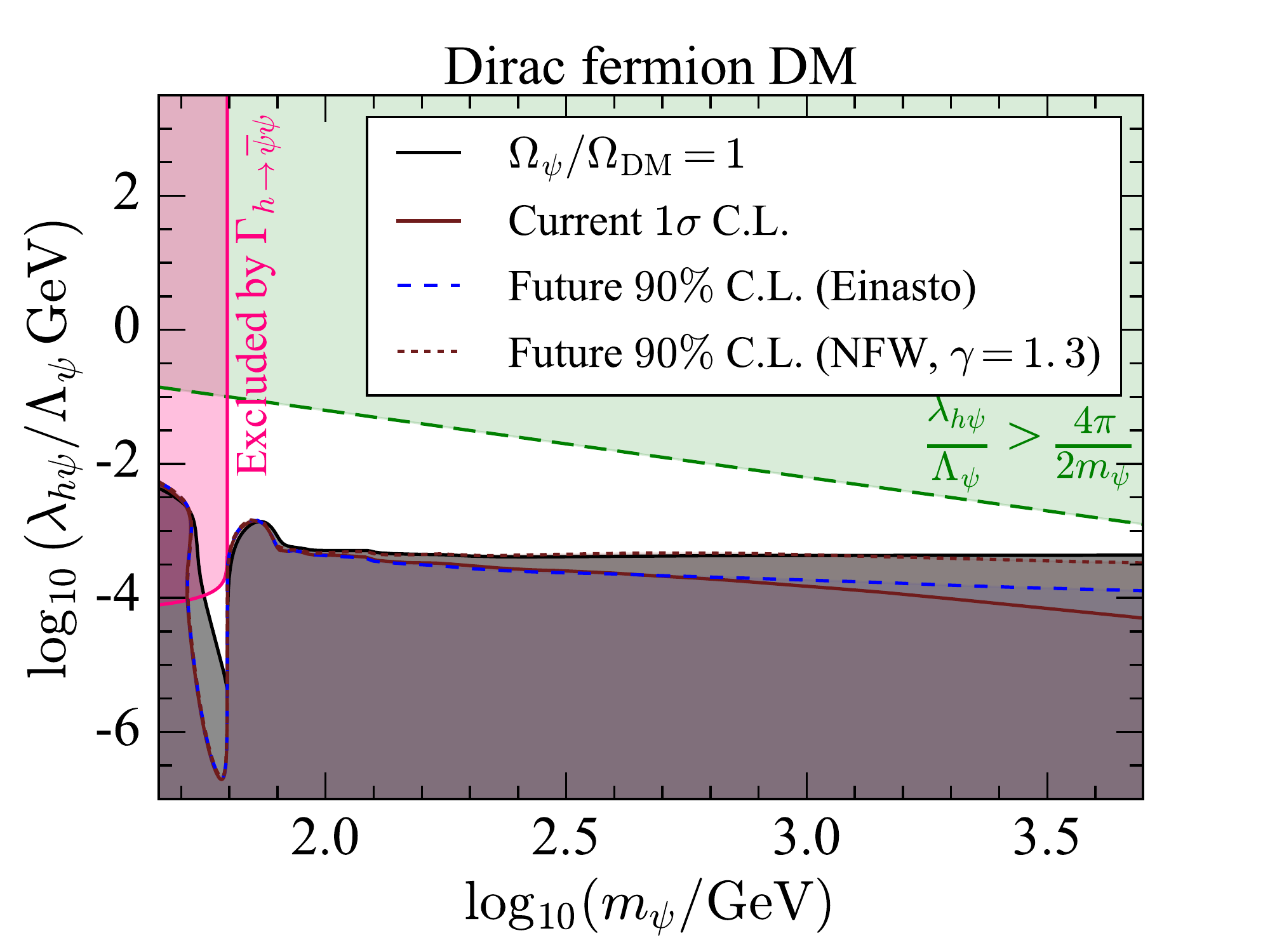}
    \end{minipage}
	
	\small{\emph{b}). Equally mixed scalar-pseudoscalar interaction, $\cos \xi = 1/\sqrt{2}$.}     
	
    \begin{minipage}[t]{0.48\textwidth}        
      \centering
      \includegraphics[width=\textwidth]{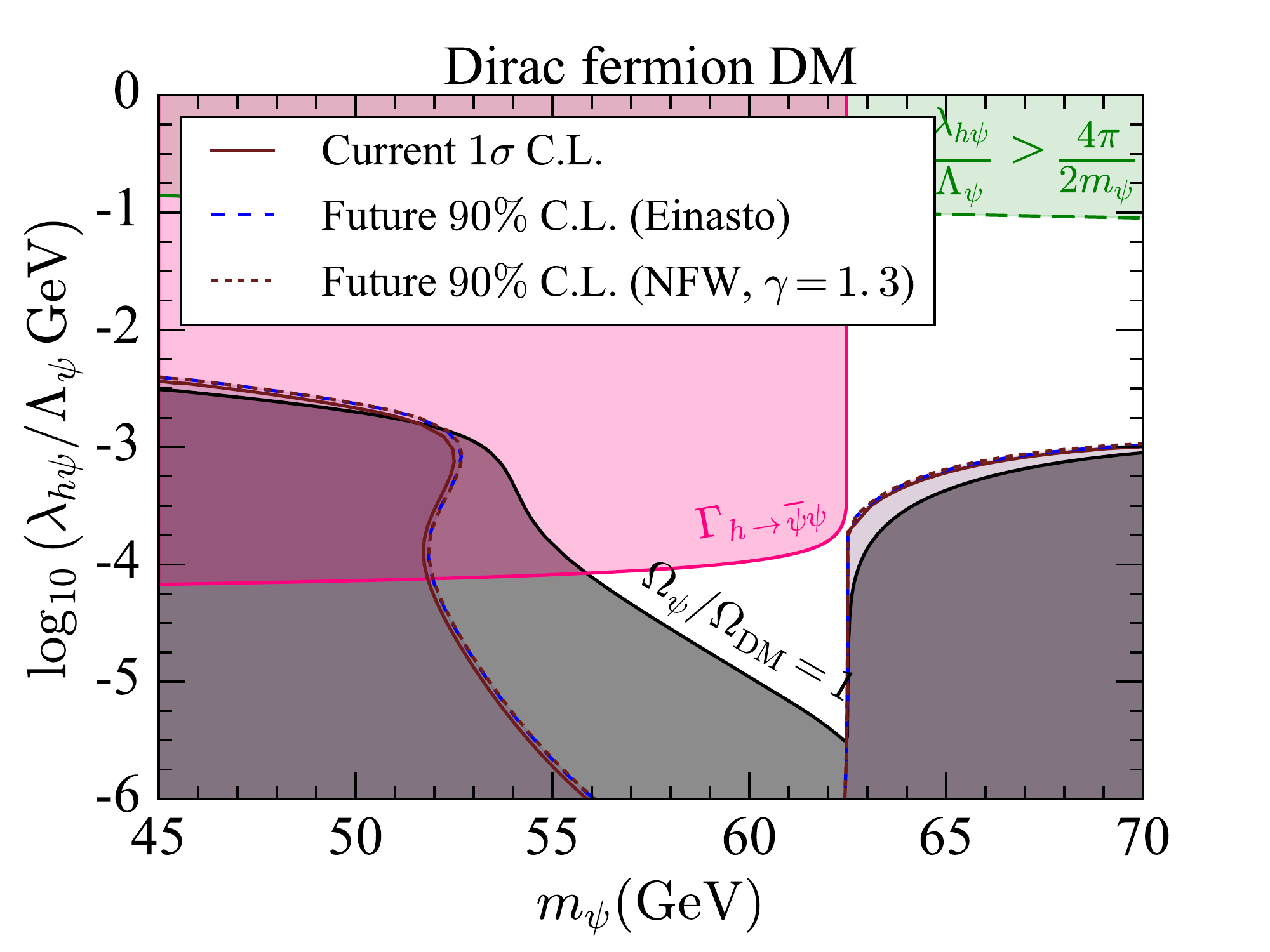}
    \end{minipage}
    \hfil
    \begin{minipage}[t]{0.48\textwidth}
      \centering
      \includegraphics[width=\textwidth]{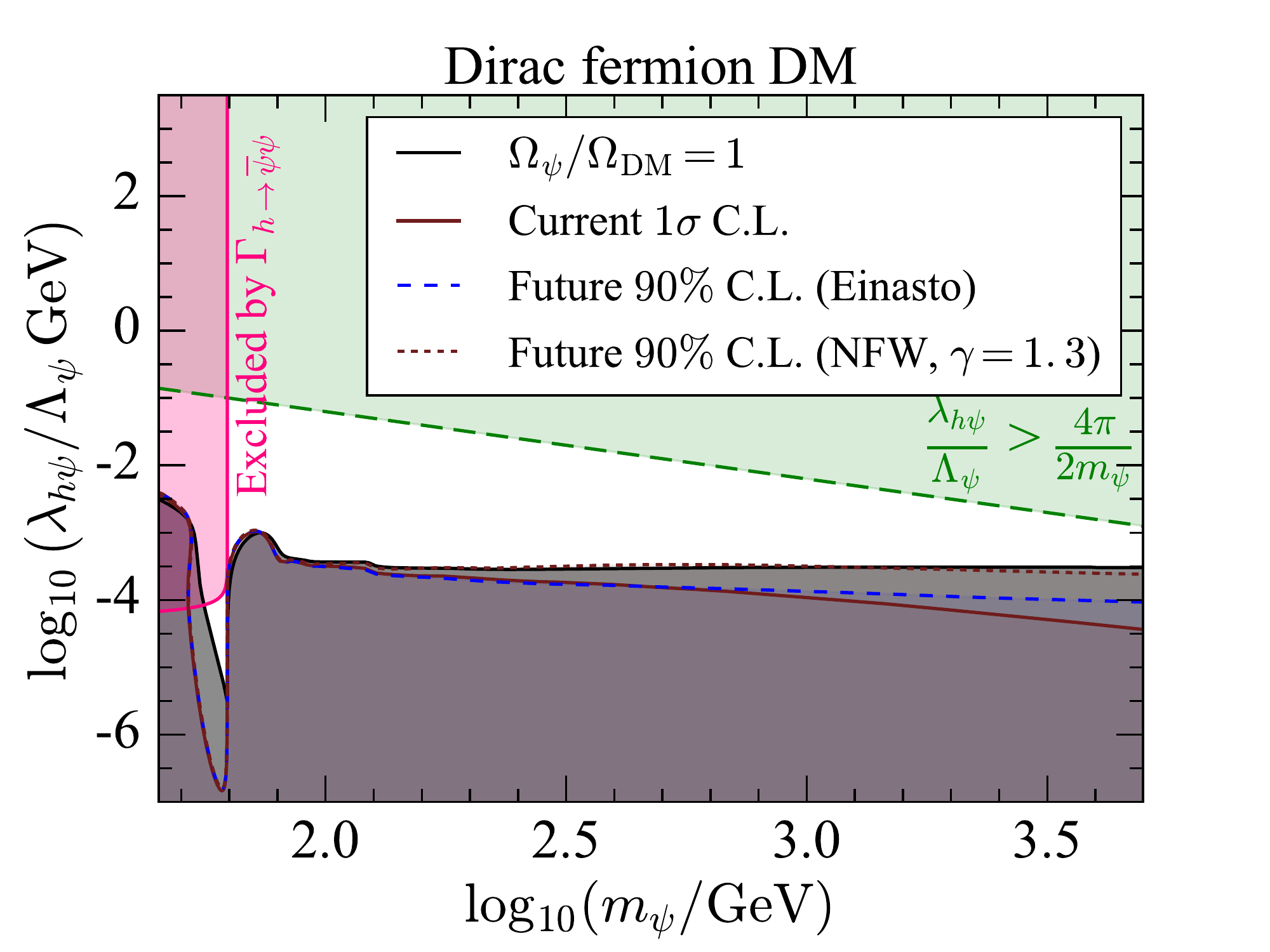}
    \end{minipage}
	
	\small{\emph{c}). Pure pseudoscalar interaction, $\cos \xi = 0$.}
	
    \caption{Indirect search limits on the Dirac model parameter space. The grey and pink shaded regions are excluded respectively by the relic density and the Higgs invisible width constraints. The green shaded region is where the EFT approximation of the full theory breaks down for $\lambda_{h\psi}/\Lambda_\psi > 4\pi/2m_\psi$. Values of $\lambda_{h\psi}/\Lambda_\psi$ below the current 1$\sigma$\,C.L. (brown solid) curve are excluded at more than $1\sigma$\,C.L. Regions below the future 90\%\,C.L. curve with the Einasto (blue dashed) and contracted NFW (brown dotted) profile will be excluded. Left: A close-up of the resonantly enhanced annihilation region, $m_\psi \sim m_h/2$. Right: The full range of $m_\psi$.}
    \label{dirac-indirect}
\end{figure}

The sensitivity of indirect searches to various regions in the $(m_\psi,\lambda_{h\psi}/\Lambda_{\psi})$ plane when $\cos\xi = 1$ (top row), $1/\sqrt{2}$ (middle row) and 0 (bottom row) are shown in Fig.~\ref{dirac-indirect}. Again, in the case of $\cos\xi = 1$, the annihilation cross section $\sigma v_{\textrm{rel}}$ is velocity suppressed. Consequently, the resulting indirect search limits are weak. In fact, the relic density constraint alone is strong enough to exclude the entire region probed by the indirect search experiments in this case.

As interactions become pure pseudoscalar, the velocity suppression of $\sigma v_{\textrm{rel}}$ is lifted. Therefore, parts of the model parameter space can be excluded by the indirect search experiments. With the strongest indirect limits in the case of a pure pseudoscalar interaction, Dirac masses between $m_h/2$ and 70\,GeV for $\log_{10} (\lambda_{h\psi}/\Lambda_\psi \textrm{\,GeV}) \in [-4,-3]$  can be seen to be in tension with the current indirect search limits which incorporates the combined analysis of 15 dwarf galaxies by Fermi-LAT and 7 years of CMB observation by WMAP (WMAP7). Similarly, Dirac masses below $\sim$\,51.5\,GeV are in tension with the current indirect searches at more than $1\sigma$\,C.L. On the other hand, future indirect searches will be able to exclude Dirac masses up to $\sim$\,74\,GeV for the Einasto profile and between 166\,GeV and 1.3\,TeV for a contracted NFW profile in the case of a pure pseudoscalar interaction, if $\psi$ makes up all of the dark matter.

\begin{figure}
    \begin{minipage}[t]{0.48\textwidth}        
      \centering
      \includegraphics[width=\textwidth]{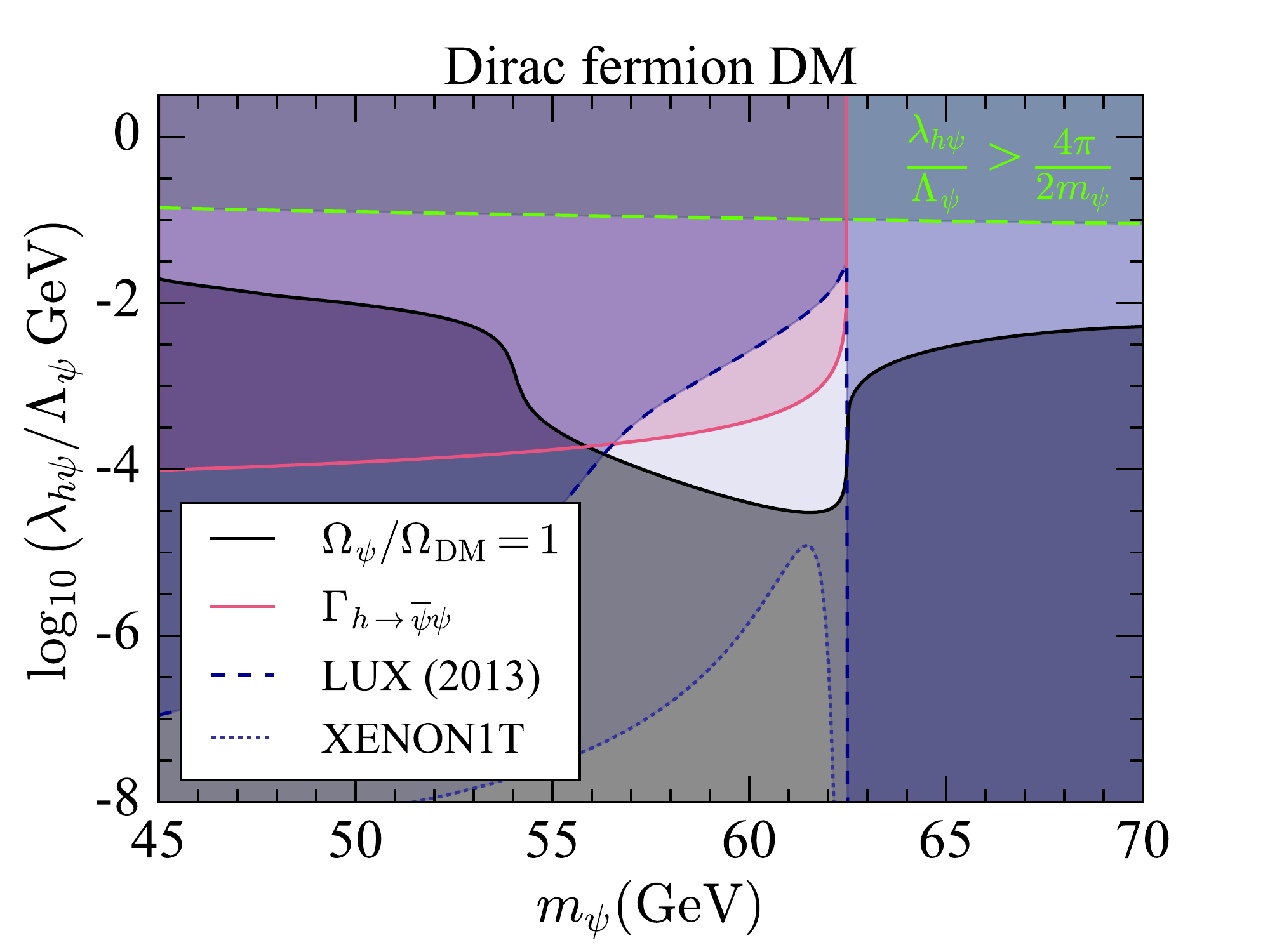}
    \end{minipage}
    \hfil
    \begin{minipage}[t]{0.48\textwidth}
      \centering    
      \includegraphics[width=\textwidth]{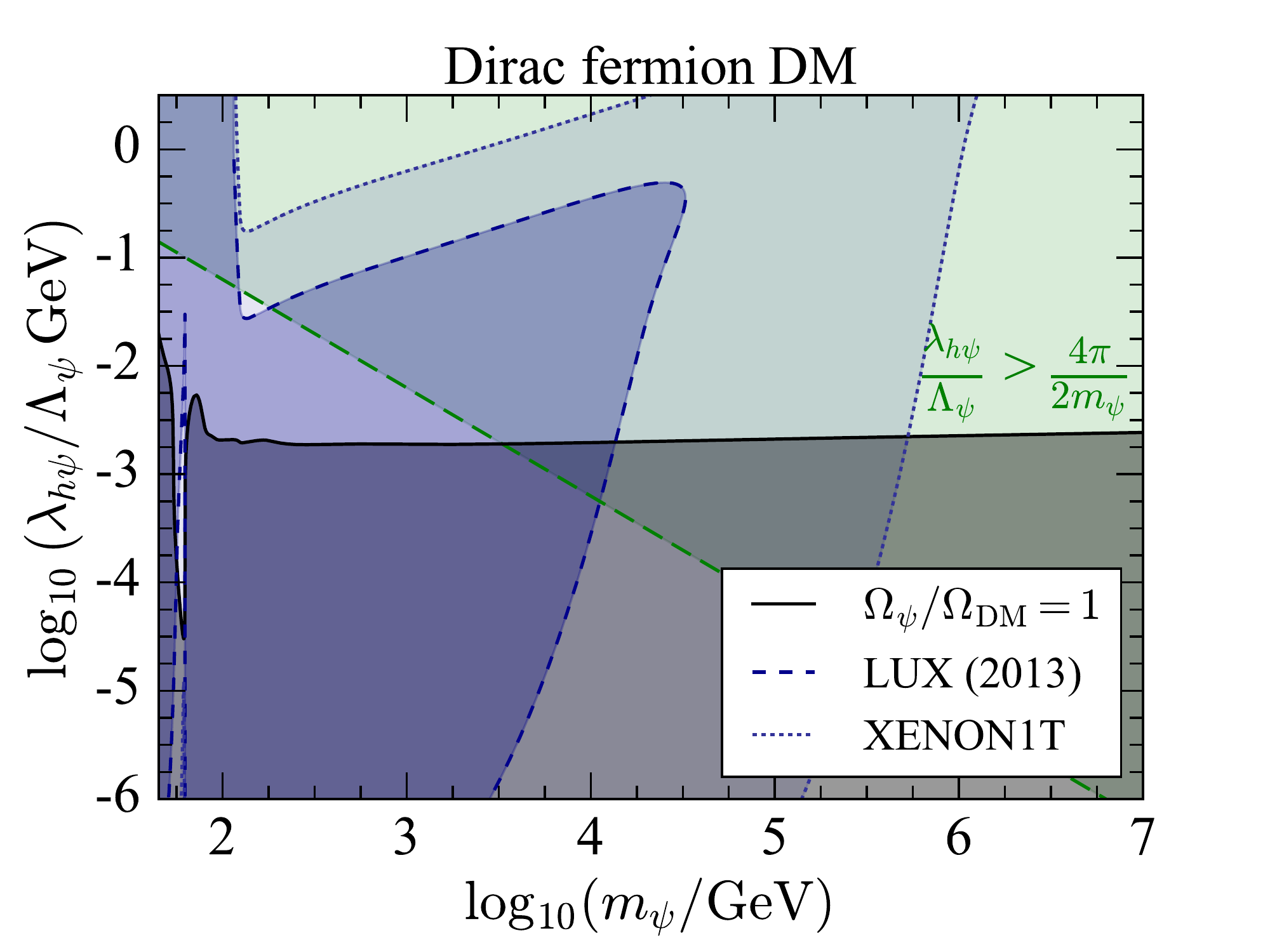}  
    \end{minipage} 
	
	\small{\emph{a}). Pure scalar interaction, $\cos \xi = 1$.}      
	
    \begin{minipage}[t]{0.48\textwidth}        
      \centering
      \includegraphics[width=\textwidth]{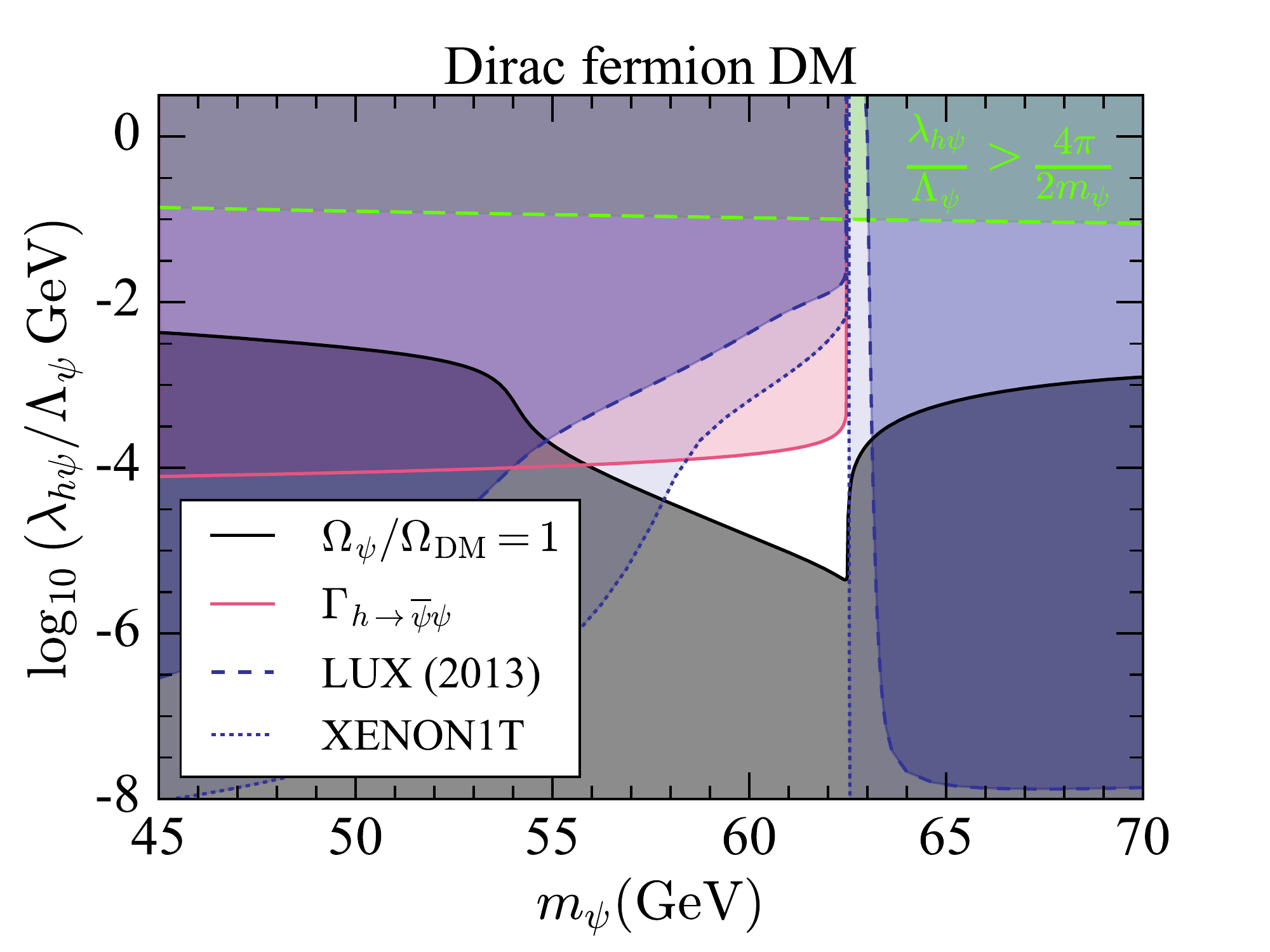}
    \end{minipage}
    \hfil
    \begin{minipage}[t]{0.48\textwidth}
      \centering
      \includegraphics[width=\textwidth]{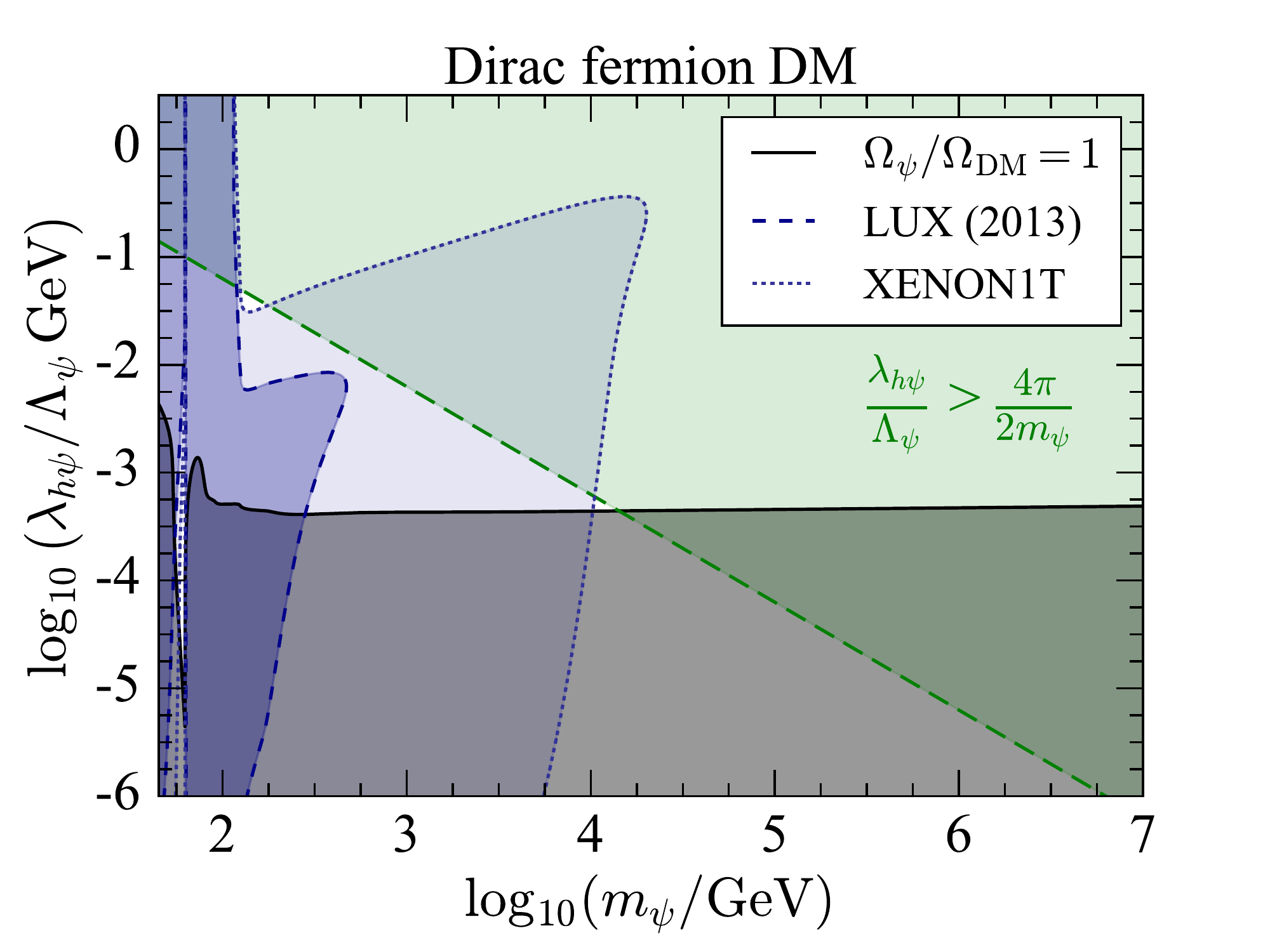}  
    \end{minipage} 
	
	\small{\emph{b}). Equally mixed scalar-pseudoscalar interaction, $\cos \xi = 1/\sqrt{2}$.}       
	
    \begin{minipage}[t]{0.48\textwidth}        
      \centering
      \includegraphics[width=\textwidth]{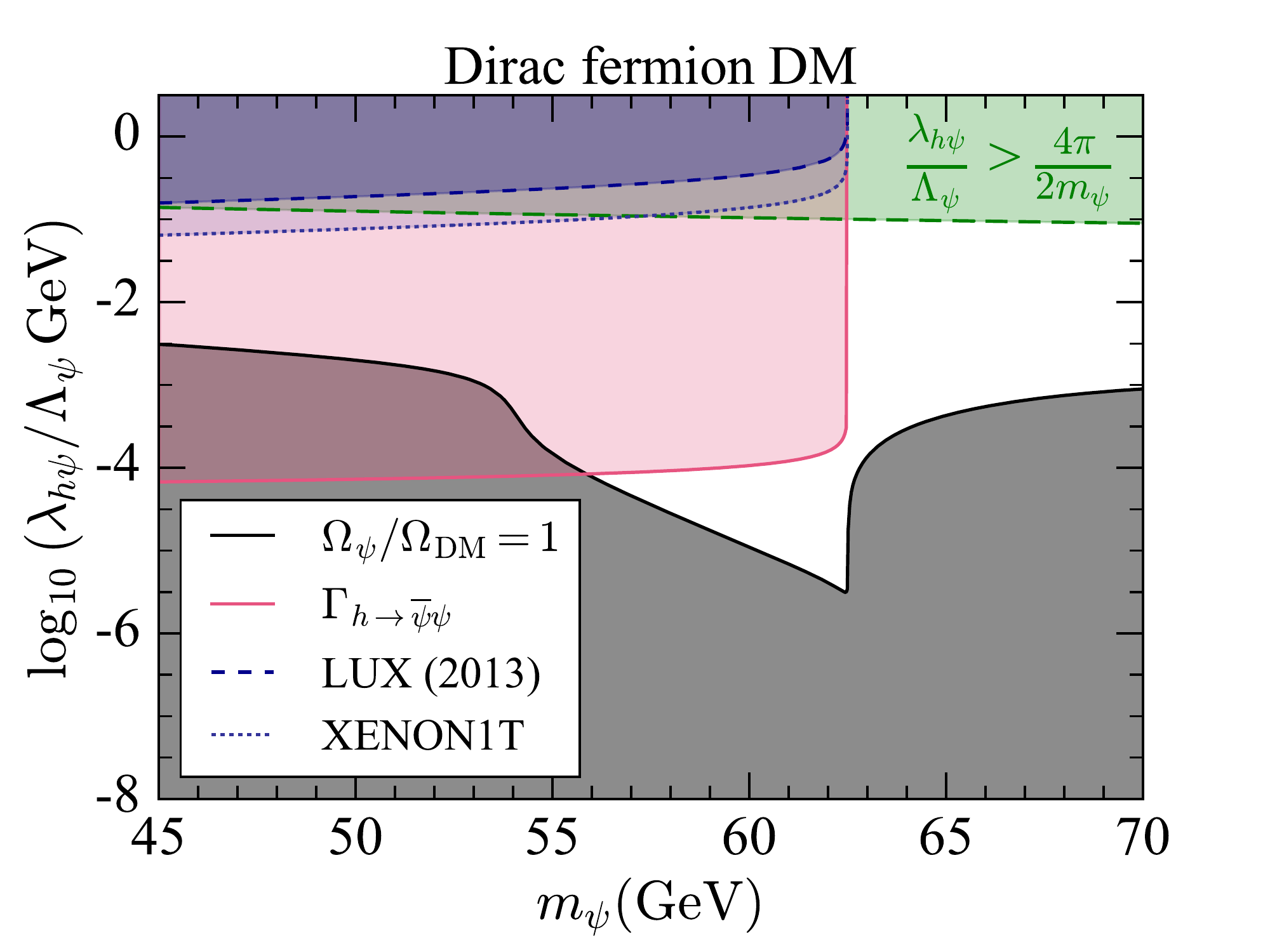}
    \end{minipage}
    \hfil
    \begin{minipage}[t]{0.48\textwidth}
      \centering
      \includegraphics[width=\textwidth]{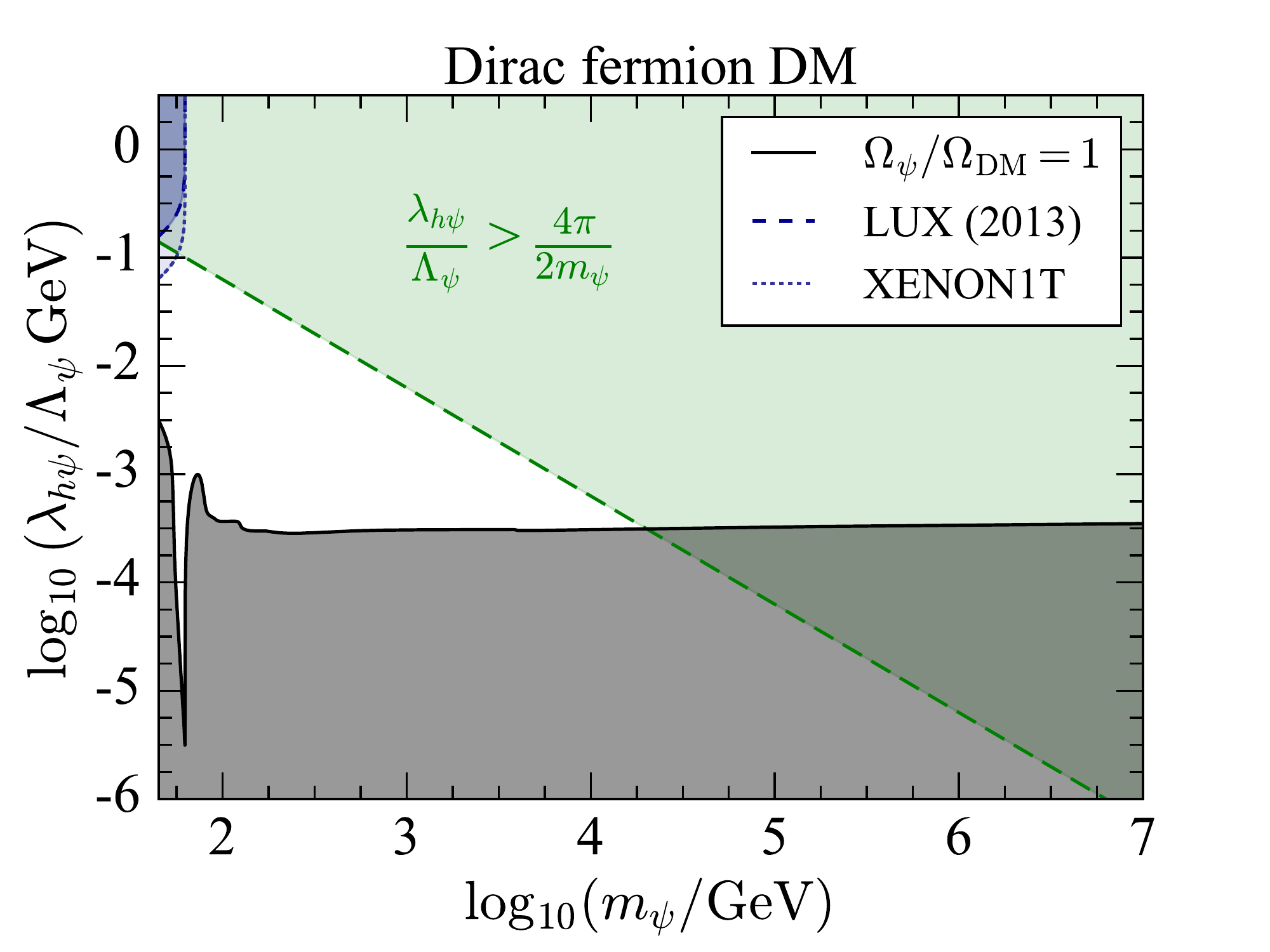}  
    \end{minipage}   
	
	\small{\emph{c}). Pure pseudoscalar interaction, $\cos \xi = 0$.}
	
    \caption{Direct search limits on the Dirac model parameter space. The grey shaded region is ruled out by the relic density constraint. The regions excluded by the LUX (XENON1T) experiment are delineated with blue dashed (blue dotted) curves and dark (light) shadings. Although, EFTs are valid at direct search experiments, our scaling of the LUX/XENON1T limits by the relic abundance parameter $f_{\textrm{rel}} = \Omega_\psi/\Omega_{\textrm{DM}}$ introduces a sensitivity to UV corrections when the EFT approximation in DM annihilations breaks down for $\lambda_{h\psi}/\Lambda_\psi > 4\pi/2m_\psi$. Left: A close-up of the resonantly enhanced annihilation region, $m_\psi \sim m_h/2$. The pink shaded region is excluded by an upper limit of $19\%$ on $\mathcal{BR}(h \rightarrow \ovr{\psi}\psi)$ at 2$\sigma$\,C.L. Right: The full range of $m_\psi$.}    
    \label{dirac-direct}
\end{figure}

In Fig.~\ref{dirac-direct}, we present direct search limits in the $(m_\psi, \lambda_{h\psi}/\Lambda_\psi)$ plane for $\cos\xi = 1$ (top row), $ 1/\sqrt{2}$ (middle row) and $0$ (bottom row). Similar to the Majorana fermion model, the LUX/XENON1T limits within the green shaded regions are subject to UV corrections solely due to our scaling of the limits by the relic abundance parameter $f_\textrm{rel} = \Omega_\psi/\Omega_{\textrm{DM}}$. For the Dirac masses above $\sim$3.3\,TeV ($\cos\xi = 1$), $\sim$14.2\,TeV ($\cos\xi = 1/\sqrt{2}$) and $\sim$\,19.7\,TeV ($\cos\xi = 0$), our limits cannot be guaranteed to keep the EFT approximation for DM annihilations valid, if $\psi$ makes up all of the dark matter. 

In the case of $\cos\xi = 0$, the SI cross section is momentum suppressed by a factor of $q^2/4m_\psi^2$. Consequently, the expected event rates are small. In analogy with the Majorana model when $\cos\xi = 0$, better prospects from indirect searches (bottom row in Fig.~\ref{dirac-indirect}) make the latter class of observation the \emph{only} type of experiment capable of probing the higher DM mass range should nature have chosen to have DM interact with the SM Higgs boson only by a pure pseudoscalar coupling.

As interactions become pure scalar $(\cos \xi = 1)$, the momentum suppression of the SI cross section is lifted, which results in significant direct search rates. The LUX experiment excludes Dirac masses up to $\sim$\,3.3\,TeV when $\cos\xi = 1$ and $\sim$\,280\,GeV when $\cos\xi = 1/\sqrt{2}$. In contrast with the Majorana model when $\cos\xi = 1$, low Dirac fermion masses between 45\,GeV and $m_h/2$ will be \emph{entirely} excluded by the  combined limits on the Dirac fermion relic density, the Higgs invisible width and projected XENON1T experiment. Furthermore, the projected XENON1T experiment will reach higher sensitivity in excluding TeV-scale Dirac masses, if $\psi$ makes up all of the dark matter. 

\section{Conclusions}\label{sec5}
In this work, we have performed an updated and combined analysis of effective scalar, vector, Majorana and Dirac fermion Higgs portal models of dark matter. For the fermion models, we investigated cases where interactions are either pure scalar, pure pseudoscalar or an equal mixture of the scalar and pseudoscalar terms. The presence of the pseudoscalar term and the requirement of a quadratic DM term corresponding to a real mass lead us to redefine the post-EWSB fermion fields through a chiral rotation. 

In nonfermion models, the combined constraints on the DM relic density and the Higgs invisible width exclude most of the low mass parameter space in all models apart from a small triangular region close to the resonance $m_X \sim m_h/2$. However, in the special case of a Dirac fermion DM interacting with the SM Higgs boson via a pure scalar coupling, the entire resonance region will be excluded by the projected XENON1T experiment. 

For the first time, we have performed a consistent study of the indirect detection prospects in all four effective Higgs portal models. Using current and future gamma-ray astronomy data, we looked for viable regions of the parameter space that can be probed by the existing or future indirect searches. Below the resonance $m_X \sim m_h/2$, the indirect search limits are weaker than the combined constraints on the relic density and the Higgs invisible width. Depending on the assumed DM density profile, indirect search experiments can provide strong exclusions at higher DM masses, if the model makes up all of the dark matter. The forthcoming CTA experiment will be very useful in searching for signs of DM annihilation. 

In agreement with similar studies performed elsewhere, direct search experiments will continue to provide the strongest limits on the parameter space in all models. The projected XENON1T experiment, for instance, will have the sensitivity to probe TeV-scale DM masses. Since previous studies on fermionic DM models with a pure pseudoscalar interaction have naively applied XENON100 and/or LUX limits without properly taking account of the momentum-suppressed SI cross section, we have rederived these limits by including the proper momentum dependence of the form $q^2/4m_X^2$ in our calculations. The resulting limits, although weak, are important in order to perform a consistent study. In such cases, indirect search experiments are our \emph{only} hope of accessing the higher DM mass range of these models.

When both the indirect and direct search limits are available, as in the portal models we have considered in this paper, a joint observation in both channels is a very realistic possibility at higher WIMP masses. Such detection in multiple experiments would provide a far more robust discovery than a single signal alone. 

For the fermion models, the EFT approximation that we have employed in this paper breaks down when DM annihilations are considered. A proper examination of the fermionic Higgs portal DM in these parameter regions requires a detailed and systematic study of possible UV completions.  We plan to do this in a future paper.

\begin{acknowledgments}
We thank Csaba Balazs, Jim Cline, Alexander Pukhov and Genevi\a`eve B\a'elanger for useful discussions. A.\,B. would like to thank Andre Scaffidi for his assistance during the \lcalc package setup. The work of A.\,B., F.\,R. and A.\,G.\,W. is supported by the ARC Centre of Excellence for Particle Physics at the Terascale (CoEPP) and the Centre for the Subatomic Structure of Matter (CSSM). M.\,W. is supported by the Australian Research Council Future Fellowship FT140100244. P.\,S. is supported by the STFC (Grants No. ST/K00414X/1 and No. ST/N000838/1). 
\end{acknowledgments}

\section*{Appendix: Chiral rotation}
\subsection{Dirac fermion model}
We start with the post-EWSB Lagrangian for the Dirac fermion DM field $\psi$, 
\begin{align*}
  \lagr_{\psi} &= \lagrsm + \ovr{\psi} i\slashed{\partial}\psi - \left[\mu_\psi \ovr{\psi}\psi + \frac{1}{2}\frac{\lambda_{h\psi}}{\Lambda_\psi}\vo^2 \Big(\cos\theta \, \ovr{\psi}\psi + \sin\theta \, \ovr{\psi}i\gamma_5 \psi \Big) \right] \\
  &\hspace{5cm} - \frac{\lambda\indices{_{h\psi}}}{\Lambda_\psi} \, \Big(\cos\theta \, \ovr{\psi}\psi + \sin\theta \, \ovr{\psi}i\gamma_5 \psi \Big) \left(\vo h + \frac{1}{2}h^2 \right).
\end{align*}
The square bracket term in the above expression contains a DM mass-type term and a pseudoscalar coupling of the DM field to the SM Higgs VEV $\vo$. However, it is convenient to remove the latter term ($\propto \sin\theta$) using a field redefinition through a chiral rotation 
\begin{equation}\label{chiralrot}
  \psi \longrightarrow \exp(i\gamma_5 \alpha/2) \psi, \quad \ovr{\psi} \longrightarrow \ovr{\psi} \exp(i\gamma_5 \alpha/2),
\end{equation} 
where $\alpha$ is a real, space-time-independent parameter.

After expanding the exponential in terms of $\cos\alpha/2$, $\sin\alpha/2$ and using the properties of the $\gamma_5$ matrix in the Dirac-representation, the exponentials become
\begin{align*}
  \exp(i\gamma_5 \alpha/2) &= \cos(\gamma_5 \alpha/2) + i\sin(\gamma_5 \alpha/2) \\
  &= \cos(\alpha/2) + i\gamma_5 \sin(\alpha/2),
\end{align*}
where the property $\gamma_5^n = \gamma_5$ ($\mathbb{I}_{\,4\times 4})$ for $n$ odd (even) is used. \\
Under the chiral rotation in Eq.~\eqref{chiralrot}, the term $\ovr{\psi}i\slashed{\partial}\psi$ is invariant since
\begin{align*}
  \ovr{\psi} i\slashed{\partial} \psi &\longrightarrow \ovr{\psi} \exp(i\gamma_5 \alpha/2) i\gamma^\mu \partial_\mu \big(\exp(i\gamma_5 \alpha/2) \psi\big) \\
  &= \ovr{\psi} \exp(i\gamma_5 \alpha/2) \gamma^\mu \exp(i\gamma_5 \alpha/2) i\partial_\mu \psi \\
  &= \ovr{\psi} \exp(i\gamma_5 \alpha/2) \exp(-i\gamma_5 \alpha/2) i\gamma^\mu \partial_\mu \psi \quad  \Big(\{\gamma_5,\gamma^\mu\} = 0\Big) \\ 
  &= \ovr{\psi} i\slashed{\partial} \psi. 
\end{align*}
On the contrary, $\ovr{\psi} \psi$ and $\ovr{\psi}i\gamma_5 \psi$ terms transform to

\begin{minipage}[H]{0.5\textwidth}
\begin{align*}
  \ovr{\psi} \psi &\longrightarrow \ovr{\psi} \exp(i\gamma_5 \alpha/2) \exp(i\gamma_5 \alpha/2) \psi \\
  &= \ovr{\psi} \exp(i\gamma_5 \alpha) \psi \\
  &= \cos\alpha \, \ovr{\psi} \psi + \sin\alpha \, \ovr{\psi} i\gamma_5 \psi
\end{align*}
\end{minipage}
\begin{minipage}[H]{0.5\textwidth}
\begin{align*}
  \ovr{\psi} i\gamma_5 \psi &\longrightarrow \ovr{\psi} \exp(i\gamma_5 \alpha/2) i\gamma_5 \exp(i\gamma_5 \alpha/2) \psi \\
  &= \ovr{\psi} \exp(i\gamma_5 \alpha) i\gamma_5 \psi \\
  &= \cos\alpha \, \ovr{\psi} i\gamma_5 \psi - \sin\alpha \, \ovr{\psi}\psi   
\end{align*}
\end{minipage}
Therefore, terms in the Dirac Lagrangian transform to 
\begin{align*}
  \mu_\psi \ovr{\psi}\psi &\longrightarrow \mu_\psi \cos\alpha \, \ovr{\psi} \psi + \mu_\psi \sin\alpha \, \ovr{\psi}i\gamma_5\psi  \\
  \frac{1}{2}\frac{\lambda_{h\psi}}{\Lambda_\psi} \vo^2 \Big(\cos\theta \, \ovr{\psi}\psi + \sin\theta \, \ovr{\psi}i\gamma_5\psi \Big) &\longrightarrow \frac{1}{2}\frac{\lambda_{h\psi}}{\Lambda_\psi} \vo^2 \Big(\cos\theta \cos\alpha \, \ovr{\psi}\psi + \cos\theta \sin\alpha \, \ovr{\psi}i\gamma_5 \psi \Big. \\
  &\hspace{2cm} \Big. + \sin\theta \cos\alpha \, \ovr{\psi}i\gamma_5\psi - \sin\theta \sin\alpha \, \ovr{\psi}\psi \Big) \\
  &= \frac{1}{2}\frac{\lambda_{h\psi}}{\Lambda_\psi} \vo^2 \Big(\cos\xi \, \ovr{\psi}\psi + \sin\xi \, \ovr{\psi}i\gamma_5 \psi \Big) \\
  \frac{\lambda\indices{_{h\psi}}}{\Lambda_\psi} \, \Big(\cos\theta \, \ovr{\psi}\psi + \sin\theta \, \ovr{\psi}i\gamma_5 \psi \Big) &\longrightarrow \frac{\lambda_{h\psi}}{\Lambda_\psi} \Big(\cos\theta \cos\alpha \, \ovr{\psi}\psi + \cos\theta \sin\alpha \, \ovr{\psi}i\gamma_5 \psi \Big. \\
  &\hspace{2cm} \Big. + \sin\theta \cos\alpha \, \ovr{\psi}i\gamma_5\psi - \sin\theta \sin\alpha \, \ovr{\psi}\psi \Big) \\
  &= \frac{\lambda_{h\psi}}{\Lambda_\psi} \Big(\cos\xi \, \ovr{\psi}\psi + \sin\xi \, \ovr{\psi}i\gamma_5 \psi \Big)
\end{align*}
where we define $\xi \equiv \theta + \alpha$ for future convenience and use the trigonometric identities $ \cos\xi = \cos\theta \cos\alpha - \sin\theta \sin\alpha$ and $\sin\xi = \sin\theta \cos\alpha + \cos\theta \sin\alpha$. Therefore, the mass and interaction terms expand to
\begin{align}
  -\left[\mu_\psi \ovr{\psi}\psi + \frac{1}{2}\frac{\lambda_{h\psi}}{\Lambda_\psi}\vo^2 \Big(\cos\theta \, \ovr{\psi}\psi + \sin\theta \, \ovr{\psi}i\gamma_5 \psi \Big) \right] &\longrightarrow - \Big[\mu_\psi \cos\alpha \, \ovr{\psi}\psi + \mu_\psi \sin\alpha \, \ovr{\psi}i\gamma_5\psi  \Big. \notag \\
  &\hspace{1cm} \Big. + \frac{1}{2}\frac{\lambda_{h\psi}}{\Lambda_\psi} \vo^2 \Big(\cos\xi \, \ovr{\psi}\psi + \sin\xi \, \ovr{\psi}i\gamma_5 \psi \Big)\Big] \notag \\
  &= -\left[\mu_\psi \cos\alpha + \frac{1}{2}\frac{\lambda_{h\psi}}{\Lambda_\psi} \vo^2 \cos\xi \right] \ovr{\psi} \psi \notag \\
  &\hspace{0.5cm} -\left[\mu_\psi \sin\alpha + \frac{1}{2}\frac{\lambda_{h\psi}}{\Lambda_\psi} \vo^2 \sin\xi \right] \ovr{\psi}i\gamma_5 \psi \label{diracMass} \\
  - \frac{\lambda\indices{_{h\psi}}}{\Lambda_\psi} \, \Big(\cos\theta \, \ovr{\psi}\psi + \sin\theta \, \ovr{\psi}i\gamma_5 \psi \Big) \left(\vo h + \frac{1}{2}h^2 \right) &\longrightarrow - \frac{\lambda\indices{_{h\psi}}}{\Lambda_\psi} \, \Big(\cos\xi \, \ovr{\psi}\psi + \sin\xi \, \ovr{\psi}i\gamma_5 \psi \Big) \notag \\
  &\hspace{1cm} \times \left(\vo h + \frac{1}{2}h^2 \right) \label{diracInt}.
\end{align}
If $\alpha = \pi$, a sign change of the mass and interaction terms in the above expressions occurs. We can therefore, without any loss of generality, take $\mu_\psi > 0$ as long as we preserve the relative signs between the terms. 

After the chiral rotation, we demand that the coefficient of $\ovr{\psi}i\gamma_5 \psi$ in Eq.~\eqref{diracMass} vanish in order to go to the real mass basis. This defines a proper chiral rotation and gives a field mass after EWSB in terms of the Lagrangian parameters. Setting the coefficient of $\ovr{\psi}i\gamma_5\psi$ to zero gives
\begin{equation*}
  \mu_\psi \sin\alpha = -\frac{1}{2} \frac{\lambda_{h\psi}}{\Lambda_\psi} \vo^2 \sin\xi = -\frac{1}{2} \frac{\lambda_{h\psi}}{\Lambda_\psi} \vo^2 \left(\cos\theta \sin\alpha + \sin\theta \cos\alpha \right) .
\end{equation*}
Dividing the above expression on both sides by $\cos\alpha$ and rearranging for $\tan\alpha$ gives 
\begin{equation}
  \tan\alpha = \left(-\frac{1}{2} \frac{\lambda_{h\psi}}{\Lambda_\psi}\,\vo^2 \sin\theta \right) \left(\mu_\psi + \frac{1}{2}\frac{\lambda_{h\psi}}{\Lambda_\psi} \, \vo^2 \cos\theta \right)^{-1},
\end{equation}
as stated for the Dirac fermion model in Sec.~\ref{sec2}. Using the expression for $\tan\alpha$, we can determine $\sin^2\alpha$ and $\cos^2 \alpha$ 
\begin{align*}
  \cos^2 \alpha &= \frac{1}{1 + \tan^2 \alpha} = \frac{\left(\mu_\psi + \frac{1}{2}\frac{\lambda_{h\psi}}{\Lambda_\psi}\vo^2 \cos\theta\right)^2}{\left(\mu_\psi + \frac{1}{2}\frac{\lambda_{h\psi}}{\Lambda_\psi}\vo^2 \cos\theta \right)^2 + \left(\frac{1}{2}\frac{\lambda_{h\psi}}{\Lambda_\psi}\vo^2 \sin\theta \right)^2}, \\
  \sin^2 \alpha &= \frac{\tan^2 \alpha}{1 + \tan^2 \alpha} = \frac{\left(\frac{1}{2}\frac{\lambda_{h\psi}}{\Lambda_\psi}\vo^2 \sin\theta\right)^2}{\left(\mu_\psi + \frac{1}{2}\frac{\lambda_{h\psi}}{\Lambda_\psi}\vo^2 \cos\theta \right)^2 + \left(\frac{1}{2}\frac{\lambda_{h\psi}}{\Lambda_\psi}\vo^2 \sin\theta \right)^2}.
\end{align*}
After EWSB and the field rotation, we define the mass $m_\psi$ as the coefficient of the term $-\ovr{\psi}\psi$. Using Eq.~\eqref{diracMass}, $m_\psi$ is given by
\begin{align*}
  m_\psi = \mu_\psi \cos\alpha + \frac{1}{2}\frac{\lambda_{h\psi}}{\Lambda_\psi} \vo^2 \cos\xi &= \mu_\psi \cos\alpha + \frac{1}{2}\frac{\lambda_{h\psi}}{\Lambda_\psi} \vo^2 \Big(\cos\alpha \cos\theta - \sin\alpha \sin\theta \Big), \\
  &= \Big[\mu_\psi + \frac{1}{2}\frac{\lambda_{h\psi}}{\Lambda_\psi} \vo^2 \left(\cos\theta - \tan\alpha \sin\theta \right) \Big]\cos\alpha.
\end{align*}
Making use of the expression for $\tan\alpha$, $m_\psi$ expands to 
\begin{equation*}
  m_\psi = \left[\frac{\left(\mu_\psi + \frac{1}{2}\frac{\lambda_{h\psi}}{\Lambda_\psi} \vo^2 \cos\theta \right)^2 + \left(\frac{1}{2}\frac{\lambda_{h\psi}}{\Lambda_\psi} \vo^2 \sin\theta \right)^2 }{\mu_\psi + \frac{1}{2}\frac{\lambda_{h\psi}}{\Lambda_\psi} \vo^2 \cos\theta} \right]\cos\alpha.
\end{equation*}
Squaring both sides of the above equation and using the relation for $\cos^2 \alpha$ gives the following expression for the physical DM mass:
\begin{align*}
  m_\psi^2 = \left[\frac{\left(\mu_\psi + \frac{1}{2}\frac{\lambda_{h\psi}}{\Lambda_\psi} \vo^2 \cos\theta \right)^2 + \left(\frac{1}{2}\frac{\lambda_{h\psi}}{\Lambda_\psi} \vo^2 \sin\theta \right)^2 }{\mu_\psi + \frac{1}{2}\frac{\lambda_{h\psi}}{\Lambda_\psi} \vo^2 \cos\theta} \right]^2 \cos^2 \alpha &= \frac{\left[\left(\mu_\psi + \frac{1}{2}\frac{\lambda_{h\psi}}{\Lambda_\psi} \vo^2 \cos\theta \right)^2 + \left(\frac{1}{2}\frac{\lambda_{h\psi}}{\Lambda_\psi} \vo^2 \sin\theta \right)^2 \right]^2}{\left(\mu_\psi + \frac{1}{2}\frac{\lambda_{h\psi}}{\Lambda_\psi}\vo^2 \cos\theta \right)^2 + \left(\frac{1}{2}\frac{\lambda_{h\psi}}{\Lambda_\psi}\vo^2 \sin\theta \right)^2} \\
  &= \left(\mu_\psi + \frac{1}{2}\frac{\lambda_{h\psi}}{\Lambda_\psi}\vo^2 \cos\theta \right)^2 + \left(\frac{1}{2}\frac{\lambda_{h\psi}}{\Lambda_\psi}\vo^2 \sin\theta \right)^2.
\end{align*}
The signs of $m_\psi$, $\cos\alpha$ and $\sin\alpha$ are common; hence, we choose the common sign to be ``+'' for $m_\psi$, $\cos\alpha = + \sqrt{\cos^2 \alpha}$ and $\sin\alpha = \sqrt{\sin^2 \alpha}$. Therefore, the physical DM mass after EWSB is 
\begin{equation}\label{massmF}
m_\psi = \sqrt{\left(\mu_\psi + \frac{1}{2}\frac{\lambda_{h\psi}}{\Lambda_\psi}\vo^2 \cos\theta \right)^2 + \left(\frac{1}{2}\frac{\lambda_{h\psi}}{\Lambda_\psi}\vo^2 \sin\theta \right)^2}.
\end{equation}
With the sign convention chosen above, the expressions for $\cos\xi$ and $\sin\xi$ reduce to 
\begin{align*}
  \cos\xi = \cos\alpha \left(\cos\theta - \sin\theta \tan\alpha \right) &= \frac{\mu_\psi \cos\theta + \frac{1}{2}\frac{\lambda_{h\psi}}{\Lambda_\psi} \vo^2}{\sqrt{\left(\mu_\psi + \frac{1}{2}\frac{\lambda_{h\psi}}{\Lambda_\psi}\vo^2 \cos\theta \right)^2 + \left(\frac{1}{2}\frac{\lambda_{h\psi}}{\Lambda_\psi}\vo^2 \sin\theta \right)^2}} \frac{}{}  \\
  &= \frac{\mu_\psi}{m_\psi} \left(\cos\theta + \frac{1}{2}\frac{\lambda_{h\psi}}{\Lambda_\psi} \frac{\vo^2}{\mu_\psi} \right) \\[2mm]
  \sin\xi = \cos\alpha \left(\sin\theta + \cos\theta \tan\alpha \right) &= \frac{\mu_\psi \sin\theta}{\sqrt{\left(\mu_\psi + \frac{1}{2}\frac{\lambda_{h\psi}}{\Lambda_\psi}\vo^2 \cos\theta \right)^2 + \left(\frac{1}{2}\frac{\lambda_{h\psi}}{\Lambda_\psi}\vo^2 \sin\theta \right)^2}} \\
  &= \frac{\mu_\psi}{m_\psi} \sin\theta
\end{align*}
where we have used Eq.~\eqref{massmF} to simplify the expressions. Therefore, the final Lagrangian after EWSB and field redefinition is given by
\begin{align*}
  \lagr_\psi = \lagrsm + \ovr{\psi} i\slashed{\partial} \psi - m_\psi \ovr{\psi} \psi - \frac{\lambda_{h\psi}}{\Lambda_\psi} \Big[\cos\xi \, \ovr{\psi} \psi + \sin\xi \, \ovr{\psi}i\gamma_5 \psi \Big] \left(\vo h + \frac{1}{2}h^2 \right),
\end{align*}
where 
\begin{align*}
	\cos\xi &= \frac{\mu_\psi}{m_\psi} \left(\cos\theta + \frac{1}{2}\frac{\lambda_{h\psi}}{\Lambda_\psi} \frac{\vo^2}{\mu_\psi} \right), \quad \sin\xi = \frac{\mu_\psi}{m_\psi}\sin\theta, \\  	
	m_\psi &= \sqrt{\left(\mu_\psi + \frac{1}{2}\frac{\lambda_{h\psi}}{\Lambda_\psi}\vo^2 \cos\theta \right)^2 + \left(\frac{1}{2}\frac{\lambda_{h\psi}}{\Lambda_\psi}\vo^2 \sin\theta \right)^2}.
\end{align*}

\subsection{Majorana fermion model}
Consider the Majorana fermion DM field $\chi$ Lagrangian after EWSB,
\begin{align*}
  \lagr_{\chi} &= \lagrsm + \frac{1}{2}\ovr{\chi} i\slashed{\partial}\chi - \frac{1}{2}\left[\mu_\chi \ovr{\chi}\chi + \frac{1}{2}\frac{\lambda_{h\chi}}{\Lambda_\chi}\vo^2 \Big(\cos\theta \, \ovr{\chi}\chi + \sin\theta \, \ovr{\chi}i\gamma_5 \psi \Big) \right] \notag \\
  &\hspace{5cm} - \frac{1}{2}\frac{\lambda\indices{_{h\chi}}}{\Lambda_\chi} \, \Big(\cos\theta \, \ovr{\chi}\chi + \sin\theta \, \ovr{\chi}i\gamma_5 \chi \Big) \left(\vo h + \frac{1}{2}h^2 \right).
\end{align*}
With the conventional factor of $1/2$ in front of each Majorana field bilinear, the form of the Majorana model Lagrangian after EWSB and chiral rotation is analogous to that of the Dirac model. 
The proper chiral rotation is again given by
\begin{equation*}
  \tan\alpha = \left(-\frac{1}{2} \frac{\lambda_{h\chi}}{\Lambda_\chi}\,\vo^2 \sin\theta \right) \left(\mu_\chi + \frac{1}{2}\frac{\lambda_{h\chi}}{\Lambda_\chi} \, \vo^2 \cos\theta \right)^{-1}, 
\end{equation*}
which leads to a real, physical DM mass of
\begin{equation*}
  m_\chi = \sqrt{\left(\mu_\chi + \frac{1}{2}\frac{\lambda_{h\chi}}{\Lambda_\chi}\vo^2 \cos\theta \right)^2 + \left(\frac{1}{2}\frac{\lambda_{h\chi}}{\Lambda_\chi}\vo^2 \sin\theta \right)^2}.
\end{equation*}
Finally, the Majorana fermion Lagrangian after EWSB and chiral rotation is given by
\begin{align*}
  \lagr_\chi = \lagrsm + \frac{1}{2}\ovr{\chi} i\slashed{\partial} \chi - \frac{1}{2}m_\chi \ovr{\chi} \chi - \frac{1}{2}\frac{\lambda_{h\chi}}{\Lambda_\chi} \Big[\cos\xi \, \ovr{\chi} \chi + \sin\xi \, \ovr{\chi}i\gamma_5 \chi \Big] \left(\vo h + \frac{1}{2}h^2 \right),
\end{align*}
where 
\begin{align*}
  \cos\xi &= \frac{\mu_\chi}{m_\chi} \left(\cos\theta + \frac{1}{2}\frac{\lambda_{h\chi}}{\Lambda_\chi} \frac{\vo^2}{\mu_\chi} \right), \quad \sin\xi = \frac{\mu_\chi}{m_\chi}\sin\theta, \\
  m_\chi &= \sqrt{\left(\mu_\chi + \frac{1}{2}\frac{\lambda_{h\chi}}{\Lambda_\chi}\vo^2 \cos\theta \right)^2 + \left(\frac{1}{2}\frac{\lambda_{h\chi}}{\Lambda_\chi}\vo^2 \sin\theta \right)^2}.
\end{align*}
as stated for the Majorana fermion model in Sec.~\ref{sec2}.

\bibliographystyle{jhepbibstyle}
\bibliography{paper-A-prd}

\end{document}